\documentclass[%
reprint,
superscriptaddress,
amsmath,amssymb,
aps,
pra,
longbibliography,
]{revtex4-2}

\usepackage{graphicx}
\usepackage{dcolumn}
\usepackage{bm}

\usepackage{physics}
\usepackage{braket}
\usepackage{siunitx}
\usepackage{booktabs}
\usepackage{floatrow}
\usepackage[shortlabels]{enumitem}
\usepackage[version=2]{mhchem}
\usepackage{upgreek,ulem}
\usepackage{subcaption}
\usepackage{xcolor}
\usepackage{lineno}

\usepackage[
	colorlinks=true,
    citecolor=blue,
    linkcolor=blue,
    filecolor=blue,      
    urlcolor=blue,	
    breaklinks=true
	]{hyperref}
\newcommand{\al}{\alpha}
\newcommand{\be}{\beta}
\newcommand{\p}{\psi}

\newcommand{\REF}{\Phi_{\mathrm{ref}}}
\newcommand{\sigg}{\upsigma_{\text{g}}}
\newcommand{\sigu}{\upsigma_{\text{u}}}
\newcommand{\bsigg}{\bar{\upsigma}_{\text{g}}}
\newcommand{\bsigu}{\bar{\upsigma}_{\text{u}}}
\newcommand{\piux}{\uppi_{\text{u,x}}}
\newcommand{\piuy}{\uppi_{\text{u,y}}}
\newcommand{\pigx}{\uppi_{\text{g,x}}}
\newcommand{\pigy}{\uppi_{\text{g,y}}}
\newcommand{\bpigx}{\bar{\uppi}_{\text{g, x}}}

\newcommand{\bpigy}{\bar{\uppi}_{\text{g, y}}}

\newcommand{\upx}{\text{x}}
\newcommand{\upy}{\text{y}}
\newcommand{\upz}{\text{z}}

\newcommand{\vn}{\ket{\mathcal{O}_{0, 0}^{N, 1} } }
\newcommand{\vsix}{\ket{\mathcal{O}_{0, 0}^{6, 1} } }
\newcommand{\vfour}{\ket{\mathcal{O}_{0, 0}^{4, 1} } }
\newcommand{\vtwo}{\ket{\mathcal{O}_{0, 0}^{2, 1} } }

\newcommand{\HO}{\hat{H}_0}
\newcommand{\Hf}{\hat{H}_\text{F}}
\newcommand{\bth}{\bm{\theta}}
\newcommand{\bTh}{\bm{\Theta}}
\newcommand{\bC}{\bm{C}}

\newcommand{\Uref}{U_\text{ref}}
\newcommand{\Uans}{U_\text{ansatz}}

\UseRawInputEncoding
\begin{document}


\title{Spin coupling is all you need:\\ Encoding  strong electron correlation in molecules on quantum computers}

\author{Daniel~Marti-Dafcik}
\email{dmartidafcik@gmail.com}
\affiliation{Physical and Theoretical Chemistry Laboratory, University of Oxford, South Parks Road, Oxford, OX1 3QZ, U.K.}
\author{Hugh~G.~A.~Burton}
\affiliation{Yusuf Hamied Department of Chemistry, University of Cambridge, Lensfield Road, Cambridge, CB2 1EW, U.K.}
\author{David~P.~Tew}
\affiliation{Physical and Theoretical Chemistry Laboratory, University of Oxford, South Parks Road, Oxford, OX1 3QZ, U.K.}

\date{\today}

 \begin{abstract}
	The performance of quantum algorithms for eigenvalue problems, such as computing Hamiltonian spectra,
	depends strongly on the overlap of the initial wavefunction and the target eigenvector.
	In a basis of Slater determinants, the representation of energy eigenstates of systems with $N$ strongly correlated electrons
	requires a number of determinants that scales exponentially with $N$.
	On classical processors, this restricts simulations to systems where $N$ is small.
	Here, we show that quantum computers can efficiently simulate strongly correlated molecular systems
	by directly encoding the dominant entanglement structure in the form of spin-coupled initial states.
	This avoids resorting to expensive classical or quantum state preparation heuristics and instead exploits
	symmetries in the wavefunction.
	We provide quantum circuits for deterministic preparation of a family of spin eigenfunctions with ${N \choose N/2}$ Slater determinants with depth $\mathcal{O}(N)$ and $\mathcal{O}(N^2)$ local gates.
	Their use as highly entangled initial states in quantum algorithms reduces the total runtime
	of quantum phase estimation and related fault-tolerant methods by orders of magnitude.
	Furthermore, we assess the application of spin-coupled wavefunctions as initial states for a range of
	heuristic quantum algorithms, namely the variational quantum eigensolver, adiabatic state preparation,
	and different versions of quantum subspace diagonalization (QSD) including QSD based on real-time-evolved states.
	We also propose a novel QSD algorithm that exploits states obtained through adaptive quantum eigensolvers.
	For all algorithms, we demonstrate that using spin-coupled initial states drastically reduces the quantum resources required
	to simulate strongly correlated ground and excited states.
	Our work provides a crucial component for enabling scalable quantum simulation of classically challenging electronic systems.
\end{abstract}

\maketitle
\raggedbottom

\section{Introduction}\label{sec:intro}

Computing the low-lying eigenstates and energies of electronic Hamiltonians remains a fundamental challenge in
the physical sciences and scientific computing.
Despite the success of classical computational approaches for quantum chemistry in simulating a wide range of molecules,
their application to systems exhibiting strong electron correlation is limited due to the difficulty of efficiently representing highly entangled quantum states.\cite{Evangelista2018}

Quantum computers can generate and transform vectors whose dimension scales
exponentially in the number of qubits with remarkable efficiency.\cite{Gilyen2019}
This suggests that they might provide a solution to the curse of dimensionality. The reality is more nuanced.
Despite the plethora of algorithms developed for computing energy eigenvalues or preparing eigenstates of many-body systems,
complexity-theoretic results rule out exponential quantum speedups for worst-case versions of the electronic Schr{\"o}dinger equation.\cite{Kempe2006, OGorman2022}
Although this does not prohibit practical speedups for realistic physical systems,
a fundamental problem lies at the core of quantum computing for electronic structure:
the performance of nearly all quantum algorithms for eigenvalue problems strongly depends on the accuracy of the initial state.

A random vector in the many-electron Hilbert space is expected to have exponentially small overlap with any exact eigenstate.\cite{Kohn1999, VanVleck1936}
Therefore, the success of quantum algorithms relies on finding an initial state which approximately contains the structure of the true eigenstate.
This can be achieved by computing approximate wavefunctions through classical heuristics, such as Hartree--Fock, configuration interaction,\cite{Helgaker}
or density matrix renormalization group (DMRG) approaches,\cite{Chan2011} and loading such states on quantum hardware.\cite{Tubman2018}
On fault-tolerant quantum hardware,
quantum heuristics---such as the variational quantum eigensolver (VQE),\cite{Peruzzo2014, McClean2016, Grimsley2019, Higgott2019, Huggins2020b, Cai2020, Yordanov2021, Vanstraaten2021,
Anselmetti2021, Hu2022a, Anschuetz2022, Arrazola2022, Gonthier2022, Baek2023, DCunha2023, Gustiani2023, Burton2023, Dalton2024}
adiabatic state preparation (ASP),\cite{Farhi2000, Aspuru-Guzik2005, Veis2014, Reiher2017, Albash2018, Kremenetski2021, Sugisaki2022a, Lee2023}
and quantum subspace diagonalization
(QSD)\cite{McClean2017, Colless2018, Motta2019, Parrish2019b, Huggins2020b, Stair2020a, Seki2021, Klymko2022, Epperly2022, Cortes2022, Shen2023, Stair2023, Kirby2023, Kirby2024}
---could be used  to prepare initial states for quantum phase estimation,
a method for which the runtime rigorously depends on the overlap between the initial state and the target eigenstate.\cite{Kitaev1995, Abrams1999, Nielsen2010}

The challenge is that, for systems where such heuristics are accurate, the problem can often be solved entirely using classical algorithms (to sufficient accuracy).
This casts doubts on the advantage of using quantum computers over classical machines for quantum chemistry.\cite{Lee2023}
To obtain quantum speedups, the approximate initial state found through heuristics must be of insufficient accuracy for chemistry applications.
At the same time, its structure must be such that further refining the wavefunction through classical algorithms
is hard, while encoding it as an initial state for quantum algorithms is easy.
In this work, we tackle the strong correlation problem in quantum chemistry by leveraging a recently-developed classical heuristic
that satisfies these requirements and is ideally suited for initial state preparation on quantum computers.\cite{Marti-Dafcik2024a}

To understand what features make an initial state preparation method
advantageous for quantum computation,
consider the success and limitations of scalable classical algorithms. All polynomially-scaling methods for quantum chemistry
rely on identifying qualitatively accurate, yet approximate, initial states with a simple, compact wavefunction description.
Restricted Hartree--Fock theory provides good initial states for systems with weak electron correlation,
because it accurately encodes the mean-field character of delocalized molecular wavefunctions.\cite{Tew2007}
The many-body Hartree--Fock state is simply an antisymmetrized product of delocalized molecular orbitals,
which corresponds to a single Slater determinant.
Since this state has a compact representation in the single-particle basis of Hartree--Fock orbitals,
one can expand around it in a controlled manner, e.g. with methods based on many-body perturbation theory,
to improve the state without incurring exponential cost.\cite{Bartlett2007}

For strongly correlated systems, the Hartree--Fock state is inaccurate because the molecular orbital picture underpinning it breaks down.
Even a qualitatively accurate wavefunction for such systems requires a superposition of Slater determinants that scales exponentially with the number of strongly correlated electrons,
regardless of the choice of single-particle basis.
In such scenarios, no efficient classical heuristics exist,
and one resorts to brute-force algorithms which in general scale exponentially as they require
a linear combination of all the relevant Slater determinants.\cite{Booth2009, Holmes2016, Tubman2016}

The high dimension of such states also poses a challenge for quantum algorithms due to the initial state dependency.
Suppose that an approximate wavefunction is found through a state-of-the-art classical algorithm,
such as selective configuration interaction or DMRG.
Even then, its use in quantum algorithms is inefficient,
because preparing a superposition with arbitrary coefficients through a quantum circuit
requires a number of steps (gates) proportional to the number of basis states (determinants).\cite{Tubman2018, Low2018}
Quantum heuristics could in principle offer an alternative, but often also suffer in strongly correlated regimes.\cite{Hu2022a}
Therefore, there appears to be a trade-off between the accuracy of the initial state used in quantum algorithms and the cost of state preparation.\cite{Wang2022, Gratsea2022, Gratsea2024}
Fundamentally, the challenge is that quantum computers need to exploit structure in the problem,
and while such structure is present in Hartree--Fock states,
it is washed away in strongly correlated wavefunctions due to the brute-force nature of algorithms that are used for their approximation.

In this work, we present a heuristic state preparation method that solves the initial state problem for a range of strongly correlated molecules.
This relies on our recently-developed generalized molecular orbital theory,\cite{Marti-Dafcik2024a}
where we showed that molecular electronic wavefunctions that arise during the stretching of chemical bonds
are often highly structured and can be approximated to high accuracy through few spin-coupled states.
These spin-coupled states can be deduced from chemical intuition and spatial-spin symmetry arguments,
and their representation is given directly from the standard Clebsch--Gordan coefficients.\cite{Pauncz1979}
This challenges the common notion that strongly correlated states are inevitably complex.
While they might be highly-dimensional when expressed as vectors expanded in a single-particle basis,
they encode a relatively small amount of information.

Here, we enable their efficient application in quantum algorithms by providing quantum circuits
that prepare a family of spin-coupled states with ${N \choose N/2}$ determinants in depth $\mathcal{O}(N)$ and using $\mathcal{O}(N^2)$ gates.
We achieve this by exploiting the symmetry structure in spin-coupled wavefunctions and connecting them to Dicke states,
a well-known family of entangled states.\cite{Bartschi2019} This approach avoids the exponential scaling of generic,
black-box state preparation methods.\cite{Tubman2018, Low2018}
This is possible because our spin-coupled molecular orbital theory assigns a bespoke set of orbitals to each reference state,
ensuring each wavefunction has a highly symmetric and compact representation.
The states can then be rotated to a common basis using standard, linear-depth quantum circuits.\cite{Kivlichan2018}

We numerically assess the application of this state preparation method in VQE, ASP, and QSD, where the quantum algorithms are initialized with spin-coupled states.
We also propose a new QSD algorithm, ADAPT-QSD, which is of interest in its own right.
It builds a subspace from states obtained through adaptive quantum eigensolvers such as ADAPT-VQE.\cite{Grimsley2019}
For all algorithms, we demonstrate that the use of spin-coupled initial states greatly reduces the quantum resources (circuit depth and gate counts) and the number of degrees of freedom (variational parameters) required to achieve a given accuracy, compared to using the Hartree--Fock reference. This confirms our spin-coupled framework as a low-cost approach for improving the performance of heuristic quantum algorithms.
 
We highlight QSD as a class of algorithms that can most strongly benefit from access to different reference states,
in particular for \textit{multiconfigurational} eigenstates i.e. those for which multiple dominant configurations (e.g. as provided through multiple initial states)
are required for an accurate wavefunction description.
These are generally hard to tackle using classical algorithms and therefore a promising target for quantum computation.
Combining QSD with spin-coupled states also allows computation of excited state energies at low cost.
 
Finally, we analyze the state preparation question in the context of fault-tolerant computation of electronic structure based on quantum phase estimation.
We consider the number of non-Clifford gates required to prepare initial states with high ground state overlap
for systems with many (up to $35$) spin-coupled electrons such as \ce{FeMoCo}.\cite{Reiher2017, Lee2021, Lee2023}
Our extrapolated gate count estimates suggest that one could greatly reduce the state preparation cost by
directly encoding the entanglement due to spin coupling through a circuit similar to the ones presented here,
compared to preparation of states obtained from classical black-box algorithms such as DMRG or selected CI.
This would enable efficient fault-tolerant quantum simulation of such strongly correlated molecules---precisely
the systems for which classical methods are most likely to remain insufficient.

Overall, our work provides a scalable framework with the necessary ingredients required to unlock the unique power of quantum computers for challenging chemical systems.

We note that we have only studied strongly correlated electrons in molecules, and it is unclear how one would apply similar concepts to solid state systems.
While the state preparation circuits can in principle be used for any electronic or spin-$\frac{1}{2}$ system,
our choice of a few bespoke CSFs is unlikely to provide a solution to the strong correlation problem in solid state materials.
Molecules consist of a relatively small number of electrons, which makes it possible to identify
the spin coupling within and across different subsystems, as we show below.


In Sec.~\ref{sec:background}, we introduce our wavefunction notation and encoding and the background on spin-coupled states.
Most of the content appeared in our previous work, Ref.~\citenum{Marti-Dafcik2024a}, in a language tailored to chemists; here, we present these concepts to a general quantum science audience.
In Sec.~\ref{sec:overview} we present a brief overview of our main results.
In Sec.~\ref{sec:qcircs}, we provide the quantum circuits for preparing spin eigenfunctions.
In Sec.~\ref{sec:heuristic}, we discuss the numerical results obtained from classical simulations of quantum algorithms, and introduce the ADAPT-QSD algorithm.
In Sec.~\ref{sec:fault_tolerant}, we analyze the initial state preparation task within the longer-term, fault-tolerant quantum computing context.

\section{Background and conceptual framework}\label{sec:background}

\subsection{Encoding and notation for qubit states}\label{ssec:notation_states}
We work with the second-quantized representation, where basis states are antisymmetric product states (Slater determinants)
in a fermionic Fock space and the antisymmetry is manifest through anticommuting operator algebra.
Using the Jordan-Wigner mapping,\cite{Jordan1928} an electronic wavefunction of $2M$ spin-orbitals ($M$ spatial orbitals) can be mapped to a system of $2M$ qubits, where each qubit represents the occupation of a spin-orbital ($0$ means unoccupied, $1$ means occupied). Every computational basis state $\ket{j}$ uniquely encodes a Slater determinant represented as
\begin{equation}
	\ket{j} = \bigotimes_{i=1}^{M} \Big( \ket{f_{i_\al}}\otimes \ket{f_{i_\be}} \Big),
	\label{eqn:standard}
\end{equation}
where $f_{i_\al}, f_{i_\be} \in \{0, 1 \}$ represent the occupation of the $\al$ (spin-up) and $\be$ (spin-down) spin-orbital corresponding to the $i$-th spatial orbital, and the Hamming weight (number of $1$s) is the number of electrons. We define the qubit ordering such that a spin-up orbital is followed by the corresponding spin-down orbital.
This convention allows us to introduce a more compact notation:
\begin{equation}
	\ket{j} = \bigotimes_{i=1}^{M} \ket{g_i},
	\label{eqn:compact}
\end{equation}
where $g_i \in \{0, 2, \al, \be \}$ is the occupation of a spatial orbital: $0$ if unoccupied, $2$ if doubly occupied, $\al$ or $\be$ if singly occupied by a spin-up or spin-down electron. We will typically drop the tensor product i.e. $\ket{g_i}\otimes \ket{g_j} =: \ket{g_i g_j}$, and often write products of $K$ identical single-qubit states as $\ket{g_i}^K$. The compact notation (Eq.~\eqref{eqn:compact}) relates to the standard notation (Eq.~\eqref{eqn:standard}) as follows: $\ket{0} \rightarrow \ket{00}$, $\ket{\al} \rightarrow \ket{10}$, $\ket{\beta} \rightarrow \ket{01}$, $\ket{2} \rightarrow \ket{11}$.
Since this notation only specifies the occupation of each orbital, we must specify the nature and ordering of the orbitals separately. We will do so through an ordered list which we define by curly brackets $\{ \phi_1, \phi_2, ... \}$.

\subsection{Discovering and describing strong electron correlation in molecules through symmetry}\label{ssec:n2_diss}
\begin{figure*}
	\begin{minipage}{0.61\columnwidth}
		\begin{subfigure}[t]{\textwidth}
			\centering
			\includegraphics[width=\textwidth]{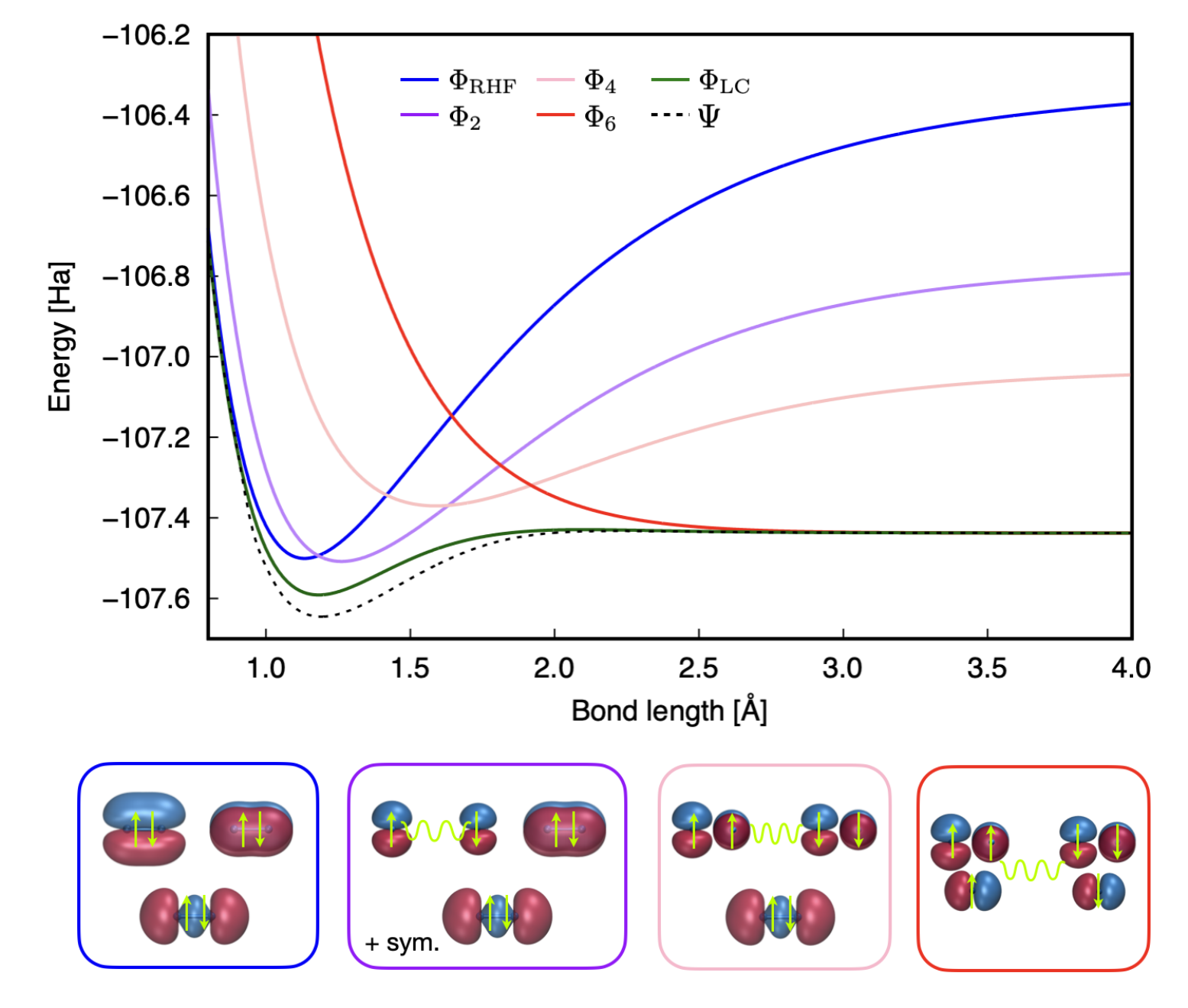}
			\caption{Energies and configurations of the valence CSFs.}
			\label{fig:n2_mainfig}
		\end{subfigure}
	\end{minipage}
	\begin{minipage}{0.38\columnwidth}
		\begin{subfigure}[t]{\textwidth}
			\centering
			\includegraphics[width=\textwidth]{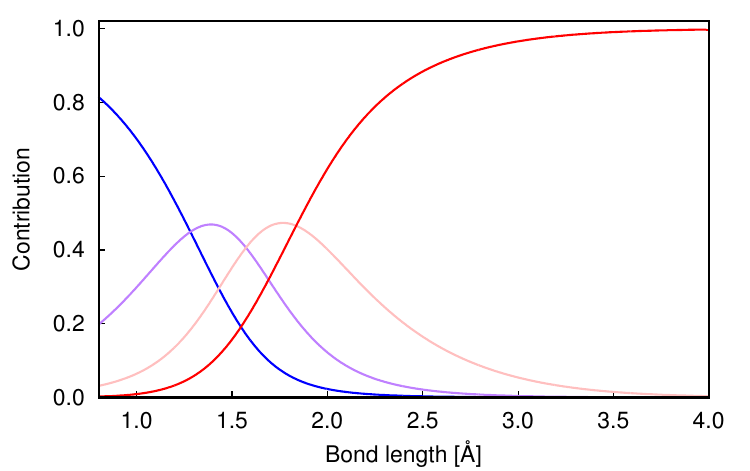}
			\caption{Coefficients of each CSF in $\Phi_\text{LC}$.}
			\label{fig:n2_csf_lin_comb}
		\end{subfigure}
		\begin{subfigure}{\textwidth}
			\centering
			\includegraphics[width=\textwidth]{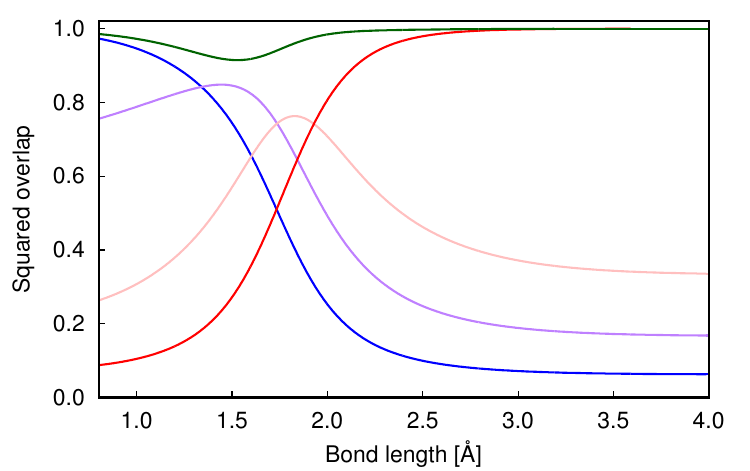}
			\caption{Squared overlaps $\abs{\braket{\Psi|\Phi_{\mathrm{i}}}}^2$.}
			\label{fig:n2_csf_fid}
		\end{subfigure}
	\end{minipage}
	\centering
	\caption{The dissociation of \ce{N2} in a STO-3G basis in the valence space of six electrons in six orbitals (twelve qubits) using four configuration state functions (CSFs) $\Phi_i$, $i=0, 2, 4, 6$, as predicted by generalized MO theory. Here, $\Phi_0$ is the RHF mean-field state, $\Phi_2$, $\Phi_4$, $\Phi_6$ are CSFs with two, four, and six spin-coupled (strongly correlated) electrons $\Psi$ is the exact ground state. The figure is reproduced with permission from Ref. \citenum{Marti-Dafcik2024a}. \label{fig:n2}}
\end{figure*}

Common examples of strongly correlated electronic wavefunctions in molecules include electronic eigenstates at dissociation.
Here, we consider the nitrogen molecule, which illustrates the challenges of both classical\cite{Booth2009, Thom2010} and quantum\cite{Lee2019b, Kremenetski2021, Hu2022a} algorithms.

In Ref.~\citenum{Marti-Dafcik2024a}, we showed that the ground state wavefunction can be represented in terms of a few states
with both mean-field and strongly-correlated character.
Each state encodes a unique entanglement pattern that corresponds to a clear physical motif.
Bonding is manifest by electrons that doubly occupy delocalized molecular orbitals,
and strong correlation corresponds to electrons occupying open-shell, localized orbitals that couple through their spin angular momentum.
A superposition of four reference states yields an accurate wavefunction approximation at any nuclear geometry (Fig.~\ref{fig:n2}),
without requiring a large number of variational parameters.
Below, we discuss the derivation of the four reference states.

In a minimal STO-3G basis, the Restricted Hartree--Fock method yields the following set of delocalized molecular orbitals: $\{ 1\sigg, 1\sigu, 2\sigg, 2\sigu, 3\sigg, 1\piux, 1\piuy, 1\pigx, 1\pigy, 3\sigu \}$.
We use a Hamiltonian that \textit{freezes} the lowest four orbitals and therefore only work with the $1\uppi$-orbitals and the $3\upsigma$ valence orbitals.\footnote{The state of the frozen core orbitals was explicitly represented as $\ket{\mathcal{C}}$ in Ref.~\citenum{Marti-Dafcik2024a}, but we drop it in this paper for compactness.}
This yields a system of six electrons in six orbitals that can be mapped to twelve qubits. We drop the shell indices from the specification of the orbitals and therefore the delocalized orbitals considered here are: $\{\sigg, \piux, \piuy, \pigx, \pigy, \sigu \}$.
In this basis, the RHF state doubly occupies the three lowest orbitals (Fig. \ref{fig:n2_mainfig}, blue), given by
\begin{equation}
	\ket{\Phi_{\mathrm{RHF}}} = \ket{222000}
\label{eqn:rhf_n2}
\end{equation}
using the compact notation introduced in Eq.~\eqref{eqn:compact}.

Although a rigorous definition of electron correlation remains an open research question,\cite{Aliverti-Piuri2024}
one natural definition is to take the view that any product state in second quantization is uncorrelated. 
In this sense, the RHF state in Eq.~\eqref{eqn:rhf_n2} is uncorrelated. This is consistent with it being a
fermionic mean-field state, and with the widely adopted definition of the correlation energy
$E_\mathrm{corr} = E - E_\mathrm{RHF}$. In this paradigm, correlation is therefore a 
measure proportional to the deviation of an entangled quantum state from a product state, using the 
Hartree--Fock orbitals as the prescribed single-particle basis.

Around the equilibrium bond length, the RHF state is a relatively accurate approximation of the true ground state (Fig.~\ref{fig:n2}, blue).
This reflects the success of molecular orbital theory in chemistry, which describes bonding as consisting of electrons occupying
delocalized orbitals, a useful model for many molecules at equilibrium.

However, this model breaks down at dissociation. The electronic wavefunction exhibits strong correlation effects, as reflected in
the poor energy of the RHF state (Fig.~\ref{fig:n2_mainfig}) and its low overlap  with the exact ground state (Fig.~\ref{fig:n2_csf_fid}).
In the basis of RHF orbitals, the exact singlet ground state is a superposition of 44 Slater determinants of similar weights.
Classical methods encounter challenges in such situations since the wavefunction is not compact in the RHF basis.
Quantum algorithms inherit these challenges if the RHF state is used as the initial state, as all the entanglement needs to be \textit{discovered} throughout the quantum computation, e.g. by optimizing the ansatz formed by a parametrized quantum circuit in a variational quantum algorithm.

Fortunately, many of the low-energy eigenstates that appear in nature are highly structured, even if they can be highly entangled.
Approximations thereof can be obtained from simple symmetry considerations.

Let us now consider the wavefunction for \ce{N2} at dissociation. The wavefunction for the dissociated state is not the direct product of the atomic states,
because the local spins of the dissociated atoms are coupled and the wavefunction is entangled across the two atomic subsystems.
The ground state of each isolated nitrogen atom is well-understood from basic chemical principles:
the three valence electrons repel each other and therefore form the maximally antisymmetric spatial wavefunction,
which maximizes the average distance between electrons. Due to the overall antisymmetry requirement
for fermionic wavefunctions,
the spin wavefunction must be fully symmetric. The wavefunction for the atom can therefore be expressed as a quartet (total spin $s=3/2$) that ferromagnetically couples three spin-$1/2$ particles (Fig.~\ref{fig:n2_mainfig}, red).
This spin-coupled state can be determined directly from the Clebsch--Gordan coefficients for angular momentum coupling:
\begin{equation}
		\ket{3, \frac{3}{2}} = 
	\begin{cases}
		\ket{\frac{3}{2}, \frac{3}{2}} = \ket{\al \al \al} \\
		\ket{\frac{3}{2}, \frac{1}{2}} = \frac{1}{\sqrt{3}} \big( \ket{\al \al \be} + \ket{\al \be \al } + \ket{\be \al \al }  \big) \\
		\ket{\frac{3}{2}, -\frac{1}{2}} = \frac{1}{\sqrt{3}} \big( \ket{\al \be \be} + \ket{\be \al \be} + \ket{\be \be \al }  \big) \\
		\ket{\frac{3}{2}, -\frac{3}{2}} = \ket{\be \be \be},
	\end{cases}
	\label{eqn:3e_quartet}
\end{equation}
where on the left we have used the notation $\ket{n, s}$ for the particle and spin quantum number,
in the center we have used $\ket{s, m_s}$ for the total spin and spin projection quantum numbers $s$ and $m_s$,
and on the right we use $\alpha$ and $\beta$ to represent the state of each site as either spin-up or spin-down.

The representation on the right hand side of Eq. \ref{eqn:3e_quartet} is equivalent to electronic Slater determinants with singly-occupied spatial orbitals using the compact notation for fermions/qubits from Eq.~\eqref{eqn:compact},
and assuming a single-particle basis of localized orbitals.
These localized orbitals can be obtained from the RHF orbitals for \ce{N_2} through a trivial unitary transformation of the molecular orbitals:
$\upz_L = \frac{1}{\sqrt{2}}(\sigg + \sigu)$, $\upz_R = \frac{1}{\sqrt{2}}(\sigg - \sigu)$, $\upx_L = \frac{1}{\sqrt{2}}(\piux + \pigx)$, $\upx_R = \frac{1}{\sqrt{2}}(\piux - \pigx)$, $\upy_L = \frac{1}{\sqrt{2}}(\piuy + \pigy)$, $\upy_R = \frac{1}{\sqrt{2}}(\pigy - \pigy)$.
Here, the delocalized RHF orbitals have been transformed to a set of atomic-like, mutually orthogonal molecular orbitals $x, y, z$ localized on the \textit{left} ($L$) or \textit{right} ($R$) atom.

In the dissociation limit, the Coulomb force between electrons localized on different atoms is zero. Thus, the \ce{N_2} ground state can be any state that couples the two atomic quartets with proper quantum numbers.
However, as the interatomic distance $R$ decreases, singlet-coupled electrons can delocalize across the bond, which reduces the kinetic energy.
This one-body delocalization lifts the degeneracy between eigenstates and the ground state requires that the two three-electron subsystems couple to a six-electron singlet.
This state, which we will refer to as $\ket{\Phi_6} = \vsix$, can also be determined from the Clebsch--Gordan coefficients and corresponds to a superposition
of 20 Slater determinants in the localized orbital basis (Eq.~\eqref{eqn:vcsf_n6} in Appendix~\ref{apdx:spin_eigenfunctions}).
In contrast to the mean-field RHF state, this spin-coupled wavefunction is exact at dissociation but very inaccurate at shorter bond lengths (Fig.~\ref{fig:n2}).
Precisely due to its symmetry origin, the state at dissociation is predictable despite being strongly correlated.
It has relatively low Kolmogorov complexity in the sense that few steps are required to specify it:
one must simply define the orbitals involved and their spin coupling pattern.
From this, the coefficients of each basis state can be computed through the standard formula for Clebsch--Gordan coefficients.

As shown in Ref.~\citenum{Marti-Dafcik2024a}, the same is true for the dissociation of a range of molecules including multiple diatomics, water, and hydrogen clusters:
in the limit of long bond length, they can all be described by a single spin-coupled wavefunction, whose general form we will define below.
For $N$ strongly correlated electrons, the spin-coupled state $\vn$ is a superposition of ${N \choose N/2}$ Slater determinants (where $N=6$ for \ce{N_2}).
This exponential scaling in the number of electrons means that it is hard to exploit such wavefunctions as reference states in classical algorithms, as it is difficult to refine highly-entangled reference states e.g. through pertubation theory or coupled cluster methods.\cite{Evangelista2018}

At intermediate bond lengths, neither the RHF state nor the fully spin-coupled state ($\vsix$ for \ce{N_2}) are an accurate description of the ground state.
Instead, multiple reference states have significant weights in the wavefunction expansion.
Such wavefunctions are referred to as multiconfigurational in quantum chemistry,\cite{Bauer2020, Izsak2023}
and methods that approximate these by using different references states are known as \textit{multireference}.

In the following, we introduce the notation used for the general configuration state functions that describe 
each of the contributing states relevant to strongly correlated regimes where bonds are partially broken.

\subsection{Configuration State Functions}\label{ssec:csfs}
The strong correlation corresponding to localized spin-coupled orbitals can be described through entangled states whose coefficients are determined from symmetry. The Hamiltonian operator commutes with both the total spin operator, $\hat{S}^2$, and the spin projection operator, $\hat{S}_z$. Therefore, energy eigenstates must simultaneously be eigenstates of the spin operators with corresponding total spin and spin projection quantum numbers $S$, $M_S$. A Slater determinant is in general an eigenfunction of $\hat{S}_z$ but not of $\hat{S}^2$.
Exceptions include fully closed-shell determinants with $S, M_S = 0, 0$, such as the state in Eq.~\eqref{eqn:rhf_n2}.

Configuration state functions (CSFs) are eigenfunctions of both $\hat{S}^2$ and $\hat{S}_z$ with quantum numbers formed by a symmetry-adapted linear combination of determinants.\cite{Helgaker}
All the reference states considered in this work are CSFs.
These are products of closed-shell delocalized orbitals as well as open-shell spin-coupled orbitals.
The spin-coupled part is defined by the pattern in which the single electrons/orbitals are coupled to form the many-body wavefunction. We define the following first-quantized representation for the CSFs used in this work (Eq.~6 in Ref.~\citenum{Marti-Dafcik2024a}):
\begin{equation}
	\ket{\Phi} = \mathcal{N}\hat{\mathcal{A}}\ket{\phi_c^2\cdots}\mathcal{O}_{S, M_S}^{N, i}(\phi_o\cdots),
	\label{eqn:csf_notation}
\end{equation}
where $S, M_S$ are the spin quantum numbers and $N$ is the number of spin-coupled electrons (and therefore the number of spin-coupled, localized spatial orbitals). Here, we have included the antisymmetrization operator $\hat{\mathcal{A}}$ and normalization constant $\mathcal{N}$, but we omit these in the remaining text.
The term $\ket{\phi_c^2\cdots}$ is a product state with doubly occupied (closed-shell) orbitals $\{ \phi_{c}, ...\}$.
The term $\mathcal{O}_{S, M_S}^{N, i}(\phi_o\cdots)$ is a state with singly-occupied (open-shell) orbitals $\{ \phi_{o}, ...\}$ that are spin-coupled. The index $i$ specifies the coupling pattern and the amplitudes of the expansion of the CSF in a product basis are simply the Clebsch--Gordan coefficients; a list of relevant spin-coupled states is provided in Appendix~\ref{apdx:spin_eigenfunctions}.

In the compact, second-quantized qubit representation (Eq.~\eqref{eqn:compact}), the closed-shell state maps as $\ket{\phi_c^2\cdots} \rightarrow \ket{2\cdots}$, the open-shell state is $\ket{\mathcal{O}_{S, M_S}^{N, i}(\phi_o\cdots)}$, and we explicitly include the occupation of the unoccupied (virtual) orbitals $\{ \phi_{v}, ...\}$ as $\ket{0 \cdots}$. Therefore, the qubit representation of any CSF with orbitals $\{\phi_{c}, ..., \phi_{o}, ..., \phi_{v}, ...\}$ takes the form
\begin{equation}
	\ket{\Phi} = \ket{2 \cdots}\ket{\mathcal{O}_{S, M_S}^{N, i}(\phi_o\cdots)}\ket{0 \cdots},
	\label{eqn:csf_notation_qubit}
\end{equation}
where the indices $c$, $o$ and $v$ run over the closed-shell, open-shell, and virtual orbitals.

\subsection{Multireference quantum chemistry}
To describe situations where multiple correlation effects occur simultaneously,
such as bond breaking in the intermediate region of the binding curve, we form hybrid, partially entangled states.
By localizing only the RHF orbitals involved in the $\uppi_{\text{x}}$-bond to obtain the basis $\{\sigg, \piuy, \upx_L, \upx_R, \pigy, \sigu \}$, we form a state
with one stretched bond (two spin-coupled electrons):
\begin{equation}
	\ket{\Phi_{2_x}} =  \ket{\sigg^2 \piuy^2}\mathcal{O}_{0, 0}^{2, 1} \ket{\pigy^2 \sigu^2}\rightarrow \ket{22} \frac{1}{\sqrt{2}} (\ket{\al \be} - \ket{\be \al})\ket{00}.
	\label{eqn:v2x_n2}
\end{equation}
This is traditionally referred to as a diradical state.
Due to the cylindrical symmetry of the molecule, whose bond points along the $z$-axis, the state has an exactly degenerate counterpart with spin coupling along the $y$ direction. Using orbitals $\{\sigg, \piux, \upy_L, \upy_R, \pigx, \sigu \}$, this state has the same qubit representation
\begin{equation}
	\ket{\Phi_{2_y}} =  \ket{\sigg^2 \piux^2}\mathcal{O}_{0, 0}^{2, 1} \ket{\pigx^2 \sigu^2} \rightarrow \ket{22} \frac{1}{\sqrt{2}} (\ket{\al \be} - \ket{\be \al})\ket{00}.
	\label{eqn:v2y_n2}
\end{equation}
In calculations, we will form a fixed, symmetric linear combination of the two (non-orthogonal) states
\begin{equation}
	\ket{\Phi_2} = \frac{1}{\sqrt{3}}(\ket{\Phi_{2_x}} + \ket{\Phi_{2_y}})
	\label{eqn:v2xy_n2}
\end{equation}
and treat this as a single reference state without any loss in circuit/state expressivity.

We also form a state describing two stretched bonds by localizing four of the orbitals. With orbitals $\{\sigg, \upx_L, \upy_L, \upx_R, \upy_R, \sigu \}$, the state is
\begin{equation}
	\ket{\Phi_4} = \ket{\sigg^2}\mathcal{O}_{0, 0}^{4, 1}(x_L, y_L, x_R, y_R) \ket{\sigu^2} \rightarrow \ket{2}\vfour\ket{0},
	\label{eqn:v4_n2}
\end{equation}
where $\vfour$ is given in the Appendix (Eq.~\eqref{eqn:vcsf_n4}).
At arbitrary bond lengths, the wavefunction is accurately described by the optimal linear combination of all CSFs,
\begin{equation}
	\ket{\Phi_{\mathrm{LC}}} = c_1 \ket{\Phi_{\mathrm{RHF}}} + c_2 \ket{\Phi_2} + c_3 \ket{\Phi_4} + c_4 \ket{\Phi_6},
	\label{eqn:lc_of_csfs}
\end{equation} which can be found by diagonalizing the Hamiltonian in the subspace built from the four (nonorthogonal) states.

The binding curve (Fig.~\ref{fig:n2_mainfig}) clearly displays the expected behaviour: the RHF state is accurate around the equilibrium distance but poor at dissociation. The spin-coupled states $\ket{\Phi_2}$, $\ket{\Phi_4}$ have increasingly longer minima and improved energies at dissociation. The fully spin-coupled state $\ket{\Phi_6}$ is exact at dissociation.

The trend seen in the energies is fully matched by the trend in the coefficients and squared overlaps (Figs.~\ref{fig:n2_csf_lin_comb} and \ref{fig:n2_csf_fid}). At short bond lengths, the RHF state dominates the linear combination with high overlap with the exact ground state. As the bond is stretched, the intermediate CSFs $\ket{\Phi_2}$ and $\ket{\Phi_4}$ become increasingly important, and at long bond lengths the fully spin-coupled state $\ket{\Phi_6}$ dominates, with $100\%$ overlap in the infinite limit.
The linear combination is least accurate at the intermediate region; nevertheless the minimum squared overlap is $92\%$. Using the linear combination state $\ket{\Phi_{\mathrm{LC}}}$ instead of $\ket{\Phi_{\mathrm{RHF}}}$ directly leads to a reduction in the runtime of QPE of more than one order of magnitude at stretched geometries (see Section \ref{sec:fault_tolerant} for an extended discussion).

\subsection{Comparison with spin-adapted classical algorithms}
We note that the techniques for construction of CSFs have long been established,\cite{Pauncz1979, Szabo2000}
and that one can choose to construct different CSF bases with different features.\cite{Serber1934a, Kotani1963, Rumer1932}
Such CSF bases enable implementations of spin-adapted versions of classical electronic structure algorithms.\cite{Sharma2012, Dobrautz2019}
The insight that working in a particular CSF and orbital basis can enable a compression of many-electron wavefunctions
for systems where spin coupling dominates has recently also been exploited in works on quantum chemistry on classical computers.
For example, Li Manni \textit{et al.} showed that one can greatly increase the sparsity in the Hamiltonian matrix if
one expresses it in a CSF basis with a set of localized orbitals,\cite{LiManni2020, Dobrautz2022}
in line with what we saw for diatomic molecules at dissociation where a single CSF is exact (Section \ref{ssec:n2_diss}).
For such systems in which a single spin-coupled CSF dominates the true eigenstate, this speeds up the classical FCI-QMC algorithm
which uses stochastic sampling of Hamiltonian terms and therefore benefits from increased sparsity.\cite{Booth2009}
However, for multiconfigurational wavefunctions \cite{Izsak2023} in which multiple configurations of different character contribute significantly to the exact eigenstate,
no many-body basis consisting of orthogonal CSFs optimally compresses the wavefunction, and therefore denser Hamiltonians and eigenvectors cannot be avoided.
This is why exponentially vanishing CSF coefficients were recently observed for more challenging molecular systems in a fixed CSF basis.\cite{Lee2023}

In contrast, our approach allows efficiently simulating challenging molecular systems by
simply preparing each of the dominant (mutually nonorthogonal) CSFs as a linear combination of Slater determinants,
where each CSF is initially represented in a different orbital basis but subsequently transformed to one common basis through simple orbital rotation circuits.
Our method does not bypass the exponential memory complexity, as this is unavoidable:
by choosing a Slater determinant basis (as opposed to a CSF basis) for the many-body Hilbert space, the RHF state has a compact description (one computational basis state) but we are forced to pay an exponential memory cost in the dimensionality of the localized, spin-coupled initial states.
For example, the spin-coupled CSF $\vn$ requires an exponential amount of Slater determinants,
whereas once could construct a CSF basis where $\vn$ would be just one basis vector.
However, exponential memory growth is not an issue when using quantum computers,
and since preparation of any of the CSFs considered here is efficient through the circuits that we provide, the time complexity of our CSF-based quantum algorithms is polynomial (Section \ref{sec:qcircs}).

\section{This work: overview of results}\label{sec:overview}
The ability to build a qualitatively accurate reference state for weakly correlated electronic systems through mean-field theory is what has driven the success of most practical approaches to computational chemistry. The key result from Ref.~\citenum{Marti-Dafcik2024a} is that a qualitatively accurate wavefunction can be built for systems with strong electron correlation by parametrizing strong correlation effects due to spin coupling. Just as MO theory predicts the Hartree--Fock state to be the dominant electron configuration, our generalized, spin-coupled MO theory predicts the dominant entangled states for strongly spin-coupled systems from chemical intuition and symmetry considerations. This type of correlation is typically referred to as strong or \textit{static} correlation in quantum chemistry.

To obtain quantitatively accurate energy estimates, the remaining (\textit{dynamic} or \textit{weak}) correlation needs to be added to the spin-coupled reference states.\cite{Helgaker2010}
This requires powerful computational methods to generate a large amount of additional Slater determinants with small weights.
Classical computational methods generally struggle to provide such accuracy due to the unavoidable memory/time complexity of simultaneously storing and processing spin-coupled as well as delocalized mean-field states, as this requires exponential wavefunction expansions regardless of the choice of single-particle basis.

On the other hand, quantum information processors can efficiently store and transform exponentially large quantum states
and therefore this limitation is not present in quantum models of computation.\cite{Gilyen2019}
Within the digital quantum circuit model, the key requirement to leverage such reference states is
the ability to efficiently prepare them by applying elementary gates on the initial product state of the qubits, typically $\ket{0} \cdots \ket{0}$.
In Section \ref{sec:qcircs}, we show how this can be achieved through quantum circuits of depth $\mathcal{O}(N)$ and using $\mathcal{O}(N^2)$ gates for systems of $N$ spin-coupled electrons, by mapping such states to the well-studied family of Dicke states.
This confirms that, in addition to their low Kolmogorov complexity,
the quantum circuit complexity of such states is also low owing to their high degree of symmetry.

This approach unlocks the power of quantum algorithms for strongly correlated electronic structure.
Through extensive classical simulations of quantum algorithms, we demonstrate how the use of spin-coupled initial states,
in particular when represented in different bases,
drastically enhances the performance of quantum algorithms for challenging electronic systems. In VQE,
it reduces the number of variational parameters, and therefore the number of parametrized quantum gates,
required to achieve chemical accuracy (Section~\ref{sec:vqe}). In quantum subspace approaches,
it greatly lowers the circuit depth and the number of measurements (Sections~\ref{ssec:qsd} and \ref{ssec:adapt-qsd}). In adiabatic state preparation,
it increases the accuracy that can be achieved through a given number of steps, effectively speeding up
the adiabatic evolution process required to reach a wavefunction of high accuracy (Section~\ref{ssec:asp}).

Finally, spin-coupled reference states reduce the cost of fault-tolerant algorithms based on QPE by orders of magnitude (Section~\ref{sec:fault_tolerant}).
Although alternative methods for initial state preparation have been proposed by other authors,\cite{Tubman2018, Low2018, Formichev2023} fault-tolerant applications relying on QPE will greatly benefit from our approach. This stems from the fact that, by directly encoding the entanglement due to spin coupling through bespoke quantum circuits, our state preparation circuits are much more efficient than generic, black-box state preparation techniques.

\section{Quantum circuits for preparation of spin eigenfunctions}\label{sec:qcircs}
To exploit spin-coupled initial states in quantum algorithms, they must be efficiently loaded onto quantum registers.
Despite their exponentially-scaling support in the computational basis,
the spin-coupled states considered in this work have a highly symmetric and well-defined structure,
and the number of distinct coefficients scales linearly with the number of electrons. 
Crucially, we can exploit this structure to derive quantum circuits that prepare the
spin eigenfunctions $\vn$, which correspond to a superposition of ${N \choose N/2}$ determinants,
with $\mathcal{O}(N^2)$ rotation and CNOT gates and depth $\mathcal{O}(N)$.
We achieve this by connecting the spin eigenfunctions $\vn$ to Dicke states, which form a different family of entangled wavefunctions.\cite{Bartschi2019}
The recursive circuit structure also reduces the cost of preparing linear combinations of spin-coupled states.
Furthermore, we consider their implementation on a fault-tolerant architecture, where there is an additional $\log(N)$ overhead in the non-Clifford gate count due to the need for synthesizing rotation gates through Clifford + T gates (Section~\ref{ssec:circ_costtot_and_examples} and Appendix~\ref{apdx:circ_cost_faulttolerant}).

We compute the exact CNOT and Toffoli gate counts and find that both are very low in practice for a range of systems sizes relevant to quantum chemistry (Table~\ref{tab:n_cnots_csfs} in Section \ref{ssec:circ_costtot_and_examples}). For example, for $N=34$, only $\sim 10^3$ gates are required to prepare the spin-coupled CSF $\ket{\mathcal{O}_{0, 0}^{34, 1}}$, which is a superposition of $L \sim 10^9$ Slater determinants.\footnote{This assumes a basis of localized orbitals, and $L$ would be significantly higher using delocalized basis functions such as Hartree--Fock orbitals.}

In contrast, there exist general algorithms for preparing arbitrary linear combinations of computational basis states (Slater determinants).\cite{Tubman2018, Shende2006, Childs2012, Babbush2018e, Low2018, Formichev2023} Such methods do not exploit any particular features of the states that are being prepared and their cost scales at best linearly with the number of Slater determinants. Therefore, the scaling is exponential for strongly correlated systems.
For the example of $N = 34$, our method is at least eight orders of magnitude more efficient than preparing the same linear combination of computational basis states one determinant at a time using the algorithm presented in Ref.~\citenum{Formichev2023} (see Section \ref{ssec:resource_estimation} for more details on the comparison between our method and others).

Other works have focused on the preparation of certain spin eigenfunctions.\cite{Sugisaki2016, Sugisaki2019, Carbone2022b}. However, these either scale exponentially\cite{Carbone2022b} or only prepare trivial geminal products that couple two-electron singlets\cite{Sugisaki2016, Sugisaki2019} (Eqn. \ref{eqn:csf_geminals}) and therefore do not have the structure necessary to encode the electronic correlation present in many molecular systems such as the ones considered in this work ( Section \ref{sec:background} and Ref.~\citenum{Marti-Dafcik2024a}).

The work in Ref.~\citenum{Sugisaki2019a} is similar in spirit to ours in that it considers superpositions of states with a varying number of spin-coupled electrons.
However, it defines the states in terms of delocalized orbitals and relies on symmetry-broken mean-field solutions.
By choosing appropriate localized orbitals and obtaining the spin-coupled states directly from the Clebsch--Gordan coefficients,
our work uses spin-pure wavefunctions that are derived from a clear physical picture,
and provides a clean mathematical framework for systematic construction of the relevant CSFs.

In Section~\ref{sec:vcsfs}, we analyze the general form of a family of spin eigenfunctions that captures the entanglement structure of molecular bonds. We discuss the connection to Dicke states in Section~\ref{ssec:mapping}.
In Section~\ref{ssec:circs_dicke}, we review the circuits for Dicke state preparation from Ref.~\citenum{Bartschi2019}. We introduce a more explicit notation for operators and circuits in Section~\ref{ssec:notation_ops}, discuss the circuits for preparation of the states $\vn$ in Section~\ref{ssec:qcircs_csfs_bondbreaking}, the mapping from the representation in the spin Hilbert space to the Fock space in Section~\ref{ssec:mapping_spintofock}, the total state preparation cost in Section~\ref{ssec:circ_costtot_and_examples}, the preparation of linear combinations of CSFs in Section~\ref{ssec:circs_lc}, as well as generalizations to other spin eigenfunctions in Section~\ref{ssec:circuits_generalization}.

\subsection{Spin eigenfunctions for bond dissociation}\label{sec:vcsfs}
Each spin eigenfunction presented in Section~\ref{sec:background} includes a set of doubly occupied, delocalized orbitals (typically, Hartree--Fock orbitals), and a set of singly occupied, localized orbitals which capture the strong electron correlation (Eqs.~\eqref{eqn:csf_notation} and \eqref{eqn:csf_notation_qubit}). The weakly correlated component can be trivially prepared  by applying Pauli-$X$ gates on each of the qubits corresponding to the closed-shell Hartree--Fock orbitals. This is the circuit structure used in most quantum algorithm implementations for preparing the Hartree--Fock state. The entanglement due to spin coupling occurs in the remaining singly-occupied orbitals, which are localized in this work. The state preparation task thus reduces to preparing a particular type of spin eigenfunction $\ket{\mathcal{O}_{S, M}^{N, i}}$ for $N$ strongly correlated electrons in $2N$ spin orbitals (qubits).

The physical model that we propose requires a partitioning of the molecular electronic system into subsystems, and considers the coupling within the subsystem and across the subsystems.
For the systems considered in Ref.~\citenum{Marti-Dafcik2024a}, these subsystems always contain the same number of electrons. Therefore, for a total number of $N$ strongly correlated electrons in $N_S$ subsystems, each subsystem has $n = N/N_S$ electrons.

In the case of diatomic bond breaking at dissociation, $N$ strongly correlated electrons localize onto $n=N/2$ subsystems, where $N = 2$ for \ce{H_2}, and $N=6$ for \ce{N_2} (Section \ref{sec:background}).
The local coupling within a molecular fragment is ferromagnetic, which corresponds to a state of maximum spin multiplicity for each subsystem. By considering the local coupling first, we can build the global wavefunction from coupling subsystem wavefunctions. We continue to use the example of stretching a triple bond as in \ce{N_2} for illustration.

The subsystem state with spin $s=3/2$ state that ferromagnetically couples three electrons is the quartet in Eq.~\eqref{eqn:3e_quartet}.
The state $\vsix$ that describes the dissociation of a triple molecular bond into two three-electron quartets is the product of linear combinations of the $2s+1$ components of each quartet subsystem that forms an overall $N$-electron state of spin $S = 0, M_S = 0$:
\begin{equation}
	\begin{split}
		& \vsix = \frac{1}{2}\Bigg[ \ket{\frac{3}{2}} \ket{-\frac{3}{2}} - \ket{-\frac{3}{2}} \ket{\frac{3}{2}} - \ket{\frac{1}{2}} \ket{-\frac{1}{2}}+\ket{-\frac{1}{2}} \ket{\frac{1}{2}}\Bigg]\ \\
		&=
		\frac{1}{2}\Big[\ket{\alpha\alpha\alpha\beta\beta\beta} - \ket{\beta\beta\beta\alpha\alpha\alpha}\Big]\\
		&+\frac{1}{6}\Big[\ket{\alpha\beta\beta\alpha\alpha\beta}
		+ \ket{\alpha\beta\beta\alpha\beta\alpha}
		+ \ket{\alpha\beta\beta\beta\alpha\alpha}
		+ \ket{\beta\beta\alpha\beta\alpha\alpha}\\
		&+ \ket{\beta\beta\alpha\alpha\beta\alpha}
		+ \ket{\beta\beta\alpha\alpha\alpha\beta}
		+ \ket{\beta\alpha\beta\beta\alpha\alpha}
		+ \ket{\beta\alpha\beta\alpha\beta\alpha}\\
		&+ \ket{\beta\alpha\beta\alpha\alpha\beta}
		- \ket{\beta\alpha\alpha\beta\beta\alpha}
		- \ket{\beta\alpha\alpha\beta\alpha\beta}
		- \ket{\beta\alpha\alpha\alpha\beta\beta}\\
		&- \ket{\alpha\beta\alpha\beta\beta\alpha}
		- \ket{\alpha\beta\alpha\beta\alpha\beta}
		- \ket{\alpha\beta\alpha\alpha\beta\beta}
		- \ket{\alpha\alpha\beta\beta\beta\alpha}\\
		&- \ket{\alpha\alpha\beta\beta\alpha\beta}
		- \ket{\alpha\alpha\beta\alpha\beta\beta}
		\Big].\end{split}
	\label{eqn:v6}
\end{equation}
In the first row, we have used the notation $\ket{m_{sL}}\ket{m_{sR}}$ for compactness to refer to the spin projection quantum number of the \textit{left} and \textit{right} spin subsystems,  i.e., we have dropped the spin quantum number $s=3/2$ and only specify the spin projection quantum numbers of the subsystems.
The general form reads:
\begin{equation}
	\vn = \frac{1}{\mathcal{N}} \sum_{m} \Big[ a_{m} \ket{m} \ket{-m} + b_{m} \ket{-m} \ket{m}(1-\delta_{m, 0}) \Big],
	\label{eqn:v_csf}
\end{equation}
where $s=n/2=N/4$ is the spin of each subsystem, $m=s, s-1, ... \geq 0$ is the absolute magnitude of the spin projection quantum number, $m=\abs{m_s}$, $\mathcal{N} = \sqrt{n + 1 }$ is a normalizing constant, and the relative signs $a_{m}, b_{m} \in \{ 1, -1\}$ are given from the Clebsch--Gordan coefficients.

\subsection{Mapping to Dicke states and symmetric states}\label{ssec:mapping}
Although Eq.~\eqref{eqn:v_csf} contains ${N \choose n} = {N \choose N/2}$ Slater determinants,
we can efficiently prepare this expansion on a digital quantum computer by mapping it to Dicke states, for which efficient quantum circuits are known.
A Dicke state $\ket{D_{k}^n}$ is defined as an equal-weight superposition of all possible states of an $n$-qubit system with Hamming weight $k$\cite{Bartschi2022}
\begin{equation}
	\ket{D_k^n} = {n \choose k}^{-1/2} \sum_\mathcal{P}
	\mathcal{P}\Big(
	\ket{1}^{ k}
	\ket{0}^{ n-k} \Big),
\end{equation}
where $\mathcal{P}$ denotes any permutation of qubits and the sum runs over all ${n \choose k}$ possible permutations.
The open-shell spin eigenfunctions discussed in this work can naturally be expressed in terms of Dicke states by rewriting $\alpha\rightarrow1$ and $\beta \rightarrow 0$.
For example $\ket{D_{2}^3} = \frac{1}{\sqrt{3}}\big( \ket{110} + \ket{101} + \ket{011} \big)$ corresponds to $\ket{\frac{3}{2}, \frac{1}{2}}$ in Eq.~\eqref{eqn:3e_quartet}.

This encoding is natural for spin systems, where the $N$-qubit states exist in the $2^N$-dimensional Hilbert space of $N$ spin-$1/2$ sites, as might be found in site-based model Hamiltonians.
However, in quantum chemistry we consider spin-$1/2$ fermions, and each \textit{site} is a localized spatial orbital that can have four possible occupations.
In this context, the appropriate encoding requires one qubit per spin-orbital (Section \ref{ssec:notation_states}).
Therefore, we first only work with in the $M$ qubits for the spin-up orbitals to prepare the $\vn$ spin eigenfunction, i.e., we map spin-coupled CSFs $\vn$
to a superposition of basis states 
\begin{equation}
	\ket{j} = \bigotimes_{i=1}^M \ket{f_{i_{\alpha}}}.
\end{equation}
Afterwards, we map the occupation from this spin space to the Fock space to recover the correct encoding (Eqs.~\eqref{eqn:standard} and \eqref{eqn:compact}), as detailed in Section \ref{ssec:mapping_spintofock}.

A symmetric state of $n$ qubits $\ket{\psi^n_{\mathcal{S}}}$ is a state that remains invariant under permutations $\mathcal{P}$ of the symmetry group of $n$ qubits, $\mathcal{S}_n$:
\begin{equation}
	\ket{\psi^n_{\mathcal{S}}} = \mathcal{P}\ket{\psi^n_{\mathcal{S}}}.
\end{equation}
Furthermore, the $n+1$ Dicke states $\ket{D_k^n}$ of an $n$ qubit-system ($k\in \{0, 1, ..., n\}$) are orthonormal and form a complete basis for the $n+1$ symmetric states of $n$ qubits.\cite{Bastin2009} Therefore, any symmetric state can be expressed as a linear combination of Dicke states:
\begin{equation}
	\ket{\psi^n_{\mathcal{S}}} = \sum_{k=0}^{n}c_k \ket{D_k^n}.
	\label{eqn:symmetry_basis}
\end{equation}

The states $\vn$ are constructed from the products of states of highest multiplicity, which are symmetric (Eqs.~\eqref{eqn:3e_quartet}, \eqref{eqn:v6}, and \eqref{eqn:v_csf}). Thus, the construction of $\vn$ is equivalent to constructing linear combinations of products of Dicke states with different Hamming weights. For example, the state in Eq.~\eqref{eqn:v6} can be rewritten as
\begin{equation}
	\begin{split}
		\vsix =
		\frac{1}{2}\Bigl( &\ket{D_3^3}\ket{D_0^3}
		- \ket{D_0^3}\ket{D_3^3}\\
		- & \ket{D_2^3}\ket{D_1^3}
		+ \ket{D_1^3}\ket{D_2^3}\Bigl).
	\end{split}
	\label{eqn:dicke_chi6_wfn}
\end{equation}
The general $N$-electron case (Eq.~\eqref{eqn:v_csf}) expressed in the Dicke basis reads:
\begin{equation}
	\begin{split}
		\vn = 
		\frac{1}{\mathcal{N}}\sum_{m}\Bigl[
		& a_m \ket{D_{s+m}^n}\ket{D_{s-m}^n} \\
		+ & b_m \ket{D_{s-m}^n}\ket{D_{s+m}^n}(1-\delta_{m, 0})
		\Bigl].
	\end{split}
	\label{eqn:dicke_vcsf}
\end{equation}
We now discuss the preparation of Dicke states and symmetric states following the approach proposed in  Ref.~\citenum{Bartschi2019}.

\subsection{Preparation of Dicke and symmetric states}\label{ssec:circs_dicke}
Dicke states have long been created on experimental platforms such as ion traps,\cite{Hume2009} but
their efficient preparation through quantum circuits was a longstanding challenge.\cite{Bacon2006}
Recently, B{\"a}rtschi and Eidenbenz provided a protocol for efficiently preparing any Dicke state $\ket{D_k^n}$ or any symmetric state using shallow circuits and without requiring any ancilla qubits.\cite{Bartschi2019}
Dicke circuits have been benchmarked in experimental implementations on quantum hardware.\cite{Aktar2022, Aktar2023}
Dicke states have direct applications in combinatorial optimization, where they can be used as initial states of the Quantum Alternative Operator Ansatz algorithm,\cite{Bartschi2020} but their connection to spin eigenfunctions has previously not been explored. In quantum simulation, Dicke circuits have been considered for preparing antisymmetrized geminal power states.\cite{Khamoshi2023}
Here, we use the Dicke state preparation method from Ref~\citenum{Bartschi2019} as a circuit primitive to prepare spin eigenfunctions. We summarize the relevant results below.

\begin{enumerate}[wide, labelindent=0pt]
	\item Efficient preparation of Dicke states: The Dicke state $\ket{D_{k}^n}$, for $k \leq n$, can be prepared deterministically by applying a unitary $U_{n, k}$ on the input state $\ket{0}^{ n-k}\ket{1}^{ k}$. A decomposition of $U_{n, k}$ in terms of single-qubit and CNOT gates is provided. The resulting quantum circuit has depth $\mathcal{O}(n)$ and uses $\mathcal{O}(kn)$ gates.
	\item Efficient preparation of symmetric states: Given any input state of the same number of qubits $n$ but with a smaller Hamming weight, $\ket{0}^{ n-l}\ket{1}^{l}$ with $l\leq k$, applying the operator $U_{n, k}$ onto said state outputs the Dicke state $\ket{D_l^n}$. Consequently, any superposition of Dicke states of equal $n$ and different $l$ can be generated by applying the unitary $U_{n, n}$ on the appropriate superposition of input states:
	\begin{equation}
		\ket{\p} = \sum_l c_l \ket{D_l^n} =  U_{n, n} \sum_l c_l \ket{0}^{ n-l}\ket{1}^{l}.
		\label{eqn:dicke_superposition}
	\end{equation}
	Since the $n+1$ Dicke states $\ket{D_l^n}$ form an orthonormal basis of the fully symmetric subspace of all $n$-qubit states (Eq.~\eqref{eqn:symmetry_basis}), the circuits for unitaries $U_{n, n}$ can be used to prepare any symmetric state $\ket{\psi_{\mathcal{S}}^n}$, provided that they act on the appropriate input states. The input state $\sum_l c_l \ket{\psi_l^n}$ can be prepared by a quantum circuit of depth and gate count $\mathcal{O}(n)$. The overall complexity for preparation of any symmetric state is thus depth $\mathcal{O}(n)$ and $\mathcal{O}(n^2)$ gates. These circuits can be implemented directly on a qubit architecture with linear connectivity, which is beneficial e.g. for applications on superconducting quantum hardware.
\end{enumerate}

The key insight that enables this efficient protocol is the following recursive formula:
\begin{equation}
	\ket{D_k^n} = \sqrt{\frac{k}{n}}\ket{D_{n-1}^{k-1}}\ket{1}+\sqrt{\frac{n-k}{n}}\ket{D_k^{n-1}}\ket{0}.
\end{equation}
Given a unitary operator $U_{n, k}$ that prepares the Dicke state $\ket{D_k^n}$ when acting on the input state $ \ket{0}^{n-k}\ket{1}^{k}$, one can recursively decompose the unitary operator $U_{n, k}$ into products of two types of elementary operators, $M_{l, l-1}$ and $M_{l, k}$ (Lemma 2 in Ref.~\citenum{Bartschi2019}):
\begin{equation}
	U_{n, k}=\prod_{l=2}^k M_{l, l-1}\prod_{l=k+1}^n M_{l, k}.
\label{eqn:dicke_recursion}
\end{equation}

For preparation of symmetric states we set $k=n$ and the unitary reduces to
\begin{equation}
	U_{n, n} = \prod_{l=2}^n M_{l, l-1} :=S_n.
	\label{eqn:lemma2}
\end{equation}
Here, the operators $M_{l, l-1}$ can be constructed from products of two primitive building blocks: one two-qubit gate block (Fig.~\ref{fig:two_qubit_circuit})
followed by $l-2$ three-qubit gate blocks (Fig.~\ref{fig:three_qubit_circuit}). This enables the preparation of an $n$-qubit Dicke state in terms of subcircuits acting on $l<n$ qubits.
An example circuit for $S_{4}$ is shown in Fig.~\ref{fig:circ_S_{4}}.
We discuss the decomposition of such circuits into gates consisting of only CNOT and single-qubit rotations in Appendix~\ref{apdx:circ_cost}, to show that the number of CNOT gates required is,
\begin{equation}
	C_{S_{n}} = \frac{5}{2}n^2+ \frac{9}{2}n + 2
\label{eqn:cost_sn}
\end{equation}
as derived in Eq.~\eqref{eqn:cost_S_n}.

One attractive feature of this protocol is the fact that only one unitary $S_{n}$ needs to be applied to generate a linear combination of multiple Dicke states, and therefore no special techniques such as Linear Combination of Unitaries \cite{Childs2012, Berry2015} are required for state preparation. Instead, the desired superposition of Dicke states is achieved by preparing a linear combination with the appropriate weights in the relatively simple input state. This is much more efficient than applying state preparation unitaries controlled by ancilla qubits.\cite{Childs2012, Tubman2016, Low2018} By slightly modifying the circuit for input state preparation, we can efficiently use this circuit structure to prepare the CSFs relevant for chemical bond breaking.
For reference, the cost for up to $n = 6$ is shown in Table~\ref{tab:n_cnots_u_nn} in Appendix~\ref{apdx:circ_cost}.

\subsection{Notation for operators and circuits}\label{ssec:notation_ops}
\begin{figure*}[!htb]
	\centering
	\begin{subfigure}[b]{0.7\textwidth}
		\includegraphics[width=\linewidth]{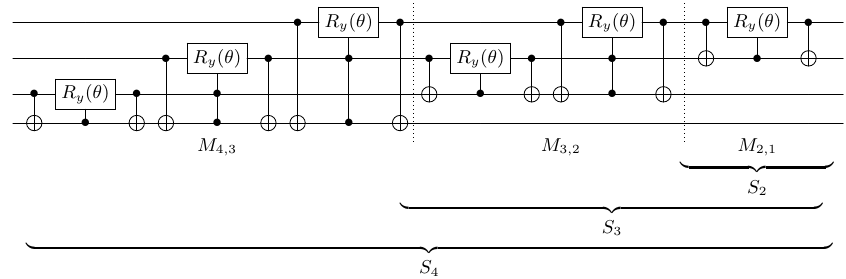}
		\caption{Circuit for $S_{4}=S_3 M_{4, 3} = (S_2 M_{3, 2})M_{4, 3}  = M_{2, 1}M_{3, 2}M_{4, 3}$.}
		\label{fig:circ_S_{4}}
	\end{subfigure}
	\hfill
	\begin{minipage}[b]{0.2\textwidth}
		\begin{subfigure}[b]{\linewidth}
			\includegraphics[width=\linewidth]{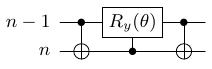}
			\caption{Two-qubit gate.}
			\label{fig:two_qubit_circuit}
		\end{subfigure}\\[\baselineskip]
		\begin{subfigure}[b]{\linewidth}
			\includegraphics[width=\linewidth]{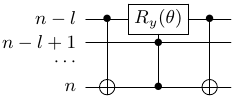}
			\caption{Three-qubit gate.}
			\label{fig:three_qubit_circuit}
		\end{subfigure}
	\end{minipage}
	\caption{(a) Quantum circuit proposed in Ref.~\citenum{Bartschi2019} to implement the symmetric state preparation unitary $S_{4}=M_{2, 1}M_{3, 2}M_{4, 3}$ (see Eq.~\eqref{eqn:lemma2}). (b), (c) Gate blocks in Dicke state preparation circuits such as (a), also from Ref.~\citenum{Bartschi2019}. The angles are $\theta = 2\arccos(\sqrt{\frac{1}{n}})$ for the two-qubit gate blocks and $\theta = 2\arccos(\sqrt{\frac{l}{n}})$ for the three-qubit gate blocks.}
	\label{fig:circ_S4}
\end{figure*}

In the above, the qubits on which $M_{l, k}$ acts are specified in the circuit diagrams (Fig.~\ref{fig:circ_S4}). For further clarity, we define the following more explicit notation for all quantum circuits below.

Superscripts denote the qubits on which gates act, e.g. $X^i$ is a Pauli-$X$ gate on qubit $i$, where $1$ is the top qubit in the circuit diagram and its state is specified by the leftmost bit in the ket: $\ket{j} = \ket{j_1, j_2, ...}$. We use $U^{i, j}_{j-i+1}$ to denote an operator acting on $(j-i+1)$ qubits with indices $i, i+1, ..., j-1, j$, e.g. $S_n^{1, n}$ means $S_n$ acting on qubits $1$ through $n$.
For two unitary operators $U$ and $V$, the product is $UV$ (unlike for states where products of kets are tensor products, see Section~\ref{ssec:notation_states}),
therefore the tensor product of two unitaries must be written explicitly as $U\otimes V$. Tensor products with identity are implied whenever the unitaries act on a subset of the qubits.

For two-qubit controlled gates, we use the following notation: $CU^{c, t}$ applies the operator $U$ on target qubit $t$ controlled by qubit $c$, i.e. $CU^{c, t} = P_0^c \otimes I^t + P_1^c\otimes U^t$, where the projectors are $P_0  = \ket{0}\bra{0}$ and $P_1 = \ket{1}\bra{1}$. On circuit diagrams, this corresponds to a full black dot on qubit $i$. We use $\overline{CU}^{c, t}$ for a gate that controls application of the operator $U$ on target qubit $t$ such that $U$ is applied if $c$ is in the state $\ket{0}$, i.e. $\overline{CU}^{c,t} = P_0^c \otimes U^t + P_1^c\otimes I^t$. This corresponds to a white dot on the control qubit $c$ in the circuit diagram. We often use the CNOT gate with $\overline{CX}^{c,t} = X^c CX^{c, t}X^c$.

\subsection{Preparation of singlet states with locally ferromagnetic coupling}\label{ssec:qcircs_csfs_bondbreaking}

\begin{figure*}[!htb]
	\includegraphics[width=\linewidth]{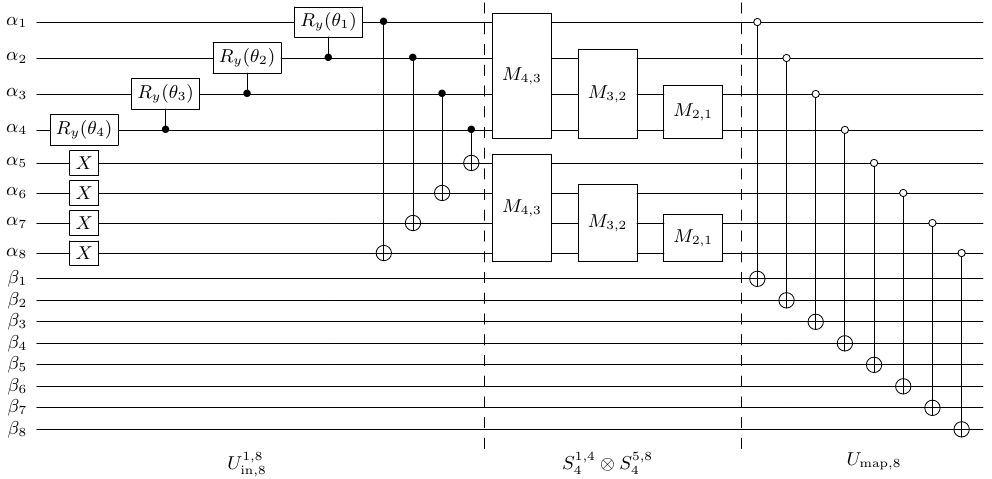}
	\caption{Quantum circuit that implements the unitary $V_8$ to prepare the spin-coupled state $\ket{\mathcal{O}_{0, 0}^{8, 1}} = V_8 \ket{00}^{8}$ (Eq.~\eqref{eqn:dicke_v8_csf}). It consists of three parts, delimited by the dashed vertical lines: preparation of the input state (left), preparation of the spin eigenfunction through Dicke state preparation circuits (center, see Fig.~\ref{fig:circ_S4}), and mapping from spin to Fock space occupation (right). Qubits labelled $\alpha_i$ and $\beta_i$ correspond to the up and down spin-orbital with the same spatial orbital $\phi_i$.
	}
	\label{fig:circuits_dicke_v8}
\end{figure*}

Equipped with the circuits for Dicke state preparation, we can rewrite the CSF for bond breaking (Eq.~\eqref{eqn:dicke_vcsf}) in terms of the corresponding unitaries acting on carefully chosen input states:
\begin{widetext}
	\begin{equation}
	\begin{split}
		\ket{\mathcal{O}_{S, M}^{N, 1}} =
		&\frac{1}{\mathcal{N}} \sum_{m}\Bigl[
		a_m \ket{D_{s+m}^n}\ket{D_{s-m}^n}
		+ b_m \ket{D_{s-m}^n}\ket{D_{s+m}^n}(1-\delta_{m, 0})
		\Bigl] \\
		= S_{n}^{1, n} \otimes S_{n}^{n+1, 2n} &\frac{1}{\mathcal{N}}\sum_{m}\Bigl[
		a_m \ket{0}^{s-m}\ket{1}^{s+m}\ket{0}^{s+m}\ket{1}^{s-m}
		+ b_m \ket{0}^{s+m}\ket{1}^{s-m}\ket{0}^{s-m}\ket{1}^{s+m}(1-\delta_{m, 0})
		\Bigl]
	\end{split}
	\label{eqn:dicke_vcsf_circ}
\end{equation}
For example, to prepare the spin-coupled state $\ket{\mathcal{O}_{0, 0}^{8, 1}}$, we only need to apply
the operator $S_{4}$ twice, in parallel, on an input state which is a linear combination of five states:
\begin{equation}
	\begin{split}
		\ket{\mathcal{O}_{0, 0}^{8, 1}} =
		&\frac{1}{\sqrt{5}}\Bigl(\ket{D_4^4}\ket{D_0^4}
		+ \ket{D_0^4}\ket{D_4^4}
		- \ket{D_3^4}\ket{D_1^4}
		- \ket{D_1^4}\ket{D_3^4}
		+ \ket{D_2^4}\ket{D_2^4}\Bigl)\\
		= S_{4}^{1,4} \otimes S_{4}^{5,8}
		& \frac{1}{\sqrt{5}}\Bigl(\ket{1111000} + \ket{00001111} - \ket{01110001} - \ket{00010111} + \ket{00110011}\Bigl).
	\end{split}
	\label{eqn:dicke_v8_csf}
\end{equation}
\end{widetext}

To prepare input states of the form
\begin{equation}
	\sum_l c_l \ket{0}^{ n-l}
	\ket{1}^{ l},
	\label{eqn:input_state_dicke}
\end{equation}
one could simply use the circuits in Fig.~4 of Ref.~\citenum{Bartschi2019}, which are composed of ladders of controlled-$R_y$ gates applied on the initial state $\ket{0}^n$. However, the spin eigenfunctions considered in this work are not symmetric and require parallel application of two circuits for $S_{n}$, with input states of the form in Eq.~\eqref{eqn:dicke_vcsf_circ}.

We define the input state $\ket{v_N}$ for preparation of the CSF $\vn = S_{n}^{1, n} \otimes S_{n}^{n+1, 2n} \ket{v_N}$ as $\ket{v_N} = \ket{L}\ket{R} = \sum_i c_i \ket{L_i}\ket{R_i}$, where the \textit{left} and \textit{right} states $\ket{L}$ and $\ket{R}$ correspond to qubits $\{1, 2, ..n\}$ and qubits $\{n + 1, n + 2, ..., 2n\}$. With $n=N/2$, the left and right qubits correspond to dividing the system into two molecular subsystems (Section~\ref{sec:background}). The following circuit produces the appropriate initial state:
\begin{enumerate}
	\item Flip the last $n$ qubits by applying $X$-gates to obtain the state $\otimes_n^{2n}X^n \ket{0}^{2n} = \ket{0}^n\ket{1}^n$.
	\item Apply  $\Big( \prod_{i=1}^{n-1}CR_y^{i+1, i}(\theta_i)\Big)R_y^n(\theta_n)$ on the first $n$ qubits to produce the state $\sum_i c_i\ket{L_i}\ket{1}^n$.
	\item Apply a CNOT \textit{accordion} $\prod_{i=0}^{N-1} CX^{1+i, N-i}$ that flips the state of the \textit{right} qubits controlled by the occupation of the \textit{left} qubits to obtain the input state $\ket{v_N} = \ket{L}\ket{R} = \sum_i c_i \ket{L_i}\ket{R_i}$.
\end{enumerate}
We define $U_{\mathrm{in}, N}^{1, N}$ as the unitary for preparation of the input state for $N$ spin-$\frac{1}{2}$ particles (spin-coupled electrons): $\ket{v_N} = U_{\mathrm{in}, N}^{1, N} \ket{0}^N$. An example for $N=8$ is shown in Fig.~\ref{fig:circuits_dicke_v8} (left panel).

\subsection{Mapping from spin to Fock space}\label{ssec:mapping_spintofock}
If one desires to prepare spin eigenfunctions for pure spin systems, e.g. for quantum simulation of Heisenberg or Ising spin models, this circuit above completes the entire state preparation procedure. However, here we are interested in systems of electrons (spin-$\frac{1}{2}$ fermions). In this context, the $\ket{\alpha}$ and $\ket{\beta}$ states in the expansion of the spin eigenfunctions (Appendix \ref{apdx:spin_eigenfunctions}) represent the spin projection of an electron occupying a localized spatial orbital (rather than of a spin orbital or qubit), where we have restricted each orbital to be occupied by exactly one electron. For a general many-electron wavefunction, any spatial orbital can be singly occupied, unoccupied or doubly occupied (Eq.~\eqref{eqn:compact}). We must therefore map the state from the spin space of $N$ spins, with dimension $2^N$, to the Fock space of $N$ electrons occupying $M$ spatial orbitals, with dimension ${2M \choose N} $. For a minimal basis set and $N$ spin-coupled valence electrons in the fully dissociated limit, we have $M=N$ spatial orbitals (Section \ref{sec:background}).

To account for this, we have so far chosen to work entirely on the qubits corresponding to $\alpha$ spin orbitals to prepare the superposition of Dicke states, leaving the qubits corresponding to $\beta$ spin-orbitals in the $\ket{0}$ state. After the spin eigenfunction is prepared through Dicke circuits, we map the occupation as $\ket{\alpha}=\ket{10}\rightarrow \ket{10}$ and $\ket{\beta}=\ket{00}\rightarrow \ket{01}$, where the first qubit corresponds to an $\alpha$ spin-orbital and the second qubit represents a $\beta$ spin-orbital.
This can be achieved by a single layer of CNOT gates that flips the state of the $i$-th $\beta$ spin-orbital if the state of the $i$-th $\alpha$ spin-orbital is $\ket{0}$: $U_{\textrm{map}, N} = \bigotimes_{i=1}^N \overline{CX}^{i_\alpha, i_\beta}$.

The full operator $V_N$ for preparing the $N$-electron CSF in two molecular subsystems ($n=N/2$) is thus
\begin{equation}
	\vn=V_N \ket{00}^{N}= U_{\textrm{map}, N}\bigl(S_{n}^{1, n}\otimes S_{n}^{n, n+1} \bigl)U_{\textrm{in}, N}^{1, N}\ket{00}^{N}.
\label{eqn:vn_op}
\end{equation}
An example for $N=8$ is shown in Fig.~\ref{fig:circuits_dicke_v8}.

\subsection{Total circuit cost and examples}\label{ssec:circ_costtot_and_examples}
The total number of CNOTs assuming all-to-all qubit connectivity is 
\begin{equation}
	C_{\textrm{tot}}^{\textrm{all}} = \frac{5}{4}N^2+2N+2
\end{equation}
(see Eq.~\eqref{eqn:cost_total_alltoall}). In Appendix \ref{apdx:circ_cost}, we provide a detailed derivation and discuss the case with linear or planar qubit connectivity. The latter might be of interest for experiments on near-term superconducting hardware.\cite{AIQuantum2020}
The depth of the circuits with all-to-all and planar connectivity is $\mathcal{O}(N)$. For linear connectivity, we obtain a depth scaling as $\mathcal{O}(N^2)$.

The CNOT counts are the most important metric for implementations on noisy quantum hardware without error correction because two-qubit gates are usually the source of the largest errors.\cite{Cerezo2021}
In Appendix \ref{apdx:circ_cost_faulttolerant}, we discuss a fault-tolerant implementation of the  state preparation circuit $V_N$ using a Clifford + Toffoli gate set. In short, the number of non-Clifford (Toffoli) gates depends on the number of rotations, $R$, and the accuracy required for synthesizing each continuous rotation gate through a discrete gate set. 
To bound the error in the state preparation unitary as $||V_N - \tilde{V}_N || \leq \epsilon$, where $\epsilon$ is the target accuracy, the number of Toffoli gates required is
\begin{equation}
	T = R \big [ 0.2875  \lceil \log(R/\epsilon) \rceil+ 4.6 \big].
\label{eqn:toff_cost_main}
\end{equation}
Here, $R = \frac{1}{4}N^2$ is the total number of rotation gates in the circuit that implements $V_N: \vn = V_N \ket{00}^N$ (including $R_y$, $CR_y$, and $CCR_y$ gates, see Eq.~\eqref{eqn:cost_toffolis_tot}).
We use $\log$ for logarithm to the base 2 throughout this manuscript.

Table \ref{tab:n_cnots_csfs} shows the total cost of preparing a spin eigenfunction $\vn$, and the number of determinants (computational basis states) in the expansion of $\vn$ in a product basis. We have chosen $\epsilon = 10^{-7}$ for the accuracy of initial state preparation in the fault-tolerant setting. It is evident that the preparation of these states is very efficient in practice, due to the $\mathcal{O}(N^2)$ scaling and the relatively small constant factors.

\ce{H_2}, \ce{H_2O}, \ce{N_2}, \ce{Cr_2} are examples of molecules for which, in a minimal basis of $N$ spatial orbitals, a single spin eigenfunction describes the exact ground state at dissociation. For $N=2$, $\vtwo= \ket{\mathcal{O}_{0, 0}^{2, 2}}$ and we can therefore directly use the more compact circuit for preparation of  $\ket{\mathcal{O}_{0, 0}^{N, 2}}$, which only contains Clifford gates (see Section \ref{ssec:circuits_generalization}).
For the \ce{FeS} systems, we hypothesize that the gate counts are within the correct order of magnitude because the number of open-shell electrons $N$ is known from existing literature,\cite{Sharma2014a, Li2019i, Li2019g} and the number of determinants as well as the cost of CSF preparation depends mainly on $N$. However, the estimates are by no means exact since the correct spin eigenfunctions are currently unknown (see Section \ref{ssec:resource_estimation} for details).

Finally, depending on the application, the total cost of the reference state preparation might require additional gates to rotate the single-particle basis, which can be done in depth $\mathcal{O}(N)$ and at worst $\mathcal{O}(N^2)$ gates, although often $\mathcal{O}(N)$ gates suffice. We exclude these from the analysis in this section since they depend on the system considered, the choice of basis, as well as the quantum algorithm in which they are being used. In Appendix~\ref{sec:circ_basis_rotation}, we present the circuits for basis rotations, their cost, and optimal implementations for the molecules considered here.

\begin{table}[htp]
\caption{Number of gates required for preparation of the state $\vn$ with $N$ spin-coupled sites through our state preparation circuits. In the quantum chemistry context discussed here and in Ref. \citenum{Marti-Dafcik2024a}, $N$ is the number of open-shell electrons occupying $N$ localized orbitals. The total CNOT cost of preparing each spin eigenfunction is given by $C_{\textrm{tot}} = C_{\textrm{in}} + 2C_{S_{N/2}} + C_{\textrm{map}}$, where the superscript indicates different hardware connectivities. 
	The number of determinants (computational basis states) with non-zero coefficients in the state $\vn$ is $L$.
	The number of Toffoli gates in a fault-tolerant implementation is given by $T$ (see Appendix \ref{apdx:circ_cost_faulttolerant}). We also report $b$, the number of digits required for the binary representation of the rotation angle.
	}	\label{tab:n_cnots_csfs} 
\begin{ruledtabular}
	\begin{tabular}{lccccccl}
		$N$ & $C^{\textrm{all}}_{\textrm{tot}}$ & $C^{\textrm{pla}}_{\textrm{tot}}$ & $C^{\textrm{lin}}_{\textrm{tot}}$ & $b$ & $T$ & $L$ & System
		\\
		\hline
		2 & 3 & 3 & 5 & - & 0 &2 & \ce{H_2}\\
		4& 14 & 19 & 63 & 13 & 49 & 6 & \ce{H_2O}\\
		6 & 35 & 49 & 163 & 14 & 114 & 20 & \ce{N_2}\\
		8 & 66 & 93 & 309 & 14 & 203 & 70 & \\
		10 & 107 & 151 & 501 & 14 & 317 & 252 & \ce{Fe_2S_2} (est.)\\
		12 & 158 & 223 & 739 & 15 & 477 & 924 & \ce{Cr_2}\\
		18 & 371 & 523 & 1729 & 15 & 1072 & 48620 & \ce{Fe_4S_4} (est.)\\
		34 & 1379 & 1939 & 6393 & 16 & 3989 & $2.3 \times 10^9$  & \ce{FeMoCo} (est.)\\
	\end{tabular}
\end{ruledtabular}
\end{table}

\subsection{Other spin eigenfunctions}\label{ssec:circuits_generalization}
We have so far discussed the preparation of CSFs for the coupling of two subsystems with maximum spin, corresponding to fully symmetric spin functions, into an overall singlet. While this spin coupling pattern represents the dominant entanglement structure of electronic eigenstates in typical covalent bonds,\cite{Marti-Dafcik2024a} which are ubiquitous in chemistry, other systems might require different types of spin eigenfunctions. For completeness, here we present circuits required to prepare the remaining CSFs in Ref.~\citenum{Marti-Dafcik2024a}.

The state of linear and square (cyclic) \ce{H_4} at stretched geometries can be expressed as a superposition of two states,\cite{Marti-Dafcik2024a} each of the form in Eq.~\eqref{eqn:vcsf_n4_2}:
\begin{equation}
	\ket{\mathcal{O}_{0, 0}^{4, 2}} = \Big[\frac{1}{\sqrt{2}} (\ket{\al \be}- \ket{\be \al}) \Big]^{2}.
\end{equation} 
The general form is a tensor product of singlets (antisymmetric functions) of $N/2$ two-spin systems
\begin{equation}
	\ket{\mathcal{O}_{0, 0}^{N, 2}} = \Bigg[ \frac{1}{\sqrt{2}} (\ket{\al \be} - \ket{\be \al}) \Bigg]^{\frac{N}{2}}.
	\label{eqn:csf_geminals}
\end{equation}
This state, also considered in Refs.~\citenum{Sugisaki2016, Sugisaki2019}, is a product of Bell states and can be implemented trivially in constant depth and using only one CNOT gate per two-electron singlet $\Big[\frac{1}{\sqrt{2}}(\ket{10}-\ket{01})\Big]^{N/2} = \bigotimes_{i=1}^{N/2} \overline{CX}^{2i, 2i-1}R_y^{2i}(-\pi/2)\ket{00}^{N/2}$.
The total CNOT cost for preparing $\ket{\mathcal{O}_{0, 0}^{N, 2}}$, including the cost of the spin-to-Fock-space mapping $U_{\textrm{map}}$, is given by (Appendix \ref{apdx:circ_cost_different_csfs}):
\begin{equation}
	C_{\textrm{tot}}^{\textrm{all}} = \frac{3}{2}N.
\end{equation}
Since $R_y(-\pi/2) = HX$, this rotation can be implemented without any non-Clifford (e.g., Toffoli) gates. On a fault-tolerant architecture, the preparation of such states has negligible cost (Appendix \ref{apdx:circ_cost_faulttolerant}).

One might wonder if the circuits can be extended to prepare any type of CSF.
The efficient preparation of arbitrary spin eigenfunctions remains an open problem, as existing approaches with rigorous performance guarantees require exponentially many gates.\cite{Carbone2022b}
In general, highly symmetric states allow efficient state preparation procedures, and one might expect that the complexity of the state preparation circuit increases with the number of distinct coefficients in the CSF.
We hypothesize that most chemical systems contain enough structure that their spin coupling is given by paths that have a significant amount of symmetry.

Crucially, for applications in electronic structure, the choice of CSFs greatly depends on the choice of the single-particle basis as well as the method used to construct the many-body CSFs.\cite{Pauncz1979}
In particular, rather than working in a fixed orthonormal basis, it is more efficient to use a bespoke basis for each physically-relevant CSF. The choice of basis should appropriately reflect the physical spin coupling pattern, e.g., by partitioning the system into molecular fragments and separately parametrizing the correlations due to local vs. global couplings, as is done in this work.
The extension of our state preparation protocols to more systems is an interesting direction for future work.

\subsection{Linear combinations of spin eigenfunctions}\label{ssec:circs_lc}
When multiconfigurational initial states are required for VQE or QPE, such as for molecular bonds at intermediate bond lengths or for transition metal clusters,
it is useful to prepare a linear combination of CSFs $\ket{\Phi_\mathrm{LC}} = \sum_{j} c_j \ket{\Phi_j}$ for $j \in \{ 0, 2, ..., N\}$. For example, $N = 0, 2, 4, 6$ for \ce{N_2} (Eq.~\eqref{eqn:lc_of_csfs}).
This can be done by controlling the state preparation unitaries $V_j: \ket{\mathcal{O}_{0, 0}^{j, 1}} = V_j\ket{00}^N$ from an ancilla register, for each $j$.

The encoding used for the ancilla register can take various forms that trade off circuit depth with space (qubits). One could either work with one ancilla qubit with a higher gate cost, as shown in Ref.~\citenum{Tubman2018}, use a one-hot encoding with $L$ ancilla qubits for $L$ CSFs, or use a compressed register of $\log(L)$ ancilla qubits as is common in general Linear Combination of Unitaries approaches.\cite{Childs2012}
The optimal encoding of the ancilla register depends on the particular hardware and molecule considered.

The gate overhead for controlling the unitaries $V_j$ scales linearly with $N$, which is lower than the cost of the circuits for $V_j$ themselves (Appendix \ref{ssec:controlling}).
Exploiting the recursive structure of the Dicke circuits enables an additional reduction of the cost of implementing $\ket{\Phi_{\mathrm{LC}}}$. Specifically, from Eq.~\eqref{eqn:lemma2} it follows that 
\begin{equation}
	S_n = M_{l, l-1}S_{n-1}.
\end{equation}
This implies that the circuit for $V_{j-2}$ can be reused as a subcircuit for $V_{j}$ rather than being applied twice.

To summarize, the cost for preparing linear combinations of CSFs is only slightly higher than the cost of preparing a single CSF, where the exact overhead depends on the particular case considered.

\section{Heuristic quantum algorithms}
\label{sec:heuristic}
While the algorithmic error of fault-tolerant quantum algorithms can be 
bound analytically,\cite{Kitaev1995, Abrams1999, Lin2020a, Dong2022, Lin2022} the factors determining the performance of 
heuristic quantum algorithms such as VQE, quantum subspace diagonalization, and adiabatic state preparation are
less well understood. 
Regardless, the numerical evidence accumulated so far suggests that the accuracy and efficiency 
of heuristic algorithms strongly depends on the quality of the initial state. 
In this section, we show that spin-coupled reference states have the correct features to provide this increase in accuracy and efficiency.

\subsection{Variational quantum eigensolver}
\label{sec:vqe}

We first consider the VQE approach, which avoids deep quantum circuits by combining quantum and
classical processors in a hybrid algorithm.
A quantum circuit with parametrized gates prepares the electronic wavefunction on a quantum computer,
and a classical computer variationally optimizes the parameters.
This is repeated in an iterative loop until convergence is reached.

At a fundamental level, VQE exploits the variational principle to rotate the initial qubit state $\ket{0 \cdots 0}$
to the true electronic ground state. Naturally, the more complex the eigenstate,
the more parametrized quantum gates are required to accurately approximate it.
In practice, finding a compact and accurate ansatz is difficult.\cite{Burton2023}
In particular, strong electron correlation exacerbates the challenges of VQE,
which reduces the potential of this quantum approach to outperform classical algorithms.\cite{Hu2022a, DCunha2023} 
For example, benchmark calculations show that many \textit{ans\"{a}tze} accurately describe
weakly-correlated wavefunctions of molecular bonds at equilibrium geometries, but their accuracy 
deteriorates significantly at strongly-correlated stretched geometries.\cite{Hu2022a}

To reduce the computational cost, VQE implementations use a classical heuristic to obtain an approximate initial state, $\ket{\Phi}$,
usually the Hartree--Fock state. Here, we show that using a spin-coupled initial state
can greatly reduce the cost required to achieve high-accuracy approximations of strongly correlated ground states.

Formally, the quantum computer is first initialized to the reference state
\begin{equation}
	\ket{\Phi} = \Uref \ket{0}^{2M}.
	\label{eqn:ref_state_unitary}
\end{equation}
Here, $2M$ is the number of spin-orbitals (qubits) and $\Uref$ is the unitary transformation required to prepare
the reference state.
To further correlate the reference state,
a unitary transformation $\Uans(\bm{\theta})$ is applied to it.
This is implemented as a sequence of parametrized quantum gates:
\begin{equation}
	\ket{\Psi(\bm{\theta})} = \Uans (\bm{\theta}) \ket{\Phi} = \prod_i U_i (\theta_i) \ket{\Phi}.
    \label{eqn:ups}
\end{equation}
The particular choice and ordering of elementary unitary transformations $U_i$ defines a variational 
\textit{ansatz}, which determines the wavefunctions that can be generated by the quantum circuit through
variation of the rotation parameters $\bm{\theta}$, starting from the reference state $\ket{\Phi}$.
The variational energy 
\begin{equation}
E(\bm{\theta}) = \braket{\Phi | \Uans^\dagger (\bm{\theta})\, \hat{H}\, \Uans(\bm{\theta})|\Phi}
\end{equation}
is estimated by measuring Hamiltonian expectation values on each qubit followed by operator averaging.\cite{McClean2016}
These estimates are then passed to a classical computer, which uses traditional optimization algorithms to 
proposes updates to $\bm{\theta}$ that minimise $E(\bm{\theta})$.

Practical VQE implementations face substantial challenges.
Firstly, although compact and accurate electronic states can in principle be prepared with short circuit 
depths,\cite{Grimsley2019,Yordanov2021, Gustiani2023,Burton2023,Burton2024}
finding these \textit{ans\"{a}tze} currently relies on computationally expensive strategies
that optimize the gate sequences.\cite{Grimsley2019,Burton2023}
Secondly, the numerical optimization appears to be fundamentally challenging.
The non-linearity of the \textit{ansatz} means that $E(\bm{\theta})$
is highly non-convex and the optimization is prone to get stuck in local minima.\cite{Anschuetz2022}
Finally, estimating the energy may incur prohibitive measurement costs.\cite{Cai2020, Gonthier2022}

Strong electron correlation greatly increases the difficulties of VQE.\cite{Hu2022a, DCunha2023}
This poor performance can be attributed to the use of a mean-field RHF initial state,
which is uncorrelated and therefore burdens the ansatz circuit $\Uans$ to recover a large amount of electron correlation.
Instead, we propose to use a spin-coupled wavefunction as the initial state, which has a greater 
overlap with the true ground state. This greatly reduces the number of parametrized ansatz operators 
required to reach quantitative accuracy. 

\begin{figure}[htp]
	\includegraphics[width=1\linewidth]{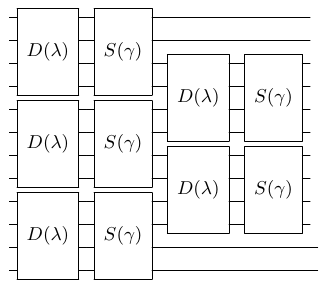}
	\caption{One layer of the QNP \textit{ansatz} ($k=1$) includes an alternating series of paired two-body operators $D(\lambda)$ and 
		spin-preserving one-body operators $S(\gamma)$. These are applied between neighboring spatial orbitals, each acting on four qubits, as
		described in Ref.~\citenum{Anselmetti2021}. The fermionic excitation operators $S(\gamma)$ and $D(\lambda)$ can be decomposed into Givens rotations that act non-trivially on a two-qubit subspace.\cite{Anselmetti2021, Arrazola2022}
	}
	\label{fig:qnp}
\end{figure}

We demonstrate this improvement by performing classical numerical simulations of VQE with
the quantum-number-preserving (QNP) \textit{ansatz},\cite{Anselmetti2021}
which uses a layered circuit structure that can be systematically improved to 
the exact result by increasing the number of repeating layers $k$.
Each layer includes a series of spin-adapted one-body rotations and paired two-body 
rotation operators acting between neighboring spatial orbitals, with the arrangement of 
these operators  depicted in Fig.~\ref{fig:qnp}.
This choice of operators preserves the spin quantum 
numbers of the initial state, which ensures that approximate wavefunctions are exact spin eigenfunctions.
Ref.~\citenum{Burton2023} proves that this type of 
\textit{ansatz} is universal, i.e. it can yield exact wavefunctions within the spin- and particle-number-preserving subspace.
In practice, it is unclear how many layers are required to achieve sufficient accuracy.

We consider two versions of VQE: a \textit{single reference} approach in which
a single unitary transformation is applied on one sole reference state composed of a linear combination of CSFs,
and a \textit{multireference} approach that applies a different unitary transformation on each individual CSF.


\begin{figure*}[t]
\includegraphics[width=\linewidth]{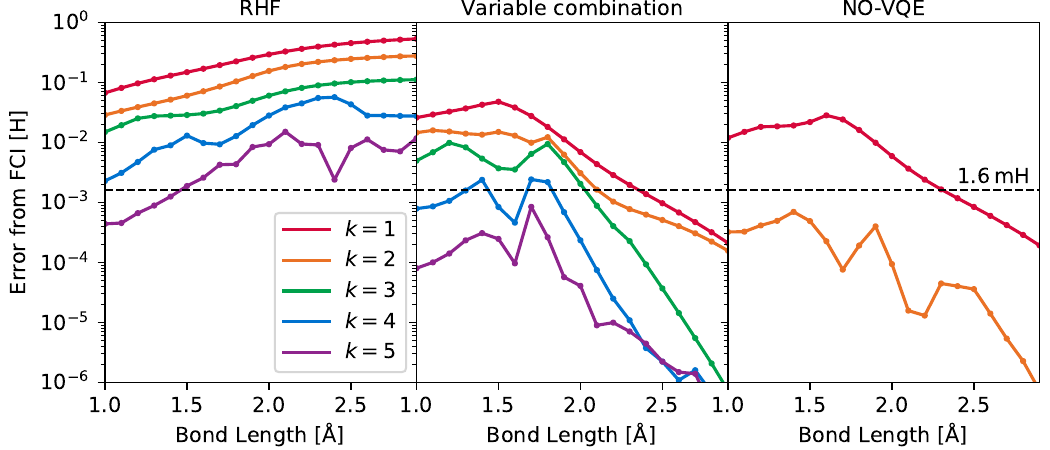}
\caption{Energy errors from the QNP \textit{ansatz} for \ce{N2} (STO-3G) using either a fixed RHF reference state or a variable linear combination of CSFs.
	The linear combination of CSFs significantly improves the accuracy in the strongly-correlated dissociation limit.
	The NO-VQE approach applies a bespoke QNP circuit to each reference state prior 
	to constructing the linear combination and reaches chemical accuracy using shallower
	circuits with $k=2$.}
\label{fig:qnp_vqe}
\end{figure*}

\subsubsection{Single-reference VQE with variable linear combination}\label{sssec:vqe_singleref}
\label{sec:variableVQE}
Our first approach is to apply the QNP \textit{ansatz} to an initial state defined 
as the variable linear combination
\begin{equation}
\ket{\Phi} = \sum_{I=1}^{L} C_I \ket{\Phi_I}.
\end{equation}
We optimize both the QNP rotation angles as well as the linear coefficients 
$C_I$ in the initial state definition, since the latter are also parametrized in terms of single-qubit rotations
through the circuits for preparing linear combinations of CSFs (Section \ref{ssec:circs_lc}).
This allows the linear combination of spin-coupled states to relax in the
presence of the correlation generated by the QNP circuit.

For the strongly correlated \ce{N2} binding curve, the variable linear combination of CSFs significantly reduces
the energetic error compared to the RHF initial state at large bond lengths (Fig.~\ref{fig:qnp_vqe}), reaching 
chemical accuracy at $R > \SI{2.5}{\angstrom}$ even with a single layer (i.e.\ $k=1$).
This result is expected since the spin-coupled wavefunction becomes exact in the dissociation limit (Section \ref{ssec:n2_diss}). 
Each layer corresponds to ten variational parameters (five for each operator in Fig.~\ref{fig:qnp}),
and there is a fixed constant number of three parameters required to prepare the linear combination of four CSFs (one is fixed through normalization).
Even around the equilibrium region, where the RHF configuration dominates the linear combination of CSFs (Fig.~\ref{fig:n2_csf_lin_comb}), 
the spin-coupled wavefunction improves the accuracy by around half an order of magnitude for each 
value of $k$. 
These results demonstrate that using a linear combination of CSFs to define the initial state reduces the 
number of parameters and unitary operators required to reach quantitative accuracy with a VQE \textit{ansatz}.

\subsubsection{Multireference ansatz with a nonorthogonal variational quantum eigensolver}
\label{ssec:novqe}
An alternative strategy is to uniquely correlate each reference state.
This is particularly useful for multiconfigurational systems where more than one CSF
dominates the wavefunction. Since each CSF is a unique vector in the Hilbert space,
separately rotating each vector enables multiple regions of the Hilbert space to be explored in a targetted manner.
Although this idea has been explored in other works e.g. using unrestricted Hartree--Fock determinants,\cite{Baek2023}
our spin-coupled wavefunction approach has the advantage that significant entanglement is already included in the reference state. Achieving the same result through single-determinant reference states would require exponentially many initial states.
Furthermore, since we strictly use CSFs, any spin-preserving ansatz is guaranteed to produce
spin-pure wavefunctions, without relying on approximately recovering the correct spin state after symmetry breaking.

On a quantum circuit implementation, the multireference approach avoids the preparation of 
a linear combination of different CSFs on the quantum register and instead relies
on the measurement of Hamiltonian and overlap matrix elements between the 
individually correlated reference states. Furthermore, the increased number of variational parameters
provides a very flexible ansatz using limited circuit depth.

Formally, a unique set of unitary operations is applied to the each 
reference state before constructing the  linear combination, giving
\begin{equation}
\ket{\Psi(\bTh, \bm{C})} = \sum_{I=1}^{L} C_I \qty(\prod_i U_{Ii}(\theta_{Ii}) \ket{\Phi_I}),
\label{eq:novqewfn}
\end{equation}
where $\bTh = (\bm{\theta}_1,\dots,\bm{\theta}_L)$.
Each reference state is correlated with a bespoke set of quantum gates that each have a unique variational 
parameter, and the gate parameters $\bTh$ are optimized simultaneously with the linear combination $\bm{C}$.
The correlated basis states in this expansion will generally not be orthogonal, and we must therefore
obtain the overlap matrix elements to guide the optimization as in any nonorthogonal quantum eigensolver.\cite{Baek2023, Huggins2020b}
Here, we run classical simulations of this nonorthogonal VQE (NO-VQE) algorithm. We describe details of our numerical implementation in Appendix~\ref{apdx:novqe}.

The multireference NO-VQE provides chemically accurate energies across
the full binding curve of \ce{N2} (STO-3G) using only two layers of the QNP 
\textit{ansatz} for each reference state (Fig.~\ref{fig:qnp_vqe}, right panel).
In contrast, five layers are required to reach an equivalent accuracy when applied on a single reference state
(Fig.~\ref{fig:qnp_vqe}, central panel), as per the approach described in Section \ref{sssec:vqe_singleref}.
This corresponds to a reduction in circuit depth by a factor of 2 to 3,
which is highly desirable on noisy quantum hardware.
Therefore, although the NO-VQE approach introduces more parameters, it 
achieves greater accuracy with a shallower circuit than applying a single unitary to the 
variable linear combination of CSFs.
The efficient exploration of the Hilbert space achieved by starting from different spin-coupled wavefunctions
makes this a natural application of our work in multiconfigurational systems.

\subsection{Quantum subspace diagonalization through real-time evolution}\label{ssec:qsd}
Although VQE might be suited for noisy hardware because typical ansatz circuits have low gate counts and depths,
severe implementation challenges in the optimization and measurement make its practical use questionable.\cite{Gonthier2022, Anschuetz2022, Dalton2024}

In this section, we consider an approach that avoids non-linear optimization and instead requires real-time-evolution of the reference states under the Hamiltonian. This falls within the broader category of quantum subspace diagonalization (QSD) methods (one could also consider the nonorthogonal VQE from Section \ref{ssec:novqe} as a QSD approach).
QSD methods are uniquely well-suited for exploiting different reference states like the ones presented in this work, as they
can explore the Hilbert space from multiple directions in parallel, through transformations applied separately on each reference state.
The ability to rotate each reference state individually relaxes the accuracy requirement for each ansatz circuit, as the classical diagonalization step gives increased variational freedom.
Here, we will show how using spin-coupled reference states can greatly improve the performance of QSD methods for both ground and excited state calculations.

\subsubsection{Background}

Quantum subspace diagonalization methods have emerged as a promising class of hybrid quantum-classical algorithms for computing Hamiltonian eigenvalues. 
Starting from a reference state $\ket{\Phi}$, the quantum device is used to generate a set of  basis states $\{ \ket{\Psi_j} \}$ using various unitary transformations $U_j$, where $\ket{\Psi_j} = U_j \ket{\Phi}$. The Hamiltonian is then diagonalized within the corresponding subspace to give the optimal linear combination
\begin{equation}
	\ket{\psi(\bm{v})}=\sum_j v_j \ket{\Psi_j}.
\label{eqn:ansatz_qsd}
\end{equation}
In general, the basis states are mutually nonorthogonal and  the subspace expansion takes the form of a nonorthogonal configuration interaction with the generalized eigenvalue problem
\begin{equation}
	\bm{H}\bm{v}=E\bm{S}\bm{v},
\label{eqn:gen_eval}
\end{equation}
where $H_{jk} = \braket{\Psi_j| \hat{H} | \Psi_k}$ and $S_{jk} = \braket{\Psi_j | \Psi_k}$. This eigenvalue problem is solved on a classical computer.
An advantage of this subspace diagonalization is that the quantum device is only used to measure the matrix elements $H_{jk}$ and $S_{jk}$, avoiding the significant overhead in gate count and circuit depth associated with explicitly preparing the linear combination in Eq.~\eqref{eqn:ansatz_qsd} as a quantum circuit. Furthermore, QSD also gives direct access to excited state energies. 
Various strategies for generating the basis states have been proposed, including: 
\begin{enumerate}
\item fermionic excitation operators applied to a ground state approximation to target excited states,\cite{McClean2017, Colless2018} 
\item application of parametrized quantum circuits $U_j(\bm{\theta_j})$ defined through optimization\cite{Huggins2020b} or perturbation theory,\cite{Baek2023}
\item application of the imaginary time-evolution operator $U_j = e^{-H\tau_j}$, where $\tau_j = j\, \Delta t$ is total duration of the evolution in imaginary time defined by increasing integers $j = 0, 1, 2, ...$,\cite{Motta2019}
\item real-time evolution operators $U_j = e^{-iH \tau_j}$, where $i$ is the imaginary unit.\cite{Stair2020a, Seki2021, Klymko2022, Cortes2022, Epperly2022, Shen2023, Stair2023, Kirby2024}
\end{enumerate}

Here, we focus on the real-time evolution approach, also referred to as variational quantum phase estimation (VQPE)\cite{Klymko2022} or quantum Krylov.\cite{Stair2020a, Stair2023, Kirby2024} This method only requires transforming the reference state through unitary time-evolution, which is a natural operation for quantum processors whose quantum circuit implementation has greatly been optimized over the past decades.\cite{Abrams1999, Aspuru-Guzik2005, Lee2021, vonBurg2021, Cortes2022} Previous work has shown that the real-time-evolved states provide a very compact subspace for diagonalizing the Hamiltonian, in terms of the number of variational parameters needed.\cite{Stair2020a, Klymko2022, Shen2023, Cortes2022} Furthermore, numerical\cite{Klymko2022} and theoretical analysis\cite{Epperly2022, Kirby2024} suggest that this algorithm is robust to noise. 
Starting from the RHF reference state, the subspace is typically built using a linear grid of $N_T$ equally-spaced time points $\tau_j = j \Delta t$ ($j=0, 1, ..., N_T$), where the fixed time step $\Delta t$ must be chosen prior to the calculation.
Naively, a total of $2M^2$ matrix elements would need to be evaluated on the quantum device in order to set up the eigenvalue problem (Eq.~\eqref{eqn:gen_eval}), where $M$ is the number of expansion states.
This computation can be reduced to $2M$ unique matrix elements if the time-evolution operator is implemented exactly, since the Hamiltonian and overlap matrices then have a Toeplitz structure, $H_{j, k} = H_{j+1, k+1}$ and $S_{j, k} = S_{j+1, k+1}$.\cite{Klymko2022}
However, if the time-evolution operator is approximated through Trotterization,\cite{Aspuru-Guzik2005} then the Toeplitz structure of the Hamiltonian matrix is lost. In this scenario, and for linear time grids, the Toeplitz structure can still be recovered if the eigenvalue problem is reformulated using the time-evolution operator instead of the Hamiltonian (see Appendix \ref{sec:vqpe_toeplitz} and Refs.~\citenum{Parrish2019b, Klymko2022}).

The main disadvantage of VQPE is that relatively deep quantum circuits are required to implement the time-evolution operator.
Specifically, the depth and gate count scale linearly with $N_T$.
The performance of VQPE worsens in the presence of strong correlation, meaning that more time steps are required.\cite{Stair2020a}
The convergence with respect to $N_T$ has also been shown to be slower for excited-state energies.\cite{Klymko2022, Cortes2022}
Furthermore, numerical results suggest that the  accuracy of ground-state calculations deteriorates when the time-evolution is implemented with low-order Trotter approximations.\cite{Stair2020a} Higher-order Trotter formulas could be applied,
but this comes at great cost since the circuit depth for the $k$-th order Trotter formula scales exponentially as $\mathcal{O}(5^{k})$.
Below, we show how combining QSD with spin-coupled reference states can significantly mitigate these limitations.

\subsubsection{QSD with spin-coupled reference states}

Rather than using more time steps, the deep circuits associated with the time-evolution operator can be reduced by building the subspace using multiple reference states.\cite{Seki2021, Stair2020a}
For example, Stair \textit{et al.}\ used a set of reference determinants that are identified using iteratively-grown subspaces.\cite{Stair2020a}
Here, we show that using the physically-inspired spin-coupled wavefunctions introduced in Ref.~\citenum{Marti-Dafcik2024a} (see Sections \ref{sec:background}, \ref{sec:qcircs}) to define the reference states allows for the accurate computation of ground- and excited-state energies at significantly reduced circuit depth compared to a single RHF reference state.
Starting from $N_R$ spin-coupled reference states $\{ \ket{\Phi_1}, \dots, \ket{\Phi_{N_R}} \}$, we construct a subspace $\{ e^{-i H \tau_j } \ket{\Phi_1}, \dots, e^{-i H \tau_j } \ket{\Phi_{N_R}} \}$ for $j = 1, \dots, N_T$, which has $N_R \times (1 + N_T)$ states. Note that each reference state is time-evolved independently.

We test this approach for \ce{N2} at $R=\SI{1.5}{\angstrom}$ using noiseless simulations on classical hardware.
We use a maximum of $N_T = 120$ steps in a linear time grid $t_j = j \Delta t$ for three different time-step values $\Delta t \in \{0.1, 1.0,  2.0\}$.
We set thresholds of $10^{-6}$, $10^{-4}$, and $10^{-2}$ for the singular values of the overlap matrix to remove the null space in the generalized eigenvalue problem. Higher thresholds might be more suited for implementation on noisy devices, whereas using lower thresholds can help to retain more expansion states and achieve faster convergence. We observed no qualitative difference for the three different thresholds and therefore limit our discussion to results with the threshold $10^{-6}$.
We apply a first-order Trotter approximation of the time-evolution operator
for a Hamiltonian given by a sum of terms $H = \sum_k h_k$ in the Pauli basis, given by
\begin{equation}
	e^{-iHt_j} \approx \prod_k e^{-ih_k t_j}.
\label{eqn:firstorderTrotter}
\end{equation}

\begin{figure*}
	\includegraphics[width=1.0\linewidth]{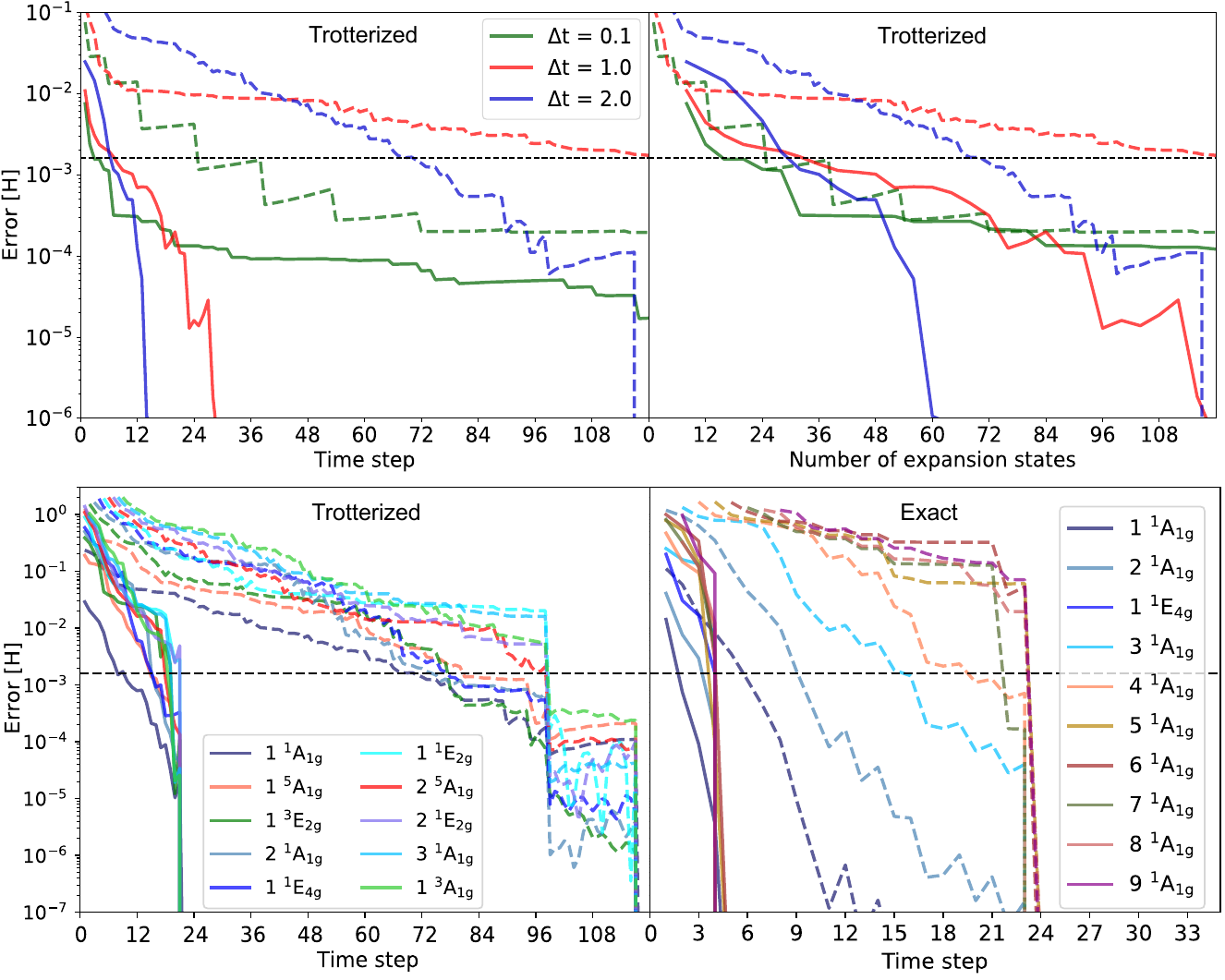}
	\centering
	\hfill
		\caption{When applied on spin-coupled reference states, real-time evolution accurately yields ground and excited state energies after subspace diagonalization, requiring very few time steps.
		Top row: Ground state energy error as a function of the number of time steps (left) and number of expansion states (right), where the time-evolved states are obtained through first-order Trotter evolution (Eq.~\eqref{eqn:firstorderTrotter}). Bottom row: Energy error for the 10 lowest eigenstates obtained from first-order Trotter evolution (left), and for the 10 lowest singlet eigenstates obtained from exact time-evolution (right), with time-evolution step size $\Delta t = 2.0$.
		Dashed lines correspond to subspaces formed from time-evolving the RHF state; solid lines correspond to time-evolving four spin-coupled reference states.
		The labels indicate the spin (superscript) and spatial symmetry of each eigenstate, where the first number is the index of the eigenstate of a particular symmetry, in increasing energy and starting at 1 for the ground state.
		All results are for N$_2$ in the STO-3G basis at bond length $R=\SI{1.5}{\angstrom}$.
		}
	\label{fig:vqpe}
\end{figure*}

Compared to a single RHF reference state, using multiple spin-coupled reference states provides significantly faster convergence with respect to the number of time steps (and thus circuit depth) for every $\Delta t$ considered (Fig.~\ref{fig:vqpe}, top left).
This implies that the multireference approach should be faster and less susceptible to hardware noise, which becomes worse for deeper quantum circuits.
One might wonder whether this comes at the cost of an increased number of measurements, since the increased number of reference states results in an additional $N_R - 1$ subspace states per time step.
Encouragingly, we find that the multireference expansion also converges more rapidly with respect to the total number of expansion states, meaning that fewer measurements would be required overall (Fig.~\ref{fig:vqpe}, top right).
The faster convergence with the number of expansion states indicates that starting from different reference states and independently time-evolving each of them might help to explore the Hilbert space more efficiently than if the time-evolution is applied onto a single same initial state.

The accuracy of VQPE strongly depends on the choice of $\Delta t$. We can understand this dependence through the phase cancellation picture discussed in Ref.~\citenum{Klymko2022}, since $\Delta t$ controls the phase applied to each eigenstate contained in the reference space. Our results suggest that larger values of $\Delta t$ enable faster cancellations and avoid the step-like plateaus that occur for $\Delta t =0.1$, which are associated with near-linear dependencies in the expansion subspace.

VQPE can also provide access to excited state energies, a key benefit that has received surprisingly limited attention so far.
When the time-evolution is Trotterized at first order,  VQPE with both the RHF or multireference expansion can compute the lowest eigenstates with spatial symmetry corresponding to $\mathrm{A_{1g}}$, $\mathrm{E_{2g}}$, and $\mathrm{E_{4g}}$, with no restriction on the spin symmetry (Fig.~\ref{fig:vqpe}, bottom left).
Like for the ground state calculation, the multireference expansion requires
an order of magnitude fewer time steps to reach the exact excited state energies compared to the single RHF reference state.
The presence of excited states with different spin symmetry to the reference state arises because the Trotter approximation of the time-evolution operator breaks the spin symmetry of two-body operators in the Hamiltonian.
Similarly, non-abelian spatial symmetries in the $\mathrm{D_{\infty h}}$ point group are also partially broken since the Trotterized time-evolution operator only conserves symmetries within the $\mathrm{D_{2h}}$ abelian subgroup. 
Since the initial states are totally symmetric, this means that we obtain excited states that reduce to $\mathrm{A_g}$ in $\mathrm{D_{2h}}$ (which correspond to the $\mathrm{D_{\infty h}}$ irreps $\mathrm{A_{1g}}$ and $\mathrm{E}_{n\mathrm{g}}$ for even $n$).
Consequently, although approximate energies can be obtained with high accuracy,
the corresponding eigenstates might not have pure spin or spatial symmetry unless they are exact (within numerical accuracy).

Contrast this with the version where we apply the exact time-evolution operator (Fig.~\ref{fig:vqpe}, bottom right).
Since the exact time-evolution operator commutes with all the symmetry operators of the Hamiltonian,
the time-evolved subspace states are guaranteed to preserve the total spin and the spatial symmetry of the reference state.
This allows us to systematically target states of a particular symmetry. Here, we consider the lowest $^1\mathrm{A_{1g}}$ states in \ce{N2}.
QSD with exact time evolution applied on multiple CSFs yields all ten eigenenergies with only 5 time steps,
whereas QSD applied on the RHF state requires 24 time steps.
(Note that our multireference approach also yields states of $^1\mathrm{E_{4g}}$ symmetry since the CSF $\ket{\Phi_2}$, defined in Eq.~\eqref{eqn:v2xy_n2}, contains spatial contaminants with azimuthal orbital angular momentum $L_z=4,8,\dots$.)
It is remarkable that our multireference approach provides such rapid convergence despite
the fact that the spin-coupled configurations are tailored explicitly for the ground state.\cite{Marti-Dafcik2024a}
This can be understood because the excited states of \ce{N_2} correspond to configurations where electrons are promoted from bonding to antibonding orbitals, which are exactly the configurations included in our CSF reference states.

In summary, defining a set of multiple reference states using spin-coupled configurations
significantly reduces the number of time steps required to converge ground- and excited-state VQPE energies compared to a single RHF reference state.
This faster convergence results in shallower quantum circuits, making the VQPE algorithm more suitable for implementation on quantum devices.
We believe that this success arises because the multireference expansion is conducive to exploring distant sectors of the Hilbert space.
Any QSD method requires the generation of linearly dependent basis states,\cite{Klymko2022}
but this is generally hard to achieve due to the redundancies that arise when applying the time-evolution operator to a single reference state.
By independently time-evolving each reference state, we generate a basis that has fewer linear dependencies and is thus more efficient at phase cancellation.

The \textit{perturb-then-diagonalize} form of QSD is naturally well-suited for multireference quantum chemistry.
Just as for the nonorthogonal VQE algorithm explored in \ref{ssec:novqe}, QSD with real-time-evolved states
is greatly enhanced by using spin-coupled initial states that contain a large part of the relevant electron correlation.

\subsection{ADAPT-QSD: QSD with adaptive ansatz}\label{ssec:adapt-qsd}
\begin{figure*}[htp!]
	\includegraphics[width=0.9\linewidth]{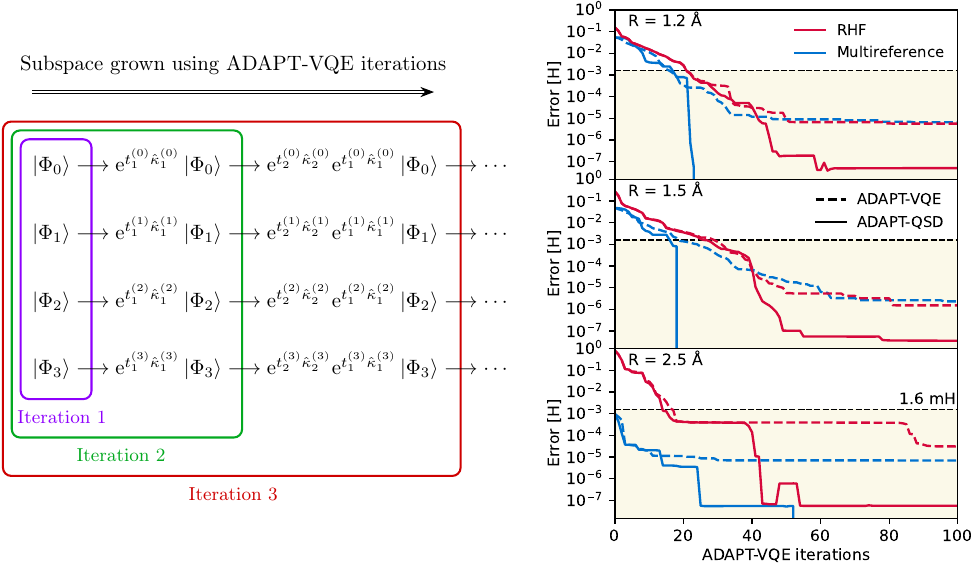}
	\caption{ADAPT-QSD: Performing QSD in the basis of sequential ADAPT-VQE states can significantly accelerate the convergence
		with respect to the number of operators and avoid stagnation, as demonstrated here for \ce{N2} (STO-3G) at 
		$R= \SI{1.5}{\angstrom}$.}
	\label{fig:adapt-qsd}
\end{figure*}

\begin{figure}[hbt!]
	\includegraphics[width=\linewidth]{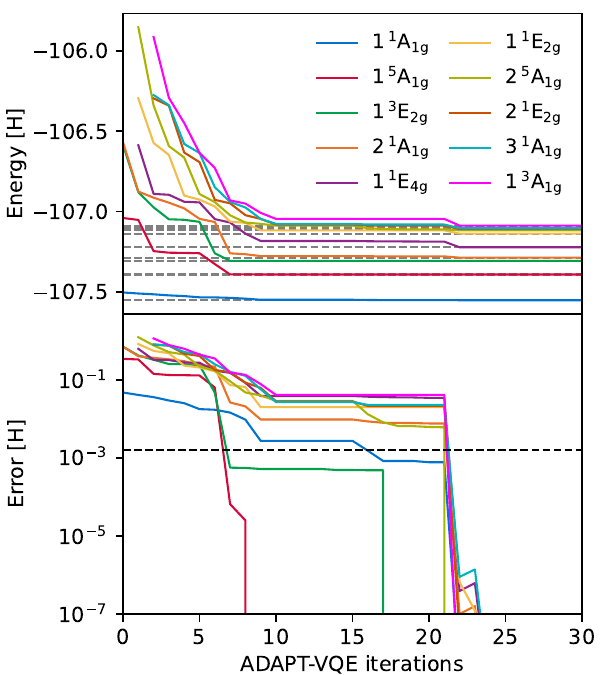}
	\caption{Multireference ADAPT-QSD captures the low-lying excited states with the same spatial symmetry as the input CSFs, as shown for \ce{N2} at $R=\SI{1.5}{\angstrom}$. The lower energy states converge more rapidly as the subspace expands, and the states considered here are converged below chemical accuracy using 22 ADAPT-VQE iterations.}
	\label{fig:adapt-qsd-ex}
\end{figure}

Despite its simplicity, the real-time evolution approach to QSD requires circuits that are significantly deeper than is feasible with near-term quantum hardware,
because each time-evolution step requires the application of every term in the Hamiltonian.
Although the circuit depth can be reduced through randomized Hamiltonian simulation,\cite{Stair2023} typically such circuits are still deeper than those for a VQE ansatz.
To mitigate this, we propose an alternative QSD approach using adaptive quantum eigensolvers.

Rather than using a fixed quantum circuit, the ADAPT-VQE algorithm dynamically grows an ansatz circuit tailored to the system of interest.\cite{Grimsley2019}
At each iteration, the gradient of the energy estimate with respect to a pool of candidate operators is measured.
The operator with the largest gradient is appended to the ansatz circuit,
and all variational parameters of the ansatz unitary are reoptimized following the usual quantum-classical VQE loop.
The ADAPT-VQE algorithm has been shown to provide very compact representations of electronic wavefunctions.\cite{Grimsley2019,Yordanov2021}
However, as the ansatz growth is only guided by a local gradient criterion,
its convergence can sometimes be slow and it can get stuck in a local minimum within the discrete operator space.\cite{Burton2023}
Furthermore, as with any VQE algorithm, the parameter optimization can converge to one of the many high-energy local minima in the continuous space.\cite{Anschuetz2022}

Here, we propose a QSD approach where a subspace is defined using the sequence of wavefunctions
obtained on each iteration of the ADAPT-VQE algorithm (Fig~\ref{fig:adapt-qsd}; left).
This can be done either using a single reference wavefunction such as the restricted Hartree--Fock (RHF) state,
or multiple reference states including different spin-coupled wavefunctions.
ADAPT-QSD performs individual ADAPT-VQE calculations on each reference state,
and uses the combined set of correlated states to define the subspace within which the Hamiltonian is diagonalized.

Our numerical simulations on \ce{N2} demonstrate that, even in the single reference case, this approach can mitigate two of the main difficulties of the ADAPT-VQE algorithm.
Firstly, ADAPT-QSD significantly accelerates the convergence of the energy estimates with respect to the number of operators,
and therefore reduces the circuit depth (Fig~\ref{fig:adapt-qsd}, right).
Secondly, when the ADAPT-VQE wavefunctions stagnate at a local energy mininum,
subspace diagonalization in ADAPT-QSD can escape that minimum,
yielding an energy estimate that is orders of magnitude lower than ADAPT-VQE at comparable circuit depth.

Comparing the role of the reference states, our results again confirm the advantage of using multiple spin-coupled reference states in QSD approaches.
We find that the multireference approach to ADAPT-QSD requires significantly fewer ADAPT-VQE iterations compared to the approach that employs a single RHF reference state.
We achieve convergence to the exact ground state energy using multireference ADAPT-QSD
with roughly half the number of operators per circuit needed in the single reference case, further reducing the circuit depth requirements.

ADAPT-QSD also systematically yields the low-lying excited states in the molecule, as shown in Fig.~\ref{fig:adapt-qsd-ex}.
This enables excited state energies to be computed using the ADAPT-VQE formalism without requiring constrained optimization.\cite{Higgott2019}
For the results presented here, we chose to work with the operators presented in the original ADAPT-VQE paper,\cite{Grimsley2019}
which are not spin-symmetry-preserving.
Furthermore, these operators also do not conserve all spatial symmetries for non-abelian point groups.
Therefore, like the Trotterized RT-QSD approach, the ADAPT-QSD expansion yields excited states that transform as the $\mathrm{A_{g}}$ irreducible representation of the $\mathrm{D_{2h}}$ abelian subgroup.
One could use symmetry-preserving operators if states of a particular symmetry were desired.

To summarize, we have considered three algorithms that perform subspace diagonalization:
the nonorthogonal quantum eigensolver, QSD with real-time-evolved states (also known as VQPE or quantum Krylov),
and our newly-proposed ADAPT-QSD. We have seen that the performance of all three algorithms significantly improves when combined with spin-coupled reference states.
This is particularly useful for multiconfigurational wavefunctions, including excited states, which are precisely the situations for which classical algorithms are known to struggle.
QSD appears particularly promising because it exploits the quantum device to represent highly entangled states in different bases,
as well as to yield matrix elements in nonorthogonal bases.
For comparison, the ability to efficiently provide such information through measurements
is also the core argument for potential quantum advantage with other quantum algorithms.\cite{Huggins2020b, Baek2023, Scheurer2024, Leimkuhler2024}
 
There is an additional reason to prefer QSD over VQE methods.
Often, VQE implementations struggle to resolve the last few digits of accuracy in the energy estimate.
Typically, this occurs either due to a lack of expressiveness in the ansatz
or because the optimization gets stuck in local traps.\cite{Vanstraaten2021, Anschuetz2022}
This becomes particularly difficult when hardware noise is present in a real quantum device implementation,
because the gate errors are usually higher than the $~ 10^{-3}\,\mathrm{E_h}$ accuracy needed for the energy estimate in chemistry applications.\cite{Dalton2024}
We expect that QSD methods can mitigate this problem since even if hardware noise corrupts the measured overlap and Hamiltonian matrix elements,
high accuracy might still be achieved through the linear QSD expansion, in particular
when the diagonalization step is regularized through thresholding the eigenvalue problem, as done in our simulations.
There is theoretical\cite{Epperly2022, Kirby2024} and numerical\cite{Klymko2022, Kirby2023} evidence
that the quantum Krylov approach discussed in Section \ref{ssec:qsd} is indeed noise resilient when combined with thresholding.
It is likely that this noise-resilience is also a feature of ADAPT-QSD, but further work is needed to establish this.

\subsection{Adiabatic state preparation from the molecular dissociation limit}\label{ssec:asp}

All the algorithms we have considered so far are hybrid quantum-classical algorithms that exploit
the variational principle to minimize the energy through iterative updates of some trial ansatz parameters.
Adiabatic state preparation (ASP) offers an interesting alternative.
It is a purely-quantum algorithm for preparing Hamiltonian ground states on digital quantum computers
that does not rely on any parametrized \textit{ansatz}.\cite{Farhi2000, Albash2018}
Instead, it exploits the adiabatic theorem, which requires using some Hamiltonian whose ground state
can be easily prepared on a quantum computer.

Typically, the initial state of choice is the Hartree--Fock state.\cite{Aspuru-Guzik2005}
As usual, this works well for weakly correlated problems, but it is inefficient for strongly correlated systems.
The CSFs considered in this work are the exact ground state of a family of
molecular Hamiltonians at dissociation (Section \ref{sec:background} and Ref.~\citenum{Marti-Dafcik2024a}).
Here, we show that by starting from the dissociation limit with a spin-coupled state,
we can speed up adiabatic state preparation of strongly correlated eigenstates.

\subsubsection{Background}

The ASP algorithm was proposed as a method to prepare initial states with high overlap with the ground state for quantum phase estimation.\cite{Aspuru-Guzik2005, Reiher2017, Lee2023}
Starting from some initial Hamiltonian $\HO$ whose ground state can be easily prepared on a quantum computer, the ground state of the full Hamiltonian $\Hf$ can be generated by slowly evolving the state along a pathway that interpolates between $\HO$ and $\Hf$, defined as
\begin{equation}
	\hat{H}\big(s(t)\big) = \big(1-s(t)\big)\HO + s(t)\Hf.
\end{equation}
Here, $s(t)$ is some continuous function with $s(0) = 0$ and $s(\tau) = 1$, where $\tau$ is the total evolution time.
The required total evolution time $\tau$ can be estimated as\cite{Albash2018}
\begin{equation}
\tau \gg \max_{s\in[0,1]}\frac{\abs{\bra{\Psi_1(s)}∣\partial_s H(s)∣\ket{\Psi_0(s)}}}{\abs{E_1(s) - E_0(s)}^2},
\end{equation}
where $\ket{\Psi_0}$ and $\ket{\Psi_1}$ are the ground and first excited states of $H(s)$, respectively.
On a digital quantum device, the path variable $s(t)$ can be discretized using constant time steps $\Delta t$.
Therefore, slower evolution, which increases the likelihood of remaining in the ground state,
requires a larger number of time steps and deeper quantum circuits.


Since the ground state of $\HO$ must be easily obtained and prepared, the obvious choice for 
electronic problems is the mean-field Fock operator, for which the (restricted) Hartree--Fock wavefunction is the ground state.\cite{Veis2014}
The accuracy of ASP is typically quantified using the squared overlap $\abs{\braket{\Psi(\tau)|\Psi_0}}^2$ between the final state $\ket{\Psi(\tau)}$ and the physical ground state $\ket{\Psi_0}$.
Numerical studies have shown that the accuracy strongly depends on the initial state,\cite{Veis2014, Kremenetski2021, Sugisaki2022a, Lee2023}
meaning that a mean-field reference is unlikely to be sufficient for strong electron correlation.
Indeed, simulations of chemical bond breaking have demonstrated that much larger $\tau$ values are required at long bond lengths due to both the inadequacy of the Hartree--Fock state and the small energy gap between eigenstates that leads to an exact degeneracy for $R\rightarrow \infty$.\cite{Veis2014, Sugisaki2022a}

To overcome the limitations of the mean-field Fock operator, alternative initial states and reference Hamiltonians 
have been explored, including 
Hamiltonians involving a subset of the molecular orbitals with complete active space configuration interaction (CASCI) wavefunctions,\cite{Veis2014, Kremenetski2021} 
active space Hamiltonians obtained from $n$-electron valence perturbation theory (NEVPT)\cite{Lee2023}, 
and unrestricted Slater determinants that break $\hat{S}^2$ symmetry.\cite{Sugisaki2022a}
The CASCI approach reduces the adiabatic evolution time, but it requires the brute-force computation of the ground state wavefunction within the active space, which is difficult to predict \textit{a priori}. Thus, active-space methods are unlikely to be applicable for systems with many strongly correlated electrons. 
Furthermore, while the symmetry-broken states analyzed in Ref.~\citenum{Sugisaki2022a} can be easily prepared on a quantum circuit, the spin symmetry must be restored by adding a penalty term to constrain the $\expval*{\hat{S}^2}$ expectation value, which increases the circuit cost of implementing time-evolution.

\subsubsection{Interpolation from the exact dissociation limit}
\begin{figure*}[htb]
	\begin{subfigure}{0.48\textwidth}
		\centering
		\includegraphics[width=\textwidth]{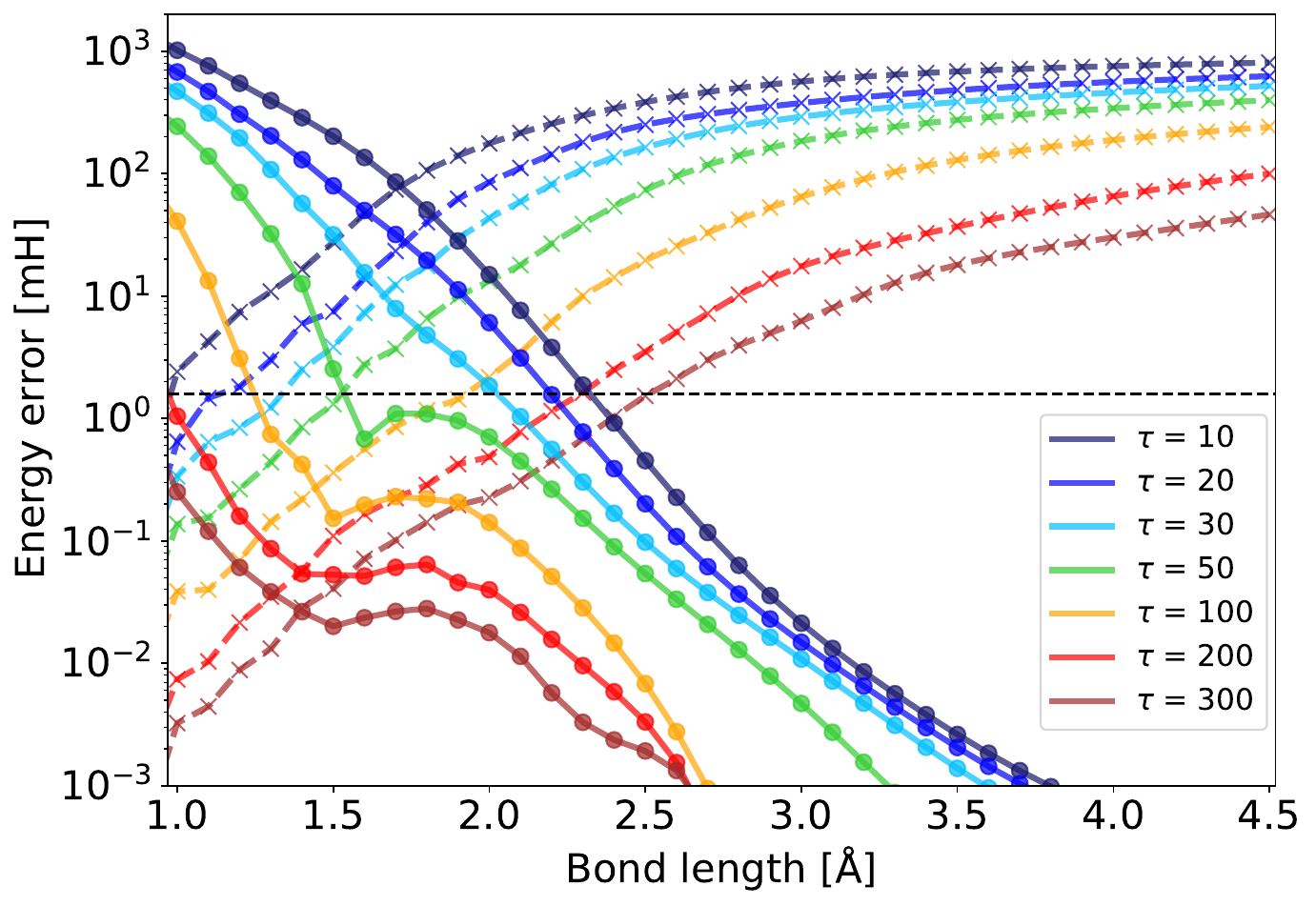}
	\end{subfigure}
	\begin{subfigure}{0.48\textwidth}
		\centering
		\includegraphics[width=\textwidth]{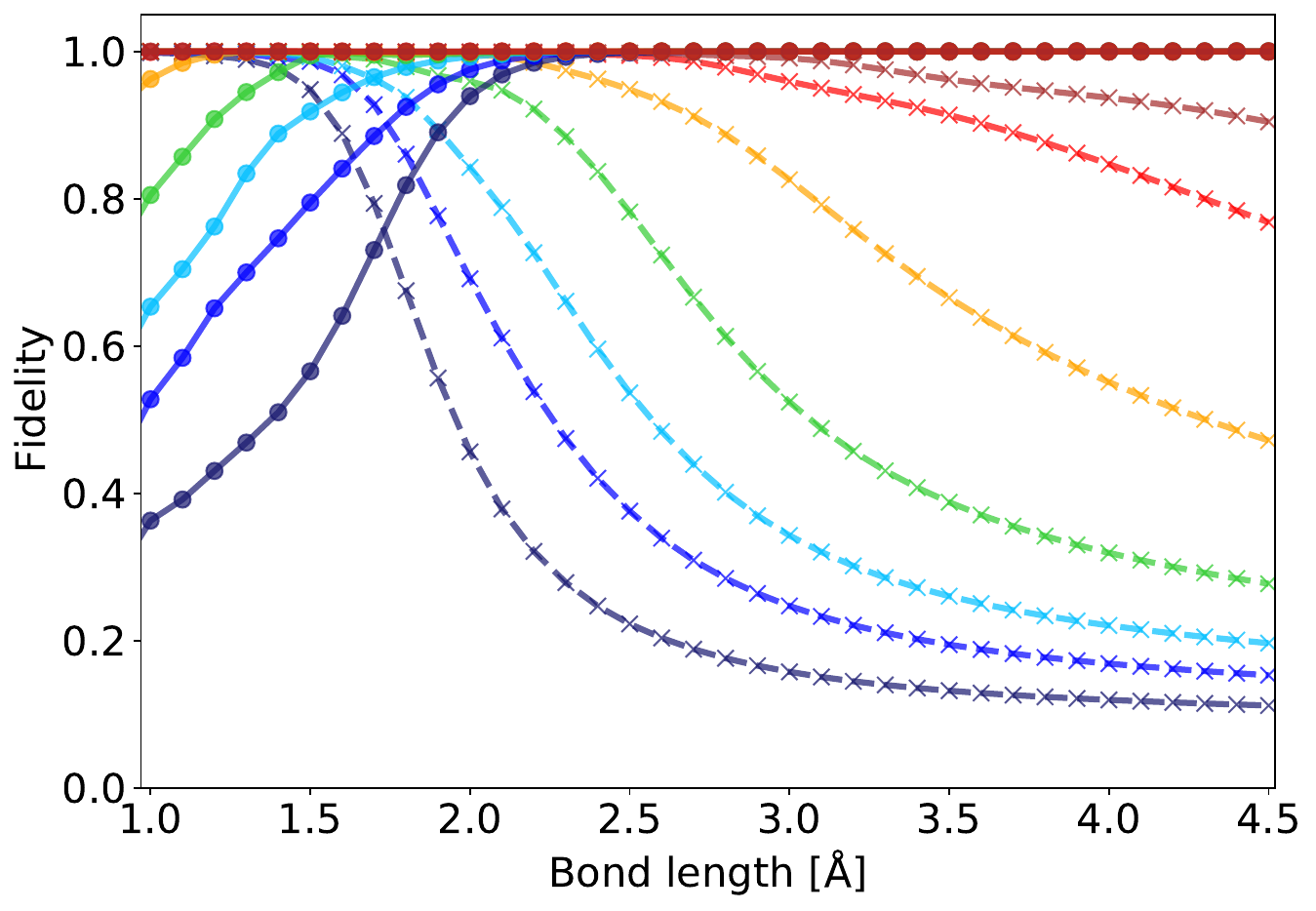}
	\end{subfigure}
	\centering
	\hfill
	\centering
	
	\caption{Adiabatic state preparation is greatly accelerated by using a spin-coupled wavefunction in the strong correlation limit. (Left)
		Error of energy estimates and (right) squared overlap between the exact ground state for N$_2$ and the adiabatically evolved states after full evolution ($s(\tau)=1$). Solid lines correspond to the version of ASP that starts with the state $\vsix$ and the Hamiltonian at dissociation, whereas dashed lines refer to ASP starting with the RHF state and the Fock operator. The total evolution time is $\tau$, and the horizontal black line denotes chemical accuracy in the energy estimate.}
	\label{fig:asp_n2}
\end{figure*}

Here, we exploit the fact that the  ground state of dissociated molecules can often be described using only one spin
eigenfunction of the form $\vn$.\cite{Marti-Dafcik2024a} This suggests an alternative approach to apply ASP in the regime of strong correlation. 
We propose using the ground state and the fully-interacting Hamiltonian at molecular dissociation as the starting point,
and applying ASP to interpolate between the dissociation limit and the target geometry to compute the entire binding curve.
We numerically investigate this approach through the \ce{N_2} molecule.

The ground state for \ce{N2} at dissociation is exactly represented by one open-shell CSF $\vsix$
which spin-couples the six valence p-electrons and can be predicted \textit{a priori}, from symmetry arguments (see Sections~\ref{sec:background}).
We define the initial Hamiltonian $\HO$ as the full interacting Hamiltonian within the (6, 6) active space at $R= \SI{4.5}{\angstrom}$,
which is representative of the dissociation limit since $1-\abs{\braket{\mathcal{O}_{0, 0}^{6, 1}|\Psi_0}}^2 \approx 4 \times 10^{-7}$,
where $\ket{\Psi_0}$ is the exact ground state.
For simplicity, we only consider the linear interpolation  $s(t) = t/\tau$. 
We use exact time evolution with a constant timestep of $\Delta t = 0.1\,\mathrm{E_h^{-1}}$ and different total evolution times up to $\tau =  300\,\mathrm{E_h^{-1}}$.
We compare CSF-based simulations with standard ASP calculations that start from the RHF reference state with $\HO$ being the Fock operator. 

The accuracy of the adiabatically-evolved RHF state for a given $\tau$ decreases as the bond length increases (Fig.~\ref{fig:asp_n2}).
For example, a very long evolution time of $\tau = 300\,\mathrm{E_h^{-1}}$ is required to achieve over $90\,\%$ fidelity (squared overlap)
with the true ground state in the dissociation limit, and the corresponding energetic error is still much larger than chemical accuracy.
In contrast, ASP starting from the fully open-shell CSF $\vsix$ achieves fidelities of $>90\%$
for all geometries above $R > \SI{1.4}{\angstrom}$ with a short evolution time $\tau \geq 30\, \mathrm{E_h^{-1}}$.
This demonstrates that the spin-coupled CSF provides a much better starting point in the strongly correlated dissociation regime. 
For a given $\tau$, simulations starting from the open-shell CSF perform better than those starting from the Hartree--Fock state for $R >  \SI{1.7}{\angstrom}$,
which corresponds to the bond length at which the open-shell CSF switches with the RHF determinant as the dominant contribution
to the ground wavefunction (Fig.~\ref{fig:n2_csf_fid}).
This observation provides further numerical evidence for the direct correlation between the accuracy of the initial state and the cost of ASP,
in line with previous research.\cite{Veis2014, Kremenetski2021, Sugisaki2022a, Lee2023}

Since the circuit depth and gate cost of implementing time-evolution on a quantum computer scale polynomially with the total time,\cite{Childs2021}  state preparation based on open-shell CSFs provides orders of magnitude more efficiency for strong correlation compared to RHF-based ASP.
Furthermore, since the CSF is explicitly an eigenfunction of the $\hat{S}^2$ operator, we do not require additional  modifications of the time-evolution circuits to enforce spin-pure wavefunction, in contrast to the symmetry-broken approach in Ref.~\citenum{Sugisaki2022a}.
Finally, because the cost of preparing the open-shell CSFs is negligible compared to the cost of implementing the time-evolution operator and we avoid the classical complexity of identifying CASCI initial wavefunctions,\cite{Veis2014, Tubman2016} our approach can be  extended to many strongly correlated electrons. 

Looking forward, although we have only considered a minimal basis representation of the \ce{N2} dissociation,
the same principles can be readily applied to larger basis sets.
For example, one possible protocol would be to start with a minimal basis representation incorporating the dominant spin coupling
and continuously evolving the state into the larger basis set at the target geometry using ASP by \textit{turning on} the matrix elements
coupling the orbitals in the minimal and large bases.
A similar approach could be applied to other strongly-correlated systems, such as \ce{FeS} clusters,\cite{Reiher2017, Li2019g}
where the fully-interacting ground state could be prepared starting from a suitable model Hamiltonian with a single CSF ground state.

\section{Avoiding the orthogonality catastrophe in quantum phase estimation}\label{sec:fault_tolerant}

\subsection{Quantum phase estimation}\label{ssec:qpe}

Large-scale error-corrected quantum hardware remains a distant prospect and near-term applications are likely to
rely on heuristic algorithms.
However, in the long term, the most anticipated application of quantum computing in chemistry
is the simulation of challenging systems through quantum phase estimation.\cite{Aspuru-Guzik2005, Reiher2017, Lee2021, vonBurg2021, Lee2023}

QPE can be used to project some initial state $\ket{\REF} = \sum_i \gamma_i \ket{E_i}$ onto one of the energy eigenstates $\ket{E_i}$ of the Hamiltonian.\cite{Kitaev1995, Abrams1999, Nielsen2010} Although the algorithm is highly promising for quantum advantage with fault-tolerant hardware architectures, its scaling depends inversely on the initial state overlap. This is problematic for strongly correlated electrons where the Hartree--Fock wavefunction has poor overlap with the exact state, which is also where classical methods are most limited.
For example, for \ce{N_2}, $|\gamma_0|^2 = |\braket{\Phi_{\mathrm{RHF}}|E_0}|^2 \approx 0.92$ at equilibrium but decreases to $|\gamma_0|^2 \approx 0.06$ at long bond lengths (Fig. \ref{fig:n2_csf_fid}).

Here, we propose using an initial state state that consists of a linear combination of the fully spin-coupled state $\vn$ and intermediate states with increasingly fewer correlated electrons ($N-2, N-4, ... 0$), including the fully uncorrelated RHF state (see Eq.~\eqref{eqn:lc_of_csfs} for \ce{N_2}, where $N=0, 2, 4, 6$).
This allows us to prepare states with high overlap with the exact ground state using $\mathcal{O}(N^2)$ Toffoli gates (Eq.~\eqref{eqn:toff_cost_main}), mitigating the initial state problem for systems where the accuracy of the Hartree--Fock state diminishes rapidly with the number of electrons.

The QPE algorithm involves three steps:
\begin{enumerate}
	\item The main register is initialized to some reference state through application of a state preparation circuit $U_\mathrm{ref}\ket{0} = \ket{\REF}$.
	\item  An invertible function of the Hamiltonian is applied in the form of a unitary operator. This is often the time-evolution operator $U = e^{-iHt}$,\cite{Abrams1999, Aspuru-Guzik2005, Reiher2017} but it can also be a qubitized quantum walk $U = e^{ i \arcsin(H/\lambda)}$,\cite{Low2019, Berry2019, Lee2021} where $\lambda$ is a rescaling parameter proportional to the norm of the Hamiltonian. Evolution of the initial state under $U$ controlled by the ancilla register accumulates a phase on the ancilla qubits corresponding to the eigenvalues of the operator.
	\item After application of the Quantum Fourier Transform on the ancilla register, the ancilla qubits are measured in the computational basis. The measurement results encode the energy eigenvalue as a binary bitstring that is read out classically.
\end{enumerate}

If $\ket{\REF} = \ket{E_i}$, the measurement yields the energy corresponding to the $i$-th eigenstate with probability $1$ (exactly, up to a desired numerical precision).
However, if the initial state is a superposition of energy eigenstates, the probability of projecting to the desired eigenstate and measuring
the corresponding energy eigenvalue is given by the squared overlap $|\gamma_i|^2 = |\braket{\REF | E_i}|^2$ between the initial state and the desired eigenstate.
Therefore, to obtain a sufficiently accurate energy estimate, the entire procedure must be repeated several times, with the number of repeats scaling as $\mathcal{O}(|\gamma_i|^{-2})$.
Post-QPE techniques can improve this scaling to $\mathcal{O}(|\gamma_i|^{-1})$.\cite{Lin2020a, Dong2022, Lin2022}
However, the overlap dependency is a fundamental feature of all projective algorithms,
and indeed the $\mathcal{O}(|\gamma_i|^{-1})$ scaling has been found to be near-optimal.\cite{Lin2020a}

Hamiltonian simulation is remarkably efficient, and the latest qubitization-based algorithms achieve quadratic scaling with the basis set size (number of orbitals).\cite{Berry2019, Lee2021} 
However, the QMA-hardness of electronic structure manifests through the initial state dependency of QPE.\cite{OGorman2022}
The reliance on the global overlap of an approximate wavefunction means that in principle,
the algorithm will suffer from the orthogonality catastrophe,\cite{Kohn1999, VanVleck1936}
requiring a number of repetitions that scales exponentially in the system size.
In practice, this worst-case scenario can be avoided by identifying approximate wavefunctions
that contain the dominant entanglement structure of the exact wavefunction.

For the \ce{N_2} example studied here, a linear combination of four CSFs (Eq.~\eqref{eqn:lc_of_csfs}) can enhance the squared overlap
of the reference state and the exact ground state by a factor of 16 when compared to the Hartree--Fock state at stretched geometries (Fig.~\ref{fig:n2_csf_fid}).
This implies an equivalent reduction in the total runtime of QPE since the cost of preparing the initial state is negligible compared to the cost of Hamiltonian simulation.
In Ref.~\citenum{Marti-Dafcik2024a}, we presented analogous results for more systems including
most second-row diatomics and the water molecule, where the overlaps of linear combinations of CSFs are always above $85\%$.
The accuracy of this linear combination state lies in the ability to capture spin coupling effects between localized electrons in open-shell orbitals.

This improved initial state removes the bias of Hartree--Fock theory towards delocalized states
and instead encodes the main physical effect that causes strong correlation in such systems:
ferromagnetic coupling within a local subsystem of electrons and global
antiferromagnetic coupling across subsystems.
We expect that diatomics with even more strongly correlated electrons, such as \ce{Cr2}, can also be treated in this way.

Nevertheless, all systems mentioned so far, including the \ce{Cr2},\cite{Larsson2022a} are tractable with classical methods,
at least if only the ground state is considered.
This can be attributed to the fact that a relatively small number of electrons are strongly spin-coupled ($N=12$ for \ce{Cr_2}),
and therefore compressed classical representations of the wavefunction can still be processed on classical devices without a big loss in accuracy.
In other words, the exponential asymptotic scaling in $N$ is not prohibitive when $N=12$.

\subsection{Resource estimation for fault-tolerant simulation of FeS systems}\label{ssec:resource_estimation}
The long-term hope is that fault-tolerant quantum devices with hundreds of logical qubits
will be able to compute energies and properties of systems that are out of reach for all classical methods.\cite{Reiher2017, vonBurg2021, Lee2021}

An often-cited class of molecules with potential technological applications,
whose simulation is classically challenging, are transition metal clusters such as systems composed of multiple iron-sulfur centers.\cite{Sharma2014a, Li2019i}
In particular, the electronic structure of \ce{FeMoCo} is considered a benchmark for QPE-based quantum advantage.\cite{Reiher2017, Li2019g, Lee2021, Lee2023}
It is strongly correlated, containing up to 35 open-shell electrons.\cite{Li2019i}
By increasing the number of \ce{FeS} centers, it has been shown that both the Hartree--Fock state and a naive single CSF
have an overlap with the best-available matrix product state that diminishes exponentially with increasing system size.\cite{Lee2023}
Although the poor Hartree--Fock overlap is expected, we hypothesize that it is possible to obtain
a much higher overlap using a linear combination of a small number of CSFs,
provided a bespoke, compact representation is chosen for each CSF.
This requires identifying the relevant spin coupling patterns through a basic understanding of the electronic structure,
and mapping these to their corresponding CSFs.
Crucially, to obtain the desired wavefunction compression,
one must choose an appropriate set of orbitals (single-particle basis) as well as CSFs (many-body basis).
A detailed study on CSFs for FeS systems is beyond the scope of this work.
Nevertheless, one can already see the potential of this CSF-based approach as a state preparation method for QPE, under reasonable assumptions.

To get a rough estimate for the cost of preparing an accurate initial state,
first assume that a CSF like the state $\vn$ is an accurate approximation of the ground state,
where $N$ is the number of open-shell, spin-coupled electrons.
We expect the number of determinants to be determined mostly by the number of open-shell electrons.
Therefore, the open-shell CSF $\vn$, whose structure and circuit representation we have studied in this work,
serves as a proxy for the cost of preparing a high-overlap approximation of the eigenstates of FeS systems.

From previous studies,\cite{Sharma2014a} it is known that $N=10$ and $N=18$ for \ce{Fe_2S_2} and \ce{Fe_4S_4}.
\ce{FeMoCo} is believed to have up to 35 open-shell electrons,\cite{Li2019i} and we thus set $N=34$.
The smaller clusters have been tackled with success using state-of-the-art DMRG programs,\cite{Sharma2014a}
and satisfactory approximations have also been obtained for the P-cluster,\cite{Li2019i}
a system believed to facilitate the nitrogen reduction process between the \ce{[Fe_4S_4]} Fe cluster and FeMoCo.
However, accurately resolving the electronic structure of FeMoCo remains an open problem.\cite{Li2019g, Lee2023}

Table \ref{tab:n_cnots_csfs} contains the number of determinants ${N \choose N/2}$ required in the expansion of the state $\vn$.
For $N=34$, this is $2.3 \times 10^9$. In this situation, it is clear that one would need post-Hartree--Fock initial states for QPE
as the overlap between the Hartree--Fock state and the ground state is very small.
Indeed, Lee \textit{et al.} \cite{Lee2023} found an overlap of $\sim 10^{-7}$ between the Hartree--Fock state and a matrix product state obtained through the DMRG algorithm.
Recent work by Ollitrault \textit{et al.} shows that the overlap of a single CSF and the ground state can be boosted by optimizing the orbitals
for the \ce{[Fe_2S_2]} and \ce{[Fe_4S_4]} Fe clusters.\cite{Ollitrault2024}
This provides a useful improvement but it does not solve the fundamental issue of exponential decay.
Therefore, more sophisticated state preparation methods are required for larger systems.

As an alternative to preparing the Hartree--Fock state or an open-shell CSF,
one could consider using the best possible selected CI state,\cite{Formichev2023} e.g., as found through FCIQMC,\cite{Booth2009, Cleland2010}
heat-bath CI,\cite{Holmes2016}, or adaptive sampling CI.\cite{Tubman2016, Tubman2018}
However, quantum circuits that prepare a generic linear combination of $L$ determinants typically require $\mathcal{O}(L)$ gates,\cite{Mottonen2004, Low2018}
usually with an additional $\mathcal{O}(\log L)$ \cite{Low2018} or $\mathcal{O}(N)$ factor for $N$ qubits,\cite{Tubman2018} as well as the use of an ancilla register.
At best, the cost can be reduced to $\mathcal{\tilde{O}}(\sqrt{L})$, but this requires trading off Toffoli gates for $\mathcal{O}(\sqrt{L})$ qubits.\cite{Low2018}

The state-of-the-art algorithm for computing the ground state energy of \ce{FeMoCo} through phase estimation requires $3.2\times 10^{10}$ Toffoli gates.\cite{Lee2021}
Using the clean qubit approach from Low \textit{et al.},\cite{Low2018} recent work gave an algorithm for preparing a CI state of $L$ determinants with cost of 
\begin{equation}
(2\log L - 2)L + 2^{\log L + 1} + L
\end{equation}
Toffoli gates.\cite{Formichev2023}
Therefore, preparing a superposition of the $L = 2.3 \times 10^9$ determinants that form the state $\ket{\mathcal{O}_{0, 0}^{34, 1}}$ would cost $1.43 \times 10^{11}$ Toffoli gates.
This would make the cost of initial state preparation higher than the cost of QPE,
which is problematic given the already-long expected runtimes for evaluation of a single energy,
which are in the order of multiple days on future superconducting quantum hardware.\cite{Lee2021}

To reduce the state preparation cost, one could consider only preparing a few of the most important determinants that dominate the wavefunction.
Examining the CSF $\vn$, while the number of determinants scales exponentially as ${N\choose N/2}$,
the determinants $\ket{\alpha}^n\ket{\beta}^n$ and $\ket{\beta}^n\ket{\alpha}^n$, where $n = N/2$, have significantly more weight that the rest.
Specifically, their amplitude is $1/\sqrt{n+1}$ (Eq.~\eqref{eqn:v_csf}).
One could therefore build a state 
\begin{equation}
\ket{\phi} = \frac{1}{\sqrt{2}} (\ket{\alpha}^n\ket{\beta}^n - \ket{\beta}^n\ket{\alpha}^n)
\end{equation}
that would have squared overlap with the $\vn$ CSF scaling inverse linearly in the number of electrons:
$|\gamma_N|^2 =|\braket{\phi|\mathcal{O}_{0, 0}^{N, 1}}|^2= 4/(N+2)$.
For example, for \ce{N_2}, this would give an initial state with squared overlap $|\gamma_6|^2 = 0.5$
with the $\vsix$ CSF (and therefore with the ground state near dissociation) at a very cheap cost.

However, this has several downsides. First, the state $\ket{\phi}$ is not a spin eigenfunction.
Furthermore, $\ket{\phi}$ has significant overlap with eigenstates other than the ground state,
which is problematic for application in QPE, as it makes it highly probable to measure an undesired eigenstate.
In the \ce{N_2} example, at $R=\SI{3.0}{\angstrom}$ the squared overlap of $\ket{\phi}$ with the singlet $(S=0)$ ground state and a quintet ($S=2$) excited state is equal at 50\%, and therefore it is equally likely to measure either state.
This situation would arise in a much more pronounced way in more challenging systems such as transition metal clusters,
where the strong correlation due to spin coupling leads to many near-degenerate eigenstates of different spin,
or eigenstates of the same spin but different spin coupling.
In such situations, it is essential to use accurate initial states to separate the different eigenstates,
which requires a much larger number of determinants.

A further alternative would be to use a classical DMRG calculation to find a good initial state.\cite{Chan2012}
The tensor product structure of the MPS can be exploited to directly encode the DMRG state in a quantum circuit, circumventing any explicit CI expansion.
To our knowledge, the only method for MPS preparation with guaranteed success is the sequential algorithm proposed by Sch{\"o}n \textit{et al.}\cite{Schon2005}
The number of gates scales as $\mathcal{O}(M\chi^2)$ for a MPS with $M$ sites and bond dimension $\chi$.\cite{Formichev2023}
For \ce{FeMoCo}, we can take $M=N=34$, i.e. the number of sites of the MPS is the number of open-shell orbitals/spin-coupled electrons.

Despite the polynomial scaling in $N$ and $\chi$ of MPS preparation circuits,
the cost of gate-level implementations have large constant factors as they require implementing non-trivial quantum arithmetic operations
and the use of $\mathcal{O}(\chi)$ ancilla qubits.\cite{Formichev2023}
For example, Formichev \textit{et al.} \cite{Formichev2023} gave a circuit decomposition
of the algorithm by Sch{\"o}n \textit{et al.} \cite{Schon2005} with approximate cost
\begin{equation}
	T_{\mathrm{MPS}} \approx (M-1)\chi [32 \chi + (b+1)\log(4\chi)]
\label{eqn:toffcost_mps}
\end{equation}
(Appendix \ref{apdx:mps}).
More fundamentally, the entanglement captured by a MPS of bond dimension $\chi$ scales as $\mathcal{O}(\log (\chi))$.
Therefore, for systems whose entanglement grows with the system size, as opposed to Hamiltonians with area-law ground states,\cite{Hastings2007}
the bond dimension can scale very steeply, at worst exponentially in $N$ to achieve high accuracy through a MPS.

In practice, the success of the DMRG algorithm when applied to strongly correlated systems, at least for smaller \ce{FeS} clusters,\cite{Sharma2014a, Li2019i}
indicates that the worst-case exponential scaling might not be strong enough to prohibit classical simulations for the system sizes considered.
Another explanation could be that locality dampens electron-electron interactions, e.g. due to charge screening,
reducing the strength of long-range entanglement between different FeS clusters. 
It remains to be seen what bond dimensions would be required to accurately resolve larger systems like \ce{FeMoCo},
and what would be the exact cost of the quantum circuits for sufficiently-accurate state preparation in the context of phase estimation.
State-of-the-art DMRG calculations for different FeS systems typically involve bond dimensions of up to several thousands ($2000-8000$)
to achieve accurate energy estimates.\cite{Sharma2014a, Li2019i, Lee2023, Ollitrault2024}
Table \ref{tab:toffcost_mps} compares the Toffoli cost of MPS state preparation for different FeS systems
as a function of the bond dimension, as per Eq.~\eqref{eqn:toffcost_mps}. Clearly, preparation of a MPS of high ($\chi > 1000$) bond dimension would
constitute a significant fraction of the overall cost of one run of QPE.

Formichev \textit{et al.}\ also proposed reducing the cost of MPS state preparation
by taking a MPS with high bond dimension (e.g., $\chi \approx 2000$) obtained from a DMRG calculation,
and compressing it to a MPS with lower bond dimension ($\chi \approx 5-10$).\cite{Formichev2023}
For some systems, including \ce{Fe_4S_4}, they found that while the energy of that MPS is very inaccurate,
compression does not significantly affect the wavefunction quality,
in the sense that there remains a strong overlap between the compressed MPS and the MPS of higher bond dimension.
Assuming that such compressed states with bond dimension in the order of $\chi = 10$ indeed provide accurate ground state approximations,
the state preparation cost would be relatively low compared to the cost of QPE (Table \ref{tab:toffcost_mps}).
Although this compression approach could be promising for cheaper state preparation,
it requires significant classical cost to initially compute the MPS with high bond dimension. 
It also remains to be seen how accurate the compression is for larger and more complex systems.

\begin{table}[htp]
\caption{Estimated number of Toffoli gates required for preparation of matrix product states for $N$ electrons in $M=N$ spatial orbitals for different choices of bond dimension $\chi$. Here, $b$ is the number of digits (ancilla qubits) that encode each rotation gate in binary representation, as required to guarantee a state preparation error of at most $\epsilon = 10^{-7}$.
For comparison, the number of Toffoli gates required for running qubitization-based quantum phase estimation circuits is $3.2 \times 10^{10}$ for \ce{FeMoCo}.\cite{Lee2021}}
\label{tab:toffcost_mps}
\begin{ruledtabular}
	\begin{tabular}{lcccc}
		$N$ & System & $\chi$ & b & $T_{\mathrm{MPS}}$
		\\
		\hline
		$10$ & \ce{Fe_2S_2} & 10 & 34 & $4.56 \times 10^4$ \\
		&  & 50 &37 & $8.51\times 10^5$ \\
		&  & 2000 & 43 & $1.16 \times 10^9$\\
	    \hline
		$18$ & \ce{Fe_4S_4} & 10 & 35 & $8.7 \times 10^5$ \\
		&  & 50 & 37 & $1.61 \times 10^6$  \\
		&  & 2000 & 44 &  $ 2.2 \times 10^9 $\\
		\hline
		$34$ & \ce{FeMoCo} & 10 & 36 & $ 1.71 \times 10^5$\\
		&  & 50 & 38 & $3.13 \times 10^6$\\
		&  & 2000 & 44 &  $4.26 \times 10^9$\\
	\end{tabular}
\end{ruledtabular}
\end{table}

In contrast, our CSF preparation circuits directly encode the relevant entanglement structure into bespoke quantum circuits,
which we expect to maximize the state preparation efficiency nearly optimally.
Preparation of $\vn$ for $N=34$ has a low cost of $\approx 4 \times  10^3$ Toffoli gates (Table \ref{tab:n_cnots_csfs}),
and does not require any ancillas, significantly reducing the space-time volume and hardware connectivity requirements
compared to the aforementioned black-box state preparation approaches.


Our current understanding of the spin coupling structure of \ce{FeMoCo} is not as precise
as that of the bond breaking examples in Ref.~\citenum{Marti-Dafcik2024a} (Section \ref{sec:background}).
However, it is well-known that the nature of electron correlation in stretched bonds
and transition metal clusters is similar.\cite{Izsak2023, Moerchen2024}
Moreover, data from state-of-the-art electronic structure simulations reveals analogous phenomena in these systems:\cite{Li2019i, Li2019g}
atoms form clusters such as \ce{Fe_4S_4} cubanes,
where the $3d$ orbitals in each FeS cluster contribute a number of unpaired electrons (spins).
These spins interact through shorter-range (local/intracubane) coupling within the cluster
as well as longer-range (global/intercubane) entanglement between the cubanes.
Typically, the local coupling is nearly maximally ferromagnetic
whereas the coupling between the cubanes is antiferromagnetic.
This locally ferromagnetic and globally antiferromagnetic alignment mirrors the mechanisms
occurring in the bond breaking examples of Ref. \onlinecite{Marti-Dafcik2024a},
including \ce{N_2} (Section \ref{ssec:n2_diss}).

FeS systems are more complex than the organic molecules we studied here and in Ref. \onlinecite{Marti-Dafcik2024a}
due to the presence of sulfur ligands, charge hopping effects,
and different oxidation states.
These complicate the picture,
as they introduce a larger number of spin coupling patterns that are energetically close,
necessitating the inclusion of a larger number of near-degenerate CSFs in the wavefunction ansatz to account for all dominant effects.
Nevertheless, the main many-body effect remains the entanglement/correlation induced by spin coupling,
and the number of dominant couplings is greatly limited by symmetries, typically in the order of the number of electrons.\cite{Li2019i, Moerchen2024}
By employing a bespoke and appropriate choice of basis for each CSF or for sets of CSFs corresponding to similar electronic configurations,
this additional complexity should only increase the number of CSFs by a moderate factor.

In conclusion, we anticipate that the proposed approach could be extended to FeS systems or similar transition metal clusters,
and that the number of relevant states required for high overlap would scale moderately (rather than exponentially) with the number of open-shell electrons.
Preparing linear combinations of CSFs would add a small overhead to the $\sim 10^3$ gate counts for state preparation in Table~\ref{tab:n_cnots_csfs} (see Section \ref{ssec:circs_lc}),
but it is unlikely to be significant enough to affect our conclusions.

\section{Conclusions and discussion}
While quantum algorithms such as QPE can in principle compute the exact eigenvalues of arbitrary electronic Hamiltonians, in practice
their performance depends critically on the overlap of the target eigenstate with the initial state.
Since the overlap between two random vectors is inversely proportional to the size of the Hilbert space,
strategies for preparing initial states with the principal features of the Hamiltonian eigenstates are
an essential component of quantum simulation. In quantum chemistry, the mean-field Hartree--Fock method provides 
an initial state with a high overlap with the exact ground state for weakly correlated electronic systems, which
pushes the orthogonality catastrophe to systems with a very large number  of electrons. Classically challenging problems, however, 
involve strongly correlated states. Here, the overlap of the RHF wavefunction with the exact eigenstate shrinks
exponentially with the number of strongly correlated electrons, which presents a severe initial state problem.

In previous work, we demonstrated that the structure of strongly correlated eigenstates in chemical systems
can be predicted from symmetry arguments using generalized spin-coupled molecular orbital theory.\cite{Marti-Dafcik2024a}
Replacing the HF wavefunction with more general spin-coupled wavefunctions, linear combinations of CSFs, 
provides high-quality initial states across the whole range of correlation regimes found in molecules. 
In this article, we have shown that the CSF-based initial states improve the performance of a wide range of 
quantum algorithms for the simulation of quantum chemistry on quantum computers, which can be applied both on 
fault-tolerant quantum hardware as well as on near-term devices. 
The CSFs directly encode the strong correlation in molecules and constitute specific patterns of entanglement. We have
presented circuits for building CSFs with depth linear in the number of electrons $N$, which generate a linear combination of states
of dimension exponential in $N$.

For VQE with a fixed-depth quantum number preserving ansatz, our numerical simulations for \ce{N2} 
show that the CSF-based initial state returns significantly more accurate energies than a HF initial state, with 
orders of magnitude improvement at stretched geometries. The multireference non-orthogonal variant of this 
VQE algorithm produces chemically accurate energies across the whole binding curve at much reduced gate depths.
Quantum subspace diagonalization through real-time evolution also benefits greatly from using CSF-based initial
states. We find that the subspace formed by independently evolving each of the chemically relevant CSF states 
rapidly spans the relevant low-energy regions of Hilbert space. This provides a powerful method for computing accurate excited states,
as well as accelerating convergence to the ground state. Since the CSFs are eigenfunctions of space and spin symmetry operators,
this has the advantage that the space and spin symmetry sectors can be treated separately,
which makes it possible to target specific molecular excited states. 
For ASP, beginning with a single spin-coupled wavefunction at dissociation enables preparation of the ground state
at stretched geometries with a much-reduced evolution time and therefore lower circuit depth.

In the language of quantum chemistry,
the initial state is constructed to contain the static, or strong correlation contributions, and the subsequent
refinement through the quantum algorithms captures the remaining dynamic correlation and orbital relaxation.
Although we have restricted the analysis to a minimal basis set, quantitatively accurate energy estimates require a more 
refined discretization of the Hilbert space in the form of a larger basis set.\cite{Helgaker2010} 
This does not present a problem since the CSFs in a minimal basis can be projected into the larger basis, and
orbital relaxation effects captured e.g. through variational quantum algorithms.

A key benefit of the QSD algorithm based on real-time evolution is that it can be used to compute excited states and has less stringent requirements on the accuracy of each ansatz circuit
when compared e.g. to VQE which relies on a single wave function ansatz and corresponding unitary.
However, implementing real-time evolution according to the Hamiltonian requires higher gate counts and depths per circuit that make it impractical with near-term quantum hardware. 
We have therefore introduced a new quantum algorithm, 
ADAPT-QSD, that combines the advantages of QSD with the benefit of the short-depth circuits of VQE.
We have demonstrated that performing QSD in the basis of sequential ADAPT-VQE states can significantly accelerate 
the convergence with respect to the number of operators and avoid stagnation, and that
ADAPT-QSD yields an energy estimate that is orders of magnitude lower than ADAPT-VQE at comparable circuit depth.
In view of the high popularity of ADAPT-VQE for ground state computation, we believe that ADAPT-QSD is a 
promising algorithm for excited state calculations.

Our state preparation heuristic avoids brute-force computation and instead exploits physical insight and symmetries to identify and prepare bespoke initial states.
Although it remains to be seen how effectively one can apply the same techniques to other systems, our framework provides a basis to tackle strongly correlated molecules where spin coupling dominates.
Previous work showed how, analogously to the \ce{N_2} molecule studied here, one can identify a small set of accurate CSFs
for a range of other systems including \ce{H_2O}, \ce{B_2}, \ce{C_2}, \ce{O_2},  \ce{F_2}, and clusters of hydrogen atoms.\cite{Marti-Dafcik2024a}
Here, we have used the example of \ce{N_2} in all our numerical simulations of quantum algorithms, but we expect a similar advantage
when enhancing quantum algorithms through CSFs for other molecules.

Looking farther ahead, the hope is that future fault-tolerant quantum devices with hundreds of logical qubits will be able to solve
the electronic Schr{\"o}dinger equation for systems that are out of reach for all classical methods.
For strongly correlated molecules, such as FeMoCo, which is considered an exemplary target for quantum simulation,
our CSF-based formalism provides a systematic approach to constructing initial states with high overlap with the exact state.
From approximate resource estimates, we conclude that our spin-coupling-based approach to initial 
state preparation will be greatly beneficial in solving challenging chemical problems through quantum phase estimation.
This is because the CSF states can be identified at the cost of mean-field classical computation and prepared with much lower numbers of gates
than expensive approximate classical wavefunctions such as DMRG, for a
similar level of initial state overlap.

We anticipate that a possible fault-tolerant workflow would be the following: 1) prepare the relevant CSFs in the minimal basis,
2) optimize the orbitals, given the CSF state, as proposed in Ref. \onlinecite{Ollitrault2024},
3) (optionally) apply a heuristic quantum algorithm to compute additional correlation arising in a larger basis, e.g. VQE or a QSD algorithm,
and/or (4) use QPE or post-QPE algorithms to obtain the final energy.


The ultimate goal of computational chemistry is to increase our understanding of complex chemical systems. 
Chemical properties are often the result of finely balanced competing effects and large parts of
modern development have focused on methods for obtaining tightly converged energy estimates.
Powerful classical algorithms such as DMRG and FCI-QMC fall into this category, as do the prevailing quantum algorithms.
While these methods can accurately compute properties directly from quantum mechanics, the brute-force nature of these algorithms obscures the interpretation of the results.
In contrast, our spin-coupled molecular orbital theory reveals configurations that correspond to the
dominant contributions to the ground and low-lying excited eigenstates.
The CSFs correspond to clear bonding motifs, providing direct insight into the underlying
electronic structure as well as a simplified, informationally-compact description of many-body
wavefunctions that otherwise appear complicated and unintelligible.

The central role of spin coupling in strongly correlated molecular states \cite{Sharma2014a, Li2019i}
	as well as the machinery for construction and application of CSFs \cite{Pauncz1979, Serber1934a, Kotani1963, Rumer1932}
	has long been known in the quantum chemistry community.
	However, the degree to which one can exploit this insight in classical algorithms has remained limited
	because it is fundamentally challenging to achieve a significant compression of truly multiconfigurational wavefunctions \cite{Izsak2023} using classical computers.
	To maximize the compactness of such wavefunctions,
	one must work in different orbital bases, and possibly different CSF bases, for each or at least some of the CSFs.
	The valence bond literature emphasizes the limitations of canonical molecular orbitals and the power of CSFs.\cite{Goddard1967}
	However, valence-bond-based ans{\"a}tze involve superpositions over all possible CSFs \cite{Goddard1967a, Goddard2003a, Goddard2003b, Ladner2003}
	and are therefore constrained to small systems by the curse of dimensionality;
	furthermore, they impose a common set of (potentially nonorthogonal) orbitals for each CSF.
	Molecular-orbital-based spin-adapted algorithms on classical computers,\cite{Sharma2012, Dobrautz2019}
	also recently proposed for state preparation on quantum computers,\cite{Lee2023, Moerchen2024}
	involve the use of a fixed single-particle basis of orthogonal orbitals,
	as well as a fixed many-body basis of CSFs, e.g. as enumerated through a geneological coupling scheme.\cite{Pauncz1979, Moerchen2024}
	These limitations have led to the view that CSF-based initial state preparation for quantum algorithms
	will also require expansions with exponentially many variational coefficients.\cite{Lee2023, Moerchen2024}
	
	Here, we take the view that, in the context of quantum computation, these limitations are unnecessary and the standard spin-adapted approach is unsuitable.
	Instead, we allow for full freedom in the choice of both the single-particle basis as well as CSF basis.
	This ensures that a compact representation of each CSF is possible in terms of its Slater determinant expansion,
	i.e. the basis is chosen such that the determinants' coefficients exhibit a highly symmetric structure, even if many Slater determinants are required for some CSFs.
	As a consequence, this enables efficient state preparation of each CSF in its corresponding orbital basis. 
	Our approach combines the unique strength of quantum computers (exponential memory scaling, ease of transforming between different bases and obtaining corresponding matrix elements)
	with the wavefunction/circuit compression and conceptual benefits (increased chemical insight) of working with CSFs.
	Note that a similar line of argument was recently taken by Leimkuhler and Whaley to motivate the potential advantage of using quantum computers
	in the context of a tensor network ansatz.\cite{Leimkuhler2024}

In summary, the ability to distil strongly correlated wavefunctions into a small number of symmetry-derived components (the dominant CSFs)
greatly enhances the performance of quantum simulation methods.
It enables designing efficient state preparation circuits to bias quantum algorithms
towards relevant, highly entangled regions of the Hilbert space at a reduced computational cost.
We have here shown how to achieve this for a range of systems and quantum algorithms and presented a blueprint for how to extend it to more challenging Hamiltonians.
As such, our work provides a conceptual framework and quantum algorithmic techniques
with the necessary ingredients for scalable and interpretable quantum simulation of classically challenging molecular electronic systems.

\acknowledgments
D.M.D. thanks Simon C. Benjamin, Nicholas Lee, Katherine Klymko and Norm Tubman for helpful discussions. D.M.D. acknowledges the use of the University of Oxford Advanced Research Computing (ARC) facility in carrying out this work (http://dx.doi.org/10.5281/zenodo.22558), as well as financial support by the EPSRC Hub in Quantum Computing and Simulation (EP/T001062/1). H.G.A.B. acknowledges funding from New College, Oxford (Astor Junior Research Fellowship) and Downing College, Cambridge (Kim and Julianna Silverman Research Fellowship).

\appendix

\section{Spin eigenfunctions}\label{apdx:spin_eigenfunctions}
Below are some examples of the spin-coupled state $\vn$. The general form is given in Eq.~\eqref{eqn:v_csf}.

\begin{subequations}
	\begin{equation}
		\vtwo= \frac{1}{\sqrt{2}}(\ket{\al\be} - \ket{\be \al}),
		\label{eqn:vcsf_n2}
	\end{equation}
	\begin{equation}
		\begin{split}
			\vfour=
			\frac{1}{\sqrt{3}}
			(\ket{\alpha\alpha\beta\beta}+\ket{\beta\beta\alpha\alpha})  \\
			- \frac{1}{2\sqrt{3}}(\ket{\alpha\beta\alpha\beta}+\ket{\beta\alpha\beta\alpha}
			+\ket{\alpha\beta\beta\alpha}
			+\ket{\beta\alpha\alpha\beta}),
		\end{split}
		\label{eqn:vcsf_n4}
	\end{equation}
	\begin{equation}
		\begin{split}
			\vsix =
			\frac{1}{2}(\ket{\alpha\alpha\alpha\beta\beta\beta}) - \ket{\beta\beta\beta\alpha\alpha\alpha})\\
			+\frac{1}{6}(\ket{\alpha\beta\beta\alpha\alpha\beta}
			+ \ket{\alpha\beta\beta\alpha\beta\alpha}
			+ \ket{\alpha\beta\beta\beta\alpha\alpha}
			+ \ket{\beta\beta\alpha\beta\alpha\alpha}\\
			+ \ket{\beta\beta\alpha\alpha\beta\alpha}
			+ \ket{\beta\beta\alpha\alpha\alpha\beta}
			+ \ket{\beta\alpha\beta\beta\alpha\alpha}
			+ \ket{\beta\alpha\beta\alpha\beta\alpha}\\
			+ \ket{\beta\alpha\beta\alpha\alpha\beta}
			- \ket{\beta\alpha\alpha\beta\beta\alpha}
			- \ket{\beta\alpha\alpha\beta\alpha\beta}
			- \ket{\beta\alpha\alpha\alpha\beta\beta}\\
			- \ket{\alpha\beta\alpha\beta\beta\alpha}
			- \ket{\alpha\beta\alpha\beta\alpha\beta}
			- \ket{\alpha\beta\alpha\alpha\beta\beta}
			- \ket{\alpha\alpha\beta\beta\beta\alpha}\\
			- \ket{\alpha\alpha\beta\beta\alpha\beta}
			- \ket{\alpha\alpha\beta\alpha\beta\beta}
			)\end{split}
		\label{eqn:vcsf_n6}
	\end{equation}
	\label{eqn:spin_eigenfunctions}
\end{subequations}

As shown in Ref. \onlinecite{Marti-Dafcik2024a}, other systems such as hydrogen clusters
are best described by a different type of CSF which is also a singlet but corresponds to a different spin coupling pattern.
This CSF is just a product of identical Bell states, whose general form is given by Eq. \ref{eqn:csf_geminals}.
The $N=4$ example used in Ref. \onlinecite{Marti-Dafcik2024a} is:
\begin{equation}
	\begin{split}
		\ket{\mathcal{O}_{0, 0}^{4, 2}}=
		\frac{1}{2}
		(\ket{\alpha\beta\alpha\beta}
		-\ket{\alpha\beta\beta\alpha}
		-\ket{\beta\alpha\alpha\beta}
		+\ket{\beta\alpha\beta\alpha}).
	\end{split}
	\label{eqn:vcsf_n4_2}
\end{equation}

\section{Quantum circuits for basis rotations}\label{sec:circ_basis_rotation}
\subsection{General fermionic basis rotations}
We can rotate the single-particle basis of any many-particle state by applying exponential unitary transformations that in general act on the entire many-body Hilbert space. A basis transformation of $K$ basis functions (here, spin-orbitals or qubits) can be represented by a $K\times K$ unitary matrix with entries $u_{pq}$:
\begin{equation}
	\tilde{\phi}_p = \sum_{q} \phi_q u_{pq},
\end{equation}
where the sets $\{\phi_p\}$ and $\{\tilde{\phi}_p\}$ are the original and the rotated basis set. Given a many-body state, this single-particle rotation is equivalent to applying a linear transformation on the second-quantized operators:
\begin{subequations}
	\begin{equation}
		\tilde{a}^\dagger_p = \sum_{q} u_{pq} a^\dagger_{pq}
	\end{equation}
	\begin{equation}
		\tilde{a}_p = \sum_{q} u_{pq}^* a_{q}
	\end{equation}
\end{subequations}
Following Thouless' theorem,\cite{Thouless1960} the action of the single-particle rotation on a many-body wavefunction $\ket{\psi}$ can be expressed by the following operator:
\begin{equation}
	U(\bm{u}) = \exp\Big(\sum_{pq}[\log(\bm{u})]_{pq}(a_p^\dagger a_q - a_q^\dagger a_p)\Big).
\end{equation}
While this operator in general acts on the entire Hilbert space (its dimension is $D\times D$, where $D=2^{K}$ for an $K$-qubit computational basis or ${K\choose N}$ for a particle-number-conserving basis of $N$ fermions in $K$ single-particle states), it can be implemented efficiently on a quantum circuit as follows.\cite{Kivlichan2018}

Following the approach outlined in Ref.~\citenum{Kivlichan2018}, the unitary operator $U(\bm{u})$ can be decomposed into a series of Givens rotations with the form 
\begin{equation}
U(\bm{\theta}) = \prod_{pq}  \exp\qty(\theta_{pq}(a_p^\dagger a_q - a_q^\dagger a_p))
	\label{eqn:big_U_product}
\end{equation}
without any trotterization error.
The corresponding rotational angles $\theta_{pq}$ can be identified by performing a QR decomposition of the orbital transformation matrix $\log(\bm{u})$, which can be solved classically.
Each exponentiated one-body operator can be implemented individually using efficient quantum circuits.\cite{McArdle2020, Kivlichan2018} In general, the indices run over all elements $p, q \in \{1, 2, .., K\}$ with $p > q$ and thus there are ${K \choose 2}$ such one-body operators. However, for many of the systems that we consider in this work, the matrix $\bm{u}$ has many zero entries since only few orbitals in $\{\phi_q\}$ contribute to each $\tilde{\phi}_p$, and therefore the cost of the basis transformations is lower than ${K \choose 2}$.

The CNOT cost of implementing the exponential of a one-body operator
\begin{equation}
	\textrm{exp}[\theta_{pq} (a_p^\dagger a_q - a_q^\dagger a_p)]
	\label{eqn:rot_op}
\end{equation}
that rotates between spin-orbitals $p$ and $q$ on a quantum circuit is \cite{Anselmetti2021, Yordanov2020b}
\begin{equation}
	C_{pq} = 
	\begin{cases}
		2(p-q)+1, & \quad p-q > 2\\
		2, & \quad p-q \in \{1, 2\} .
	\end{cases}
\label{eqn:rot_CNOT_cost}
\end{equation}
In this work, we use restricted orbitals and therefore the transformations are the same for both spin-up and spin-down orbitals that share the same spatial orbital. The cost for a transformation between two spatial orbitals is therefore $2\,C_{pq}$.
The linear scaling with the orbital indices $p-q$ stems from the requirement of implementing the Pauli-$Z$-strings that arise when mapping the fermionic operators onto qubit operators via the Jordan-Wigner encoding (see SI in Ref. \citenum{Burton2023}). To avoid this overhead, we can geometrically arrange the ordering of qubits such that the orbitals that contribute to the same orbital transformation neighbor each other.

\subsection{Basis rotations for diatomic bond breaking}\label{ssec:circ_basis_rotation_vcsfs}
For the diatomic systems discussed in Section \ref{sec:background} and Ref.~\citenum{Marti-Dafcik2024a}, transforming a delocalized orbital to a localized one requires building a linear combination of two spatial orbitals. For $N$ spin-coupled electrons, we have $N$ such transformations.

Consider the spin-coupled states $\vn$ (Appendix \ref{apdx:spin_eigenfunctions}). To obtain the localized orbitals from the Hartree--Fock orbitals, we need rotations of the form $\tilde{\phi}_L = \frac{1}{\sqrt{2}}(\phi_1 + \phi_2)$, $\tilde{\phi}_R = \frac{1}{\sqrt{2}}(\phi_1 - \phi_2)$ for each pair of spin-coupled electrons.
Each rotation involves two spatial orbitals, or four spin-orbitals. In total, it requires four one-body operators of the form in Eq.~\eqref{eqn:rot_CNOT_cost}, each with cost $C_{pq}$, to localize two electrons/spatial orbitals.
The total cost for localizing or delocalizing $N$ electrons/orbitals is 
\begin{equation}
	C_{\textrm{rot}, N} = 2N C_{pq}.
\label{eqn:cost_rot_N}
\end{equation}

For example, for \ce{N_2}, we have
\begin{equation}
\begin{split}
\upz_L &= \frac{1}{\sqrt{2}}(\sigg + \sigu), \qquad \upz_R = \frac{1}{\sqrt{2}}(\sigg - \sigu), \\
\upx_L &= \frac{1}{\sqrt{2}}(\piux + \pigx), \quad \upx_R = \frac{1}{\sqrt{2}}(\piux - \pigx), \\
\upy_L &= \frac{1}{\sqrt{2}}(\piuy + \pigy), \quad \upy_R = \frac{1}{\sqrt{2}}(\pigy - \pigy).
\end{split}
\end{equation}
as described in Section \ref{ssec:n2_diss}.
Using the qubit (spin-orbital) ordering $\{ \sigg, \bsigg, \sigu, \bsigu, \pigx, \bpigx, \pigy, \bpigy \}$, where the absence (presence) of an overbar indicates a $\alpha$ ($\beta$) spin-orbital, respectively, we always have $p-q=2$ and thus $C_{pq} = 2$ (Eq.~\eqref{eqn:rot_CNOT_cost}). Therefore, to localize $2$, $4$, and $6$ spatial orbitals, as required to apply the circuits for preparation of ($\ket{\phi_{2x}}$, $\ket{\phi_{2y}}$), $\ket{\phi_{4}}$, and $\ket{\phi_{6}}$ (Section \ref{sec:background}), the cost is $4N$ i.e. $8$, $16$, and $24$ CNOTs, respectively.

\section{Quantum circuits for spin eigenfunctions: explicit decompositions and cost}\label{apdx:circ_cost}
Below we analyze the exact cost of the CSF preparation circuit in Section \ref{sec:qcircs}. We focus on counting CNOT gates because entangling gates are typically the biggest sources of noise on devices without error correction.\cite{Cerezo2021}
In Section \ref{apdx:circ_cost_faulttolerant}, we also compute the Toffoli gate counts for longer-term fault-tolerant implementations of these circuits, since Toffoli gates dominate the cost when applying error-correcting codes.\cite{Babbush2018e}
We separately consider qubit architectures with all-to-all interactions between qubits,
as well as devices with restricted nearest-neighbor-only connectivity; specifically, linear, and planar (grid-like) connectivity.
The latter require decomposing gates between distant qubits into nearest-neigbor gates, which introduces a gate overhead.
This detailed analysis might be relevant in particular when considering implementations on near-term quantum hardware.
For this, we made some circuit design choices to get concrete numbers for the gate counts, but these are not necessarily optimal and could possibly be improved if tailored for a particular architecture or for the simulated system.

The circuit for preparing a single spin eigenfunction of the form $\vn $ contains three parts (see Sections \ref{ssec:qcircs_csfs_bondbreaking}, \ref{ssec:mapping_spintofock} and the circuit diagram in Fig.~\ref{fig:circuits_dicke_v8}):
\begin{enumerate}
	\item Preparation of the input state
	\item Application of the two $S_{n}$ unitaries
	\item Mapping from spin to Fock space.
\end{enumerate}

Below we discuss their cost in terms of the number of CNOT gates, assuming all-to-all connectivity. While the unitaries $S_n$ can be implemented using circuits with only linear (nearest-neighbor) connectivity, this is not the case for circuits that implement the input state and the mapping from spin to Fock space. We therefore compare the cost assuming all-to-all connectivity and the cost with connectivity restrictions (linear and planar connectivity).

\subsection{All-to-all connectivity}\label{ssec:circcost_alltoall}

Let $C_2$ and $C_3$ denote the number of CNOT gates required to implement the two- and three-qubit gate blocks required for the Dicke state preparation circuits (Figs. \ref{fig:two_qubit_circuit}, \ref{fig:three_qubit_circuit}).
Each two-qubit gate block consists of a controlled $R_y$-gate conjugated by two CNOTS. This can be decomposed into 3 CNOT gates: $C_2=3$ (Fig.~3 in \cite{Yordanov2020b}). The three-qubit gate consists of a $R_y$-gate controlled by two qubits, which can be decomposed into 5 CNOT gates: $C_3=5$ (Fig.~3 in \cite{Bartschi2019}).
Since each unitary $M_{l, l-1}$ consists of one two-qubit gate block and $l-2$ three-qubit gate blocks (Section \ref{ssec:circs_dicke}), the overall CNOT cost for implementation of the operator $S_{n}$ is
\begin{equation}
	\begin{split}
		C_{S_n} = \sum_{l=2}^n C_{M_{l, l-1}}=\sum_{l=2}^n\Bigl(C_2 + C_3(l-2))\Bigl) \\
		= \sum_{l=2}^n\Bigl(3+5(l-2)\Bigl) =
		\frac{5}{2}n^2 - \frac{9}{2}n + 2,
	\end{split}
	\label{eqn:cost_S_n}
\end{equation}
where $C_{M_{l, l-1}}$ denotes the CNOT cost of implementing the unitary $M_{l, l-1}$.
Table \ref{tab:n_cnots_u_nn}  shows the cost up to $n = 6$.

\begin{table}[htp]
	\caption{Number of two-qubit and three-qubit gate blocks (Figs.~\ref{fig:two_qubit_circuit}, \ref{fig:three_qubit_circuit}) required for application of symmetric state preparation unitaries $S_{n}$, and cost $C_{S_n}$ (number of CNOT gates) in the circuit for $S_n$ after decomposition into elementary gates (Eq.~\eqref{eqn:cost_S_n}).}
	\label{tab:n_cnots_u_nn}
\begin{ruledtabular}
	\begin{tabular}{cccc}
		& 2-qubit gates & 3-qubit gates & $C_{S_n}$
		\\ \hline
		$S_2$ & 1 & 0 & 3 \\ 
		$S_3$ & 2 & 1 & 11 \\
		$S_4$ & 3 & 3 & 24 \\
		$S_5$ & 4 & 6 & 42 \\
		$S_6$ & 5 & 10 & 65 \\
	\end{tabular}
\end{ruledtabular}
\end{table}

\begin{figure}[!htb]
		\centering
		\includegraphics[width=0.75\linewidth]{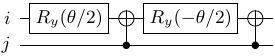}
		\caption{Decomposition of $CR_y^{j, i}(\theta)$.}
		\label{fig:cry_decomposed}
\end{figure}

The cost of the circuit for input state preparation is linear in the number of CNOT gates (Section \ref{ssec:qcircs_csfs_bondbreaking}). The first step only contains $X$ gates; the second step requires controlled rotation gates $CR_y(\theta_i)$ which can be decomposed into single-qubit rotations and two CNOTs (Fig.~\ref{fig:cry_decomposed}) to give CNOT cost of $2(n-1)=N-2$; the third step requires $n = N/2$ CNOT gates. The total CNOT count for input state preparation is therefore
\begin{equation}
	C_{\mathrm{in}}^{\mathrm{all}} = \frac{3N}{2} - 2.
	\label{eqn:cost_input_alltoall}
\end{equation}

The mapping from spin to Fock space costs
\begin{equation}
	C_{\textrm{map}}^{\textrm{all}} = N
\end{equation}
 for hardware with all-to-all connectivity, giving a total cost of
\begin{equation}
	\begin{split}
		C_{\textrm{tot}}^{\textrm{all}} &= 2C_{S_{n}}+C_{\textrm{in}}^{\textrm{all}}+C_{\textrm{map}}^{\textrm{all}} \\
		&= \Bigl(\frac{5}{4}N^2-\frac{9}{2}N + 4\Bigl) + \Bigl(\frac{3}{2}N -2\Bigl) + N \\
        &= \frac{5}{4}N^2 - 2N + 2.
	\end{split}
	\label{eqn:cost_total_alltoall}
\end{equation}

\subsection{Linear connectivity}\label{ssec:circcost_linear}

\begin{figure}[!htb]
	\includegraphics[width=\linewidth]{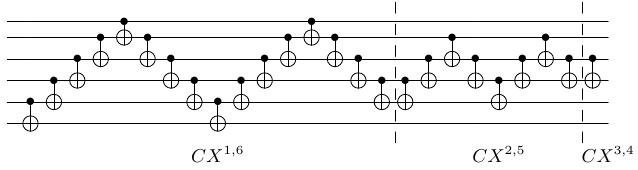}
	\caption{Recompilation of CNOT accordion for $N=6$, namely $CX^{1, 6}CX^{2,5}CX^{3, 4}$, into gates restricted to linear connectivity. This corresponds to the last part of the circuit for preparation of the input state (see Fig. \ref{fig:circuits_dicke_v8} for $N=8$ example).}
	\label{fig:cnot_accordion_decomposed}
\end{figure}

For hardware with linear connectivity, the circuit $S_n$ can be implemented directly without any modification.\cite{Bartschi2019} However, in the last step of the input start preparation circuits, some CNOT gates act between distant qubits and therefore cannot be implemented as such on a linear topology. The circuit can be adapted for hardware with nearest-neighbor connectivity via a recompilation of the CNOT accordion into nearest-neighbor CNOTs  (see Fig.~\ref{fig:cnot_accordion_decomposed}).
Evidently, some of the CNOT gates cancel out after the decomposition.
The CNOT count for the accordion with linear connectivity is
\begin{equation}
		C_{\textrm{acc}}(N)
		= N - 1 + 2 \sum_{i=0}^{\frac{N}{2}-2}(N-2i-2) = \frac{N^2}{2}-1.
\label{eqn:accordion_decomposition}
\end{equation}
The overall CNOT cost for preparation of the input state with linear connectivity is
\begin{equation}
	C_{\mathrm{in}}^{\mathrm{lin}} = (N-2) + \Big(\frac{N^2}{2}-1\Big) = \frac{N^2}{2} + N - 3. 
	\label{eqn:input_cost}
\end{equation}

To implement $U_{\textrm{map}}$, one could apply the same naive decomposition of the CNOTs into linear connectivity as in Fig.~\ref{fig:cnot_accordion_decomposed}, where in this case it is not directly obvious if any CNOTs cancel out. This has a large overhead as what was previously $C_{\textrm{map}}^{\textrm{all}}= N$ CNOTs now becomes
\begin{equation}
	C_{\textrm{map}}^{\textrm{lin}} = N \big[N + 2(N-1)+(N-2)\big] = 4N^2-4N.
\end{equation}

The total CNOT cost for hardware with linear connectivity would thus be
\begin{equation}
	\begin{split}
		C_{\textrm{tot}}^{\textrm{lin}} &= 2C_{S_{n}}+C_{\textrm{in}}^{\textrm{lin}}+C_{\textrm{map}}^{\textrm{lin}} \\
		&= \Bigl(\frac{5}{4}N^2-\frac{9}{2}N + 4\Bigl) + \Bigl(\frac{1}{2}N^2 + N - 3\Bigl) + \Big( 4N^2 - 4N \Big) \\
		&= \frac{23}{4}N^2 - \frac{15}{2}N + 1.
	\end{split}
	\label{eqn:cost_total_linear}
\end{equation}

Due to the decomposition of the CNOT accordion into nearest-neigbor gates, the depths of the circuits for input state preparation $U_{\textrm{in}, N}$ and for spin-to-Fock mapping $U_{\textrm{map}}$ scale as $\mathcal{O}(N^2)$, which is worse than the $\mathcal{O}(N)$ scaling of the depth for the Dicke circuits $S_n$. Furthermore, $C_{\textrm{map}}^{\textrm{lin}}$ dominates the gate count of CSF preparation with linear connectivity (Eq.~\eqref{eqn:cost_total_linear}).
Below, we show that the overhead can be removed in the case of planar connectivity.

\subsection{Planar connectivity}\label{ssec:circcost_planar}
Many state-of-the-art quantum devices based on superconducting qubits, such as Google's Sycamore processor,
consist of planar (two-dimensional) structures with nearest-neighbor connectivity.\cite{AIQuantum2020, Huggins2022b}
In cases with such a hardware topology, one can carefully align the spin-orbitals to avoid the overhead required when going from all-to-all to linear connectivity
associated with recompiling the two-qubit gates required by the spin-to-Fock space mapping $U_{\textrm{map}}$ (Appendix \ref{ssec:circcost_linear}).

Starting from a basis of localized orbitals where $\{i_L\}$ and $\{ i_R \}$ are spin-up orbitals localized on the left and right atom, and $\{\bar{i}_L\}$ and $\{ \bar{i}_R \}$  are the corresponding spin-down orbitals, the circuits for preparation of a single CSF $\vn$ require the following pairs of qubits/spin-orbitals to be directly connected to avoid any gate overhead due to recompiling two-qubit (CNOT) gates between non-neigboring qubits:
\begin{enumerate}
	\item Spin-up orbitals $i_L$ and $(i+1)_L$ for all $ i \in \{ 1, 2, ...N-1\}$,  must be connected to each other. This is required for the circuits for input state preparation $U_{\textrm{in}, N}$ and the circuits $S_{N/2}$ (left and middle section in Fig. \ref{fig:circuits_dicke_v8}). The ordering within the left and right subsystem is irrelevant due to the symmetry of the state within that subspace.
	\item Every spin-up orbital $i_{L/R}$ must be connected with their corresponding spin-down orbital $\bar{i}_{L/R}$. This is required for the spin-to-Fock-space mapping $U_{\textrm{map}}$ (right section in Fig. \ref{fig:circuits_dicke_v8}).
	\item For the basis rotation circuits (Appendix \ref{ssec:circ_basis_rotation_vcsfs}), each \textit{left} orbital $i_L$ or $\bar{i}_L$ must be connected to its corresponding \textit{right} orbital $i_R$ or $\bar{i}_R$.
\end{enumerate}

Assuming a rectangular grid, there is no ordering that satisfies all three conditions. If we only wish to prepare a single CSF without performing orbital rotations,
we can place all spin-up qubits $i_{L/R}$ in one line, and all the corresponding spin-down qubits $\bar{i}_{L/R}$ in a line below or above it.
With this ordering, the first two conditions are met and therefore there is no overhead in implementing the input and Dicke state preparation circuits.
However, basis rotations would require applying SWAP gates to place the relevant qubit pairs next to each other, with an overhead scaling as $\mathcal{O}(N^2)$.

If we wish to use multiple CSFs (including localized and delocalized states) in a quantum algorithm,
we can prepare the CSF in a basis of localized orbitals and then rotate the basis as in Appendix \ref{ssec:circ_basis_rotation_vcsfs}.
This is preferably done using the following arrangement to exploit planar connectivity without a large gate cost:
set all left spin-up orbitals $\{ i_L\}$ , $i = \{ 1, 2, ...n\}$ in a horizontal line,
and all right spin-up orbitals $\{ i_R\}$, $i = \{ 1, 2, ...n\}$ in parallel, below the spin-up line.
Then, set the left spin-down orbitals $\{ \bar{i}_L \}$, $i = \{ 1, 2, ...n\}$ in a line on top, and the right spin-down orbitals $\{ \bar{i}_R \}$, $i = \{ 1, 2, ...n\}$  in a line below.
This enables applying the circuits for preparation of a single CSF with nearly the same cost as if we had all-to-all connectivity.
The overhead using this grid structure comes from the need of decomposing the CNOT accordion
for preparation of the input state to a linear array (Appendix \ref{ssec:circcost_linear}).
Since the spin-up orbitals $i_L$ and $i_R$ are already connected $\forall i \in \{ 1, 2, ..., n\}$,
those rotations can be implemented directly. However, to connect the spin-down orbitals $\bar{i}_L$ and $\bar{i}_R$,
we must swap $i_L$ and $\bar{i}_L$ and $i_R$ and $\bar{i}_R$ for all $i$.
Then, we can implement the input state preparation circuit. The overhead for going from
all-to-all to planar connectivity is therefore $2n=N$ SWAP gates, or $3N$ CNOT gates.
This linear overhead is small compared to the total $\mathcal{O}(N^2)$ cost of CSF preparation (Eq.~\eqref{eqn:cost_total_alltoall}).

Using this strategy, the total CNOT count for state preparation of a single CSF with $N$ spin-coupled electrons in the localized basis on a device with planar connectivity is:
\begin{equation}
	C_{\textrm{tot}}^{\textrm{pl}} = C_{\textrm{tot}}^{\textrm{all}} + 3N = \frac{5}{4}N^2 + N + 2. 
\label{eqn:cost_tot_planar}
\end{equation}
Preparing that state and then rotating $N$ spin-orbitals orbitals (occupied by $N$ electrons) to a delocalized basis has an additional cost of $2NC_{p, q} = 4N$ (Eq.~\eqref{eqn:cost_rot_N}). This might not be needed depending on the application (e.g. when preparing linear combinations of CSFs, we must rotate only a subset of the orbitals for each CSF).

\subsection{Fault-tolerant circuits and non-Clifford cost}\label{apdx:circ_cost_faulttolerant}
Universal gate sets for fault-tolerant quantum computation include Clifford + Toffoli or Clifford + $T$ gates.
Since Clifford gates including the set of Clifford group generators $\{ H, S, CX\}$ are efficiently classically simulable,\cite{Nielsen2010} most fault-tolerant resource estimates typically assume that their cost is negligible compared to the overall cost.\cite{Reiher2017, Lee2021} On the other hand, expensive magic-state distillation protocols are required to apply noisy non-Clifford gates with high fidelity within an error-correcting code, resulting in a large space-time overhead orders of magnitude larger than the cost of fault-tolerant Clifford gates.\cite{Gidney2019}
We therefore focus on minimizing and counting non-Clifford gates. This allows direct comparison with the literature on Hamiltonian simulation for electronic structure (see \cite{Reiher2017, Berry2019, vonBurg2021, Lee2021} and Table III in Ref. \onlinecite{Lee2021} for an overview).

The circuit $V_N$ for preparation of the CSF $\vn$, presented in Section \ref{ssec:qcircs_csfs_bondbreaking}, requires converting the continuous rotation gates $R_y(\theta)$ to a discrete gate set of Clifford + Toffoli gates. Due to the finite precision in the binary representation of continuous numbers, this introduces an error $\epsilon_r$ per rotation gate that can be exponentially suppressed since the gate and qubit counts scale as $\mathcal{O}(\log(1/\epsilon_r))$. The number of rotations in $V_N$ scales as $\mathcal{O}(N^2)$ and therefore the asymptotic Toffoli complexity remains $\tilde{\mathcal{O}}(N^2)$
after suppression of polylogarithmic factors. Below, we derive the exact cost.

\subsubsection{Rotation gate synthesis}
A commonly-used approach is to synthesize rotations in terms of Clifford + $T$ gates. The dominant factor affecting the number of $T$-gates per rotation gate is
\begin{equation}
	N_{TR} = c\lceil \log(1/\epsilon_r) \rceil,
\end{equation}
where $c$ is a constant that depends on the implementation. For deterministic algorithms, $c$ is lower-bounded by $3$,\cite{Selinger2012, Ross2014} but this can be improved using randomized methods such as the repeat-until-success (RUS) approach in Ref. \citenum{Bocharov2015}, which finds an empirical average of $c = 1.15$. Here, we consider the probabilistic RUS technique, where the cost includes an extra additive factor:
\begin{equation}
	N_{TR} \approx c\lceil \log(1/\epsilon_r) \rceil + 9.2.
\end{equation}

Since we later compare the cost of initial state preparation with the cost of implementing quantum phase estimation, we translate the $T$-gate counts into Toffoli counts to compare with the Toffoli counts in state-of-the-art qubitization-based algorithms for electronic structure.\cite{Lee2021, vonBurg2021} The corresponding number of Toffoli gates is roughly $T_R \approx \frac{1}{2}N_{TR}$ in a surface code implementation, since the cost of applying a Toffoli gate is approximately twice the cost of applying a $T$-gate using the magic state factories introduced in Ref.~\citenum{Gidney2019}. Thus, the number of $T$-gates per rotation is $\frac{1}{2} (c R \lceil \log(1/\epsilon_r)\rceil + 4.6)$.

The circuits for preparation of $\vn$ also require application of controlled $R_y$ gates $CR_y$ and $CCRy$ (Section \ref{sec:qcircs}). Controlling rotation gates does not have any non-Clifford overhead if the rotation angle is given by a classical register, as we can simply use CNOT gates to change the direction of the rotation $\theta \leftrightarrow -\theta$.\cite{Sanders2020}  This is indeed the case in the circuits for preparation of the spin eigenfunction $\vn$, since the entire state preparation circuit (including the angles) can be specified classically before any quantum computation. To determine the number of Toffoli gates, we must therefore simply count the number of rotations, irrespective of whether these are controlled or not.

Finally, consider a circuit with $R$ rotations  with overall error $\epsilon$. A naive error analysis based on a triangle inequality $\epsilon \leq \sum_{r=1}^R \epsilon_r$ would
suggest allocating an error $\epsilon_r = \epsilon/R$ for each rotation, where the number of bits required to represent each rotation angle is $b = \log(1/\epsilon_r)$.
Ref. \citenum{Low2021} suggested that one can reduce this pessimistic bound if one regards errors as random rather than coherent.
The corresponding random walk complexity analysis gives an error bound that is tighter by a quadratic factor $\epsilon \rightarrow \sqrt{\epsilon}$, or equivalently halves the number of bits $b$. With this, we obtain a reduced average of
\begin{equation}
	T_R = \frac{1}{4} c R \lceil \log(1/\epsilon_r)\rceil + 4.6
	\label{eqn:error_per_rot}
\end{equation}
Toffoli gates per rotation, where $b = \frac{1}{2} \lceil \log(1/\epsilon_r)\rceil$ and $c=1.15$.

\subsubsection{Toffoli cost for preparation of spin eigenfunctions}\label{sssec:toffoli_cost}
As discussed in Section \ref{ssec:circs_dicke}, each subcircuit $M_{l, l-1}$ that implements the symmetric state preparation unitary $S_n$ (Eq.~\eqref{eqn:dicke_recursion}) consists of one two-qubit gate gate block (Fig. \ref{fig:two_qubit_circuit}) and $l-2$ three-qubit gate blocks. The total number of rotations for implementation of $S_n$ is therefore
\begin{equation}
	R_{S_{n}} 
= \sum_{l=2}^n (l-1) = \frac{1}{2}n(n-1).
	\label{eqn:nrot_sn}
\end{equation}
This must be implemented twice to implement $V_N: \vn = V_N \ket{00}^N$. The number of rotation gates required for preparation of the input state is $n$ (Section \ref{ssec:qcircs_csfs_bondbreaking}).
Thus, the total number of rotation gates for $V_N$, where $n = N/2$ is
\begin{equation}
	R = 2R_{S_{N/2}} + R_{\mathrm{in}, N} = n(n-1) + n = n^2 = \frac{1}{4}N^2.
	\label{eqn:nrot}
\end{equation}
The total (average) Toffoli cost thus becomes
\begin{equation}
	T = R \times T_R = R \big [ 0.2875  \lceil \log(R/\epsilon) \rceil+ 4.6 \big].
	\label{eqn:cost_toffolis_tot}
\end{equation}
The method requires only a single ancilla qubit to check if each rotation was implemented successfully.\cite{Bocharov2015}

Our goal is to use the spin eigenfunctions as initial states in fault-tolerant quantum algorithms. Thus, we must consider two sources of errors: the error due to state preparation $\epsilon_{\mathrm{SP}}$, and the error due to implementation of the quantum algorithm itself, e.g. quantum phase estimation, $\epsilon_{\mathrm{QPE}}$.
The total error will at worst be 
\begin{equation}
	\epsilon_{\mathrm{tot}} \leq \epsilon_{\mathrm{SP}} + \epsilon_{\mathrm{QPE}}.
\end{equation}

Since our circuits for initial state preparation are very efficient, a reasonable strategy would be to allocate most of the error to the quantum algorithm itself, rather than the state preparation task. For example, in the context of quantum phase estimation, one could choose $\epsilon_{\mathrm{tot}} = 0.0016$ (in Hartree atomic units) to achieve chemical accuracy in the energy estimation. We divide this into $\epsilon_{\mathrm{SP}} = 10^{-7}$ and $\epsilon_{\mathrm{QPE}} = \epsilon_{\mathrm{tot}} - 10^{-7}$. Inserting these values and Eq.~\eqref{eqn:nrot} into Eq.~\eqref{eqn:cost_toffolis_tot}, we get the Toffoli counts for preparation of $\vn$ states reported in Table~\ref{tab:n_cnots_csfs}. Even for the largest systems with $N=34$, the cost is only $T = 3989 \approx 4 \times 10^3$. This is indeed very low compared to the typical cost of quantum phase estimation, which is in the order of $10^{10}$ using state-of-the-art techniques.\cite{Lee2021, vonBurg2021}


Note that we have also considered the elegant phase gradient technique from Ref.~\citenum{Gidney2018} as an alternative to rotation gate synthesis. While we found that the Toffoli cost per rotation can be slightly lower depending on the choice of $\epsilon$, the advantage is washed away due to the one-off cost of preparing the phase gradient state (which in itself requires rotation gate synthesis), even for the largest system we consider, where $N=34$.

\subsection{Controlling state preparation circuits}\label{ssec:controlling}
In some quantum algorithms, e.g. when preparing initial states for VQE or phase estimation at intermediate bond lengths, it is necessary to prepare linear combinations of spin eigenfunctions (Section \ref{ssec:circs_lc}). This can be achieved by controlling the circuits for preparation of a single CSF of the form $\ket{\mathcal{O}_{0, 0}^{N, 1}}$.
It does not affect the asymptotic scaling with $N$ but introduces a constant factor overhead. The overhead is small because one only needs to control a small part of the circuit for input state preparation, $U_{\textrm{in}, N}$, as well as the circuit for $U_{\textrm{map}}$.

To control the circuit for input state preparation and the Dicke circuits $S_{N/2}$, it is sufficient to simply control the single-qubit Pauli $X$-gates at the beginning of $U_{\textrm{in}, N}$, as well as the first single-qubit rotation gate $R_y(\theta_{N/2}$). If the control qubit is in $0$, the remaining part of the circuit $U_{\textrm{in}, N}$ and the entire circuit $S_{N/2} \otimes S_{N/2}$ act like the identity, because the CNOT and rotation gates therein have no effect, and thus this effectively controls the entire circuit.
Therefore the only overhead for controlling this is $n=N/2$ CNOT gates, as well as the cost of converting the $R_y(\theta_n)$ to a $CR_y(\theta_n)$ gate. The $CR_y(\theta_n)$ can be implemented with 2 CNOTs and two $R_y$ gates (Figure \ref{fig:cry_decomposed}) and therefore has an overhead of $2$ CNOT gates and one $R_y$ gate.

Finally, consider the circuit $U_{\textrm{map}}$, consisting of $N$ CNOT gates of the form $\overline{CX}^{i, j} = X_i CX^{i, j} X_i$. This can be controlled by simply controlling the two $X$ gates and therefore requires $2N$ CNOTs gates. The overhead (cost to add on top of state preparation cost without controlled qubits) is $3N-N = 2N$.
The total CNOT overhead becomes:
\begin{equation}
	C_{\textrm{ctrl}} = \Big(\frac{N}{2} + 2\Big) + 2N = \frac{5}{2}N + 2.
	\label{eqn:cnot_cost_controlvn}
\end{equation}

In fault-tolerant hardware, the dominant cost comes from implementing the non-Clifford (Toffoli) gates, not CNOT gates. As discussed in Appendix \ref{apdx:circ_cost_faulttolerant}, rotations controlled by arbitrary qubits only have non-Clifford overhead compared to uncontrolled rotations.\cite{Sanders2020} Therefore, there is no Toffoli overhead for controlling any part of the state preparation circuit $V_N$.

\subsection{Other spin eigenfunctions}\label{apdx:circ_cost_different_csfs}
We briefly discuss the cost for preparing the state
\begin{equation}
	\ket{\mathcal{O}_{0, 0}^{N, 2}} = \Bigg[ \frac{1}{\sqrt{2}} (\ket{\al \be} - \ket{\be \al}) \Bigg]^{N/2}.
	\label{eqn:csf_geminals_appendix}
\end{equation}
(Section \ref{ssec:circuits_generalization}, Eq. \ref{eqn:csf_geminals}).
Working entirely in the spin space, where this maps to the qubit state 
\begin{equation}
	\Bigg[ \frac{1}{\sqrt{2}} (\ket{1001} - \ket{0110}) \Bigg]^{N/2},
\end{equation}
state preparation only requires one CNOT per $two$-electron singlet, therefore $N/2$ CNOTs. Two additional CNOT gates are required to implement the spin-to-Fock space mapping $U_{\textrm{map}}$. Therefore the total CNOT cost assuming all-to-all connectivity is $C_{\textrm{tot}}^{\textrm{all}} = N/2 + N = 3/2N$.

This is the same for planar connectivity, $C_{\textrm{tot}}^{\textrm{pla}} = C_{\textrm{tot}}^{\textrm{all}}$ if we carefully align the spin-orbitals on the qubit grid using the strategy in Appendix \ref{ssec:circcost_planar}. For linear connectivity, we can choose the qubit ordering $\{ \al_1, \be_1, \al_2, \be_2, ..., \al_N, \be_{N},  \}$ to minimize the cost of $U_{\textrm{map}}$ to $2 \times (N/2)$. The CNOT between orbitals $\al_i$ and $\al_{i+1}$, which are separated by the $\be_i$ qubit, can be decomposed into $3$ CNOTs using the CNOT accordion (Fig. \ref{fig:cnot_accordion_decomposed}). The total cost for linear connectivity thus becomes
\begin{equation}
	C_{\textrm{tot}}^{\textrm{lin }} = \frac{5}{2} N.
\end{equation}

Controlling these circuits only requires controlling the $R_y(-\theta/2)$ rotation for each two-electron singlet, which costs two CNOTs (Fig.~\ref{fig:cry_decomposed}). Adding the cost of controlling $U_{\textrm{map}}$, the CNOT overhead for controlling the preparation of $\ket{\mathcal{O}_{0, 0}^{N, 2}}$ is $N/2+N = 3/2N$.

On a fault-tolerant device, $R_y(-\theta/2)$ can be factorized into $HX$, therefore the state preparation and its controlled version can be implemented entirely with Clifford gates.

\section{Energy and gradients in Nonorthogonal VQE}
\label{apdx:novqe}

The nonorthogonal VQE (NO-VQE) algorithm variationally optimizes the energy of a wavefunction corresponding to
the linear combination defined in Eq.~\eqref{eq:novqewfn}.
A simultaneous optimization of the gate parameters $\bTh = (\bm{\theta}_1,\dots,\bm{\theta}_M)$ and linear coefficients
$\bm{C}$ requires the gradient of the energy with respect to each variable. 
In what follows, we define the correlated basis states as 
\begin{equation}
\ket{\Psi_I(\bm{\theta}_{I})} = \prod_i U_{Ii}(\theta_{Ii}) \ket{\Phi_I},
\label{eq:correlatedbasis}
\end{equation}
such that the full wavefunction is given by 
\begin{equation}
\ket{\Psi(\bTh,\bC)} = \sum_{I=1}^{L} C_I \ket{\Psi_I(\bth_{I})}.
\label{eq:simplernovqe}
\end{equation}
Since the states $\{ \ket{\Psi_I(\bth_{I})} \}$ are not mutually orthogonal, the VQE optimization 
requires gradients of the energy expectation value
\begin{equation}
E(\bTh,\bC) = 
\frac{\braket{\Psi(\bTh,\bC) | \hat{H} | \Psi(\bTh,\bC)}}
{\braket{\Psi(\bTh,\bC) | \Psi(\bTh,\bC)}},
\end{equation}
which are obtained through the quotient rule as 
\begin{equation}
\frac{\partial E}{\partial \bTh} = 
2 \qty[ \mel*{ \partial_{\bTh} \Psi }{ \hat{H} }{\Psi} - E \braket{ \partial_{\bTh} \Psi | \Psi} ],
\end{equation}
and likewise for $\frac{\partial E}{\partial \bC}$.
Here, $\partial_{\bTh}\equiv \frac{\partial}{\partial \bTh}$ and we have exploited the Hermitian symmetry, e.g.\
$\mel*{\Psi }{ \hat{H} }{ \partial_{\bTh} \Psi} = \mel*{\partial_{\bTh} \Psi }{ \hat{H} }{ \Psi}$.

Using Eq.~\eqref{eq:simplernovqe}, explicit expressions for the derivatives with respect
to the linear coefficients can be obtained as
\begin{equation}
\begin{split}
\frac{\partial E}{\partial C_I}
&=
2 \qty[ \braket{\Psi_I(\bth_{I}) | \hat{H} | \Psi(\bTh,\bC) } - \braket{\Psi_I(\bth_{I}) | \Psi(\bTh,\bC) }],
\\
&=
2 \sum_{J=1}^{M} \qty[ H_{IJ} - S_{IJ}] C_J
\end{split}
\end{equation}
where the Hamiltonian and overlap coupling elements are
\begin{subequations}
\begin{align}
H_{IJ}  &= \braket{\Psi_I(\bth_{I}) | \hat{H} | \Psi_J(\bth_{J}) },
\\
S_{IJ}  &= \braket{\Psi_I(\bth_{I}) | \Psi_J(\bth_{J}) },
\end{align}
\label{eq:SH}
\end{subequations}
and we assume that the linear coefficients $\bm{C}$ are real valued.
These derivatives may be computed using the established circuits for measuring nonorthogonal coupling
terms outlined in Ref.~\citenum{Huggins2020b}, which are also used in subspace diagonalization approaches.
Similarly, using Eq.~\eqref{eq:correlatedbasis}, the derivatives with respect to the gate
parameters are
\begin{equation}
\begin{split}
\frac{\partial E}{\partial \theta_{Ii} }
=
2\, C_I \sum_J \Big[ &\braket{\partial_{\theta_{Ii}} \Psi_I(\bth_{I}) | \hat{H} | \Psi_J(\bth_J) } 
\\
        &- \braket{\partial_{\theta_{Ii}} \Psi_I(\bth_{I}) | \Psi_J(\bth_J) } \Big] C_J,
\end{split}
\end{equation}
where the partial derivative of the wavefunction is 
\begin{equation}
\ket{\partial_{\theta_{Ii}} \Psi_I(\bth_{I})} 
=
\prod_{j<i} U_{Ij}(\theta_{Ij}) 
\,
\frac{\partial U_{Ii}(\theta_{Ii})}{ \partial \theta_{Ii}} 
\,
\prod_{k>i} U_{Ij}(\theta_{Ij}) \ket{\Phi_I}.
\end{equation}
For the QNP \textit{ansatz}, the circuits for these partial derivatives can be constructed using the parameter shift rules detailed in Ref.~\citenum{Anselmetti2021}.
Therefore, the coupling terms $\mel{\partial_{\theta_{Ii}} \Psi_I(\bth_{I}) }{ \hat{H} }{ \Psi_J(\bth_J) }$ and $\braket{\partial_{\theta_{Ii}} \Psi_I(\bth_{I}) | \Psi_J(\bth_J) }$ can be evaluated with the same circuit architecture used to evaluate the Hamiltonian and overlap terms in Eq.~\eqref{eq:SH} with a constant prefactor.

With these gradient expressions, the NO-VQE algorithm proceeds using the standard L-BFGS optimization approach.
The initial linear coefficients $\bm{C}$ are obtained by solving the generalized eigenvalue problem with $\bTh = 0$.
Although optimizing the expectation value of the energy means that the linear expansion [Eq.~\eqref{eq:simplernovqe}] 
does not need to be normalized, we obtain more stable optimization by normalizing $\bm{C}$ on each iteration.

\section{Matrix elements for QSD based on real-time evolution}\label{sec:vqpe_toeplitz}

For simplicity, we restrict the analysis below to the case where we only time-evolve a single reference state $\ket{\Psi_0}$, and note that the conclusions remain unchanged for cases with multiple reference states.
Choosing a linear time grid, $t_j = j \Delta t$ with $j = 0, 1, ..., N_T$, we form a subspace of $M = N_T+1$ states. The overlap matrix elements between the time-evolved states forming the subspace are
\begin{equation}
	S_{j, k} = \braket{\Psi_j |\Psi_k} = \braket{\Phi_0| e^{-iH\Delta t (k-j)} |\Phi_0}.
\label{eqn:ovlp_vqpe}
\end{equation}
Replacing the Hamiltonian operator with the time-evolution operator $U(\Delta t) := e^{-iH\Delta t}$, we can write $S_{j, k} = \braket{\Phi_0| [U(\Delta t)]^{k-j}|\Phi_0}$.
The matrix elements of $U(\Delta t)$ in the basis of expansion states are:
\begin{equation}
	\begin{split}
	U_{j, k} &= \braket{\Phi_j| U(\Delta t)|\Phi_k} = \braket{\Phi_0 |e^{-iH\Delta t(k-j+1)} |\Phi_0} 
\\ &= S_{j, k+1} = S_{j-1, k}.
	\end{split}
\label{eqn:toeplitz}
\end{equation}

These expressions show that, if we reformulate the eigenvalue problem in Eq.~\eqref{eqn:gen_eval} to use the time-evolution operator $U(\Delta t)$ rather than the Hamiltonian, the matrix elements $U(\Delta t)_{j, k}$ become equivalent to the overlap matrix elements (shifted by one row or column), which has a Toeplitz structure.
This equivalence is advantageous for implementations on quantum hardware, as it only requires measuring the overlap matrix rather than separately measuring the Hamiltonian and overlap (intuitively, the matrix elements in Eq.~\eqref{eqn:toeplitz} correspond to an autocorrelation function).\cite{Parrish2019b, Klymko2022} The number of matrix elements is reduced from $2M^2$ to $M+1$. This result holds even when the time-evolution operator is Trotterized, as long as the time grid is linear.\cite{Klymko2022}

\section{Quantum circuits for preparation of matrix product states}\label{apdx:mps}
Here, we discuss the cost of known quantum circuits for preparation of matrix product states (MPS).
An MPS of $M$ sites has the form
\begin{equation}
	\ket{\Psi} = \sum_{\bm{m}} A_{1}^{[m_1]} A_{2}^{[m_2]} \cdots A_{M}^{[m_M]} \prod_{i=1}^M\ket{m_i}.
\end{equation}
Here, $m_i \in \{ 0, 1, ..., d-1 \}$ is the physical index that runs over the possible occupations of the Hilbert space of a single site (of \textit{local} dimension $d$), and the vector $\bm{m} = (m_1, m_2, ..., m_M)$ defines the occupation numbers for each site. In the case of quantum chemistry, each site can be mapped to a spatial orbital, therefore the site's Hilbert space dimension is $d=4$, and $M$ is the number of spatial orbitals.\cite{Chan2011}. Each $A_i^{[m_i]}$ is a tensor of order $\chi_i$, where $\chi_i$ is the bond dimension, and each bond dimension is bounded by the maximal bond dimension, $\chi_i \leq \chi$.

The preparation of an MPS on a quantum computer can be achieved using the sequential method by Sch{\"o}n \textit{et al.}\cite{Schon2005} This scales as $\mathcal{O}(M)$ in the number of gates and has depth $\mathcal{O}(M)$. Although an improved scaling of depth $\mathcal{O}(\mathrm{log}(M))$ can be reached through the recent technique by Malz \textit{et al.} \cite{Malz2024}, which is provably optimal,\cite{Malz2024} this only applies to MPS with short-range correlations.
For long-range correlated states such as GHZ states, it has been proven that state preparation circuits of depth $M$ are optimal based on Lieb-Robinson bounds.\cite{Bravyi2006, Malz2024}
Since we do not expect any area-laws to apply to the entanglement of most molecular eigenstates,\cite{Chan2012} we must consider the method for general MPS.\cite{Schon2005}

The scaling of the deterministic method in Ref. \citenum{Schon2005} is also polynomial in the bond dimension. Specifically, for generic MPS, each circuit contains $M$ gate blocks, one for each of the tensors $A^{[m_i]}_i$. Each gate block is a multi-qubit unitary acting on $n_a = \log(\chi_i)$ ancilla qubits (corresponding to the virtual Hilbert space) and one physical qubit. Decomposing a block into single and two-qubit gates requires a circuit with depth exponential in $n_a$, or $\mathcal{O}(\chi_i)$.\cite{Malz2024, BenDov2024, Rudolph2023}
Therefore, the total gate complexity is $\mathcal{O}(M \chi^2)$, which is proportional to the scaling of the dimension of the overall tensor.\cite{Chan2011} Although approximate methods might ameliorate the scaling,\cite{Ran2020, BenDov2024, Rudolph2023} it is unclear if they can retain sufficient accuracy and their error cannot be theoretically bound.\cite{BenDov2024}

Formichev \textit{et al.} provided a detailed derivation of the cost and a more explicit circuit implementation of the MPS state preparation circuits from Ref. \citenum{Schon2005} using modern quantum linear algebra techniques.\cite{Formichev2023}
This implementation has Toffoli cost of $\chi_{i-1}[8\chi_i d + b \log(\chi_i d) + \log(\chi_i d)]$ for each block.
Assuming $\chi_i = \chi$ for all $i$, we obtain the following approximate cost for the entire MPS preparation circuit:
\begin{equation}
	\begin{split}
	T_{\mathrm{MPS}} = \sum_{i = 2}^M \chi_{i-1}[8\chi_i d + b \log(\chi_i d) + \log(\chi_i d)]	\\
	 \approx (M-1)\chi [32 \chi + (b+1)\log(4\chi)].
	\end{split}
\end{equation}

\section{Computational details}
We obtained the Hamiltonians using PySCF\cite{Sun2018, Sun2020} and used Openfermion\cite{McClean2020} to define the operator matrices. 
We performed the NO-VQE and ADAPT-VQE calculations using a developmental version of GMIN.\cite{gmin}
We developed in-house Python code for all other tasks including generation of the Configuration State Functions, real-time evolution, and subspace diagonalization.

\bibliographystyle{apsrev4-2-author-truncate-with-titles}

\bibliography{export}

\end{document}